\documentstyle[12pt,rotate]{article}

\input epsf.sty

\newcommand{\insertfig}[2]{\mbox{\epsfxsize=#1cm \epsfbox{#2.eps}}}

\newcommand{\CA}{\mbox{CA}}
\newcommand{\CoA}{\mbox{CoA}}
\newcommand{\CeA}{\mbox{CeA}}
\newcommand{\SSA}{\mbox{SSA}}
\newcommand{\Bx}{x_{\rm B}}
\newcommand{\cQ}{{\cal Q}}
\newcommand{\GeV}{\mbox{GeV}}
\newcommand{\ft}[2]{{\textstyle\frac{#1}{#2}}}

\begin{document}

\begin{titlepage}

\begin{flushright}
DOE/ER/40762-009 \\ [-2mm]
UMD-PP\#02-011 \\ [-2mm]
YITP-SB-01-51
\end{flushright}

\centerline{\large\bf Theory of deeply virtual Compton scattering on the nucleon}

\vspace{10mm}

\centerline{\bf A.V. Belitsky$^{a,b,c}$, D. M\"uller$^{c,d}$, A. Kirchner$^d$}

\vspace{10mm}

\centerline{\it $^a$Department of Physics}
\centerline{\it University of Maryland at College Park}
\centerline{\it MD 20742-4111, College Park, USA}

\vspace{5mm}

\centerline{\it $^b$C.N.\ Yang Institute for Theoretical Physics}
\centerline{\it State University of New York at Stony Brook}
\centerline{\it NY 11794-3840, Stony Brook, USA}

\vspace{5mm}

\centerline{\it $^c$Fachbereich Physik, Universit\"at  Wuppertal}
\centerline{\it D-42097 Wuppertal, Germany}

\vspace{5mm}

\centerline{\it $^d$Institut f\"ur Theoretische Physik,
                Universit\"at Regensburg}
\centerline{\it D-93040 Regensburg, Germany}

\vspace{10mm}

\centerline{\bf Abstract}

\vspace{0.3cm}

\noindent

We compute the cross section for leptoproduction of the real photon off the
nucleon, which is sensitive to the deeply virtual Compton scattering
amplitude with power accuracy. Our considerations go beyond the leading
twist and involve the complete analysis in the twist-three approximation. We
discuss consequences of the target and lepton beam polarizations for
accessing the generalized parton distributions from experimental
measurements of the azimuthal angular dependence of the final state photon
or nucleon. We introduce several sets of asymmetries, defined as Fourier
moments with respect to the azimuthal angle, which allow for a clear
separation of the twist-two and -three sectors. Relying on a simple ansatz
for the generalized parton distributions, we give quantitative estimates for
azimuthal and spin asymmetries, discuss the uncertainties of these
predictions brought in by radiative corrections, and compare them with
experimental data as well as other theoretical expectations. Furthermore, we
derive a general parametrization of the DVCS amplitudes in the region of
small Bjorken variable.

%\vspace{0.5cm}

%\noindent Keywords: deeply virtual Compton scattering, twist-three effects,
%asymmetries, generalized parton distribution

%\vspace{0.3cm}

%\noindent PACS numbers: 11.10.Hi, 12.38.Bx, 13.60.Fz

\end{titlepage}

\tableofcontents

%%%%%%%%%%%%%%%%%%%%%%%%%%%%%%%%%%%%%%%%%%%%%%%%%%%%%%%%%%%%%%%%%%%%%
\section{Introduction}
%%%%%%%%%%%%%%%%%%%%%%%%%%%%%%%%%%%%%%%%%%%%%%%%%%%%%%%%%%%%%%%%%%%%%

The complicated long distance dynamics of hadron constituents, not
controllable by well-defined methods of perturbative QCD, was studied over
decades by means of experimental measurements with inclusive lepton-hadron
and hadron-hadron scattering. Information on the nucleon's parton structure,
originating from them, can be used to access inclusive properties of
hadrons, like parton distributions. However, they are
insufficient to constrain the detailed picture of the hadron wave function
in full diversity of its manifestations.

The parton distributions and correlation functions encode the long-distance
dynamics of inclusive cross sections. The former accompany the interaction
of a hard probe with a single parton, while the latter are involved in the
rescattering off a multi-parton system. Both of them are given by the
interference of light-cone wave functions with equal or different numbers of
constituents in the amplitude and its conjugate, but the same total momentum
flowing through each of them. This obviously simplifies the theoretical
description on the one hand and, however, lacks a plethora of important
features of strong interaction dynamics on the other one.

As has been recently understood, a straightforward generalization of the
aforementioned parton densities via new nonperturbative nucleon
characteristics arises in exclusive two-photon processes in the generalized
Bjorken region \cite{MulRobGeyDitHor94}, e.g., in the Compton scattering
with highly virtual incoming $\gamma$-quantum \cite{Ji96,Ji98,Rad96}, and in
the hard photoproduction of mesons \cite{ColFraStr96}. These are genuinely
exclusive processes. Here one finds off-forward matrix elements, as
distinguished from the forward ones in inclusive reactions, of certain
non-local quark and gluon composite operators, which give rise to the
so-called generalized parton distribution (GPDs). They carry information on
parton distributions, distribution amplitudes, and form factors and are of
interest in their own right.  Moreover, they can shed some
light on the intriguing spin structure of the proton as they provide a way
to get access to the parton's angular momentum unreachable elsewhere
\cite{Ji96}.

The deeply virtual Compton scattering (DVCS) is the hard photoproduction of
a real photon, i.e., $\gamma^\ast N \to \gamma N'$. Being a process involving a
single hadron, it is one of the cleanest tools in the problem of
constraining GPDs from data. The theoretical efforts and achievements are
supported by pioneering experimental results from HERMES \cite{Aip01}, HERA
\cite{Sau00,Adl01}, and CLAS \cite{Ste01} collaborations, and encouraging
future plans \cite{SteBurEloGar01,Chen00,Nowak01}.

This paper is a culmination of our previous studies initiated in Ref.\
\cite{BelKirMulSch01b} and completes the theoretical background of the
theory of DVCS by an exhaustive set of analytical results for the cross
section to the twist-three accuracy for all possible hadron and lepton
polarizations involved. Numerical studies performed here are done in
the so-called Wandzura-Wilczek (WW) approximation for the twist-three
GPDs, developed by two of us in Ref.\ \cite{BelMul00b}. Our present
consideration is compatible with the one performed earlier in a fully
numerical manner \cite{KivPolVan00}, see also Ref.\ \cite{GoePolVan01}. We also
comment on the effect of multi-particle correlations, provided we cease
using the WW relations. The knowledge of power corrections to the leading
twist-two results, with the twist-three terms being the first of them, is
indispensable because the hard momentum of the incoming photon is not in the
deep Euclidean domain for the present experiments. Therefore, the
higher-twist contributions may affect the understanding of lowest order
results. There is a number of theoretical uncertainties in phenomenological
predictions for DVCS. Apart from the basically unconstrained shape of GPDs,
which one models on the ground of previous theoretical considerations, these
are higher-twist and -order effects, which alter the handbag approximation.
The former include the dynamical and target mass corrections, see
\cite{BelMul01a} for a partial result. While the latter have been computed
in next-to-leading order (NLO) of perturbation theory in the strong coupling
constant, see Refs.\ \cite{BelMul97a,JiOsb98,ManPilSteVanWei97} for coefficient
functions and Refs.\ \cite{BelMul98,BelFreMul00} for evolution kernels.
Their numerical significance was studied in Refs.\
\cite{BelMulNieSch99,FreMcd01a,FreMcd01b}. We will elaborate on these issues
in the present study as well.

Our presentation is organized as follows. The next section recapitulates
basic results for the DVCS amplitude to the twist-three accuracy and
introduces the WW relations for the twist-three quark GPDs. In section
\ref{GluonTransversity} we briefly discuss the gluon double helicity-flip
distributions, vanishing in the forward scattering off spin-1/2 targets.
Their peculiar properties propagate into a genuinely distinguishable angular
dependence in physical observables. This allows to access them,
uncontaminated by leading twist quark GPDs. The differential
leptoproduction cross section is extensively analyzed with power accuracy
in section \ref{Sec-AziAngDep} for all polarization options of the lepton
beam and target. Based on these results, we discuss a way to unravel GPDs
from cross sections and then turn to appropriate definitions of the most
favorable physical observables, i.e., asymmetries. These asymmetries allow
to extract separate components of the angular dependence of the cross
section, and in this manner to project out distributions carrying
information on the orbital momentum of constituents in the nucleon. In
section \ref{Sec-CFF}, we address the models for GPDs and discuss their
diverse properties. We then study the magnitude of radiative corrections to the
Compton form factors (CFFs). Next in section \ref{Sec-NumEst}, fitting the
model parameters to the recent H1 data in the small Bjorken-variable region,
we give phenomenological predictions for the HERMES and CLAS kinematics.
Finally, we conclude.

%%%%%%%%%%%%%%%%%%%%%%%%%%%%%%%%%%%%%%%%%%%%%%%%%%%%%%%%%%%%%%%%%%%%%
\section{DVCS amplitude in twist-three approximation}
%%%%%%%%%%%%%%%%%%%%%%%%%%%%%%%%%%%%%%%%%%%%%%%%%%%%%%%%%%%%%%%%%%%%%

Let us give a few basic definitions required for understanding of the Compton
amplitude with power accuracy. Here we mainly recapitulate the results of
our previous studies, where we have introduced the WW-approximation
\cite{BelMul00b} for the twist-three GPDs, and used it for preliminary
studies of appropriate physical observables \cite{BelKirMulSch01b}.

The DVCS hadronic tensor is given by the time-ordered product of the
electromagnetic currents $j_\mu = e \sum_i Q_i \bar\psi_i \gamma_\mu \psi_i$
of quarks, having fractional charge $Q_i$, which is sandwiched
between hadronic states with different momenta. In leading order (LO) of
perturbation theory it reads \cite{BelMul00b}
\begin{eqnarray}
\label{HadronicTensor}
T_{\mu\nu} (q, P, \Delta)
\!\!\!&=&\!\!\!
\frac{i}{e^2} \int dx {\rm e}^{i x \cdot q}
\langle P_2 | T j_\mu (x/2) j_\nu (-x/2) | P_1 \rangle \\
&=&\!\!\!
- {\cal P}_{\mu\sigma} g_{\sigma\tau} {\cal P}_{\tau\nu}
\frac{q \cdot V_1}{P \cdot q}
+ \left( {\cal P}_{\mu\sigma} P_\sigma  {\cal P}_{\rho\nu}
+ {\cal P}_{\mu\rho}  P_\sigma {\cal P}_{\sigma\nu} \right)
\frac{V_{2\, \rho}}{P \cdot q}
- {\cal P}_{\mu\sigma} i\epsilon_{\sigma \tau q \rho} {\cal P}_{\tau\nu}
\frac{A_{1\, \rho}}{P \cdot q}
\, , \nonumber
\end{eqnarray}
where we have kept all contributions up to twist-three accuracy. The
three independent four-momenta are $P = P_1 + P_2$, $\Delta = P_2 - P_1$,
and $q = (q_1 + q_2)/2$, where the vectors $P_1$ ($q_1$) and $P_2$ ($q_2$)
refer to the incoming and outgoing proton (photon) momentum, respectively.
The decomposition (\ref{HadronicTensor}) is similar to the one used in
deeply inelastic scattering. Indeed, the twist-two part of the generalized
structure functions $V_1$ and $A_1$ corresponds to the conventional one
$F_1$ and $g_1$ (see the first paper of $\cite{BelKirMulSch01b}$). The
current conservation in the tensor decomposition (\ref{HadronicTensor}) is
ensured by means of the projection operator
\begin{equation}
{\cal P}_{\mu\nu} = g_{\mu\nu} - \frac{q_{1 \mu} q_{2 \nu}}{q_1 \cdot q_2}
\, .
\end{equation}
This is consistent with an explicit calculation of the amplitude
(\ref{HadronicTensor}) to twist-three accuracy
\cite{PenPolShuStr00,BelMul00b}, see also Refs.\ \cite{AniPirTer00,RadWei00}
for spinless targets.

In Eq.\ (\ref{HadronicTensor}) we have used conventional conditions on the
kinematical invariants in the Bjorken limit, $- q^2 \sim P \cdot q =$ large,
$\Delta^2 =$ small, a generalized Bjorken variable $\xi = -q^2/ q \cdot P =$
fixed. If both photons are virtual, we would have an extra scaling variable
$\eta = \Delta \cdot q/P \cdot q$, the skewedness \cite{MulRobGeyDitHor94}.
The reality of the outgoing photon implies the presence of only one scaling
variable $\xi$, namely, for $q_2^2 = 0$ one has
\begin{eqnarray*}
\eta = - \xi \left( 1 + \frac{\Delta^2}{2 \cQ^2} \right)^{-1}
\, .
\end{eqnarray*}
It is convenient to introduce variables closely related  to the ones
used in deeply inelastic scattering, namely, the incoming photon virtuality
and conventional Bjorken variable
\begin{equation}
{\cal Q}^2 \equiv -q_1^2
\, , \qquad
\Bx \equiv \frac{\cQ^2}{2 P_1 \cdot q_1} \, ,
\end{equation}
with the latter related to $\xi$ via
\begin{eqnarray}
\xi = \Bx
\frac{ 1 + \frac{\Delta^2}{2 \cQ^2}
}{
2 - \Bx + \Bx \frac{\Delta^2}{\cQ^2} }
\, .
\end{eqnarray}
In the above Eq.\ (\ref{HadronicTensor}), $V_{2 \, \rho}$ is
expressed in terms of the vector $V_{1 \, \rho}$ and axial-vector
$A_{1 \, \rho}$ functions\footnote{
We use the conventions for Dirac and Lorentz tensors from Itzykson and Zuber
\cite{ItzZub80}, e.g., $\epsilon^{0123}= +1$.},
\begin{eqnarray}
\label{V2}
V_{2 \, \rho} = \xi V_{1 \, \rho} - \frac{\xi}{2}
\frac{P_\rho}{P\cdot q} q\cdot V_{1} + \frac{i}{2}
\frac{\epsilon_{\rho\sigma\Delta q}}{P\cdot q} A_{1 \, \sigma}
\, .
\end{eqnarray}
The amplitudes $V_1$ and $A_1$ depend on the scaling variable $\xi$,
momentum transfer $\Delta^2$, and hard momentum of the probe ${\cal Q}^2$.
The latter dependence is logarithmic and is governed by a generalized
off-forward evolution equation \cite{BukFroKurLip85,GeyDitHorMueRob88},
which is presently known to two-loop order \cite{BelMul98,BelFreMul00}. In
order to simplify notations considerably, we will drop the dependence on
$\Delta^2$ and ${\cal Q}^2$ when it is not essential for the presentation
and restore them in sections dealing with the modeling of GPDs and numerical
estimates.

A general decomposition of the vector and axial-vector amplitudes, in a complete
basis of CFFs to  twist-three accuracy, reads
\begin{eqnarray}
\label{Def-V1}
V_{1\, \rho}\!\!\!&=&\!\!\!
P_\rho  \frac{q\cdot h}{q\cdot P} {\cal H}
+
P_\rho  \frac{q\cdot e}{q\cdot P} {\cal E}
+
\Delta_\rho^\perp  \frac{q\cdot h}{q\cdot P}  {\cal H}^3_+
+
\Delta_\rho^\perp  \frac{q\cdot e}{q\cdot P}  {\cal E}^3_+
+
\widetilde{\Delta}_\rho^\perp
\frac{q\cdot \widetilde{h}}{q\cdot P} \widetilde{{\cal H}}^3_-
+
\widetilde{\Delta}_\rho^\perp
\frac{q\cdot \widetilde{e}}{q\cdot P} \widetilde{{\cal E}}^3_-
\, ,\\
\label{Def-A1}
A_{1\, \rho}\!\!\!&=&\!\!\!
P_\rho  \frac{q\cdot \widetilde h}{q\cdot P} \widetilde{\cal H}
+
P_\rho  \frac{q\cdot \widetilde e}{q\cdot P} \widetilde{\cal E}
+
\Delta_\rho^\perp  \frac{q\cdot \widetilde h}{q\cdot P}
{\cal \widetilde H}^3_+
+ \Delta_\rho^\perp  \frac{q\cdot \widetilde e}{q\cdot P}
{\cal \widetilde E}^3_+
+
\widetilde{\Delta}_\rho^\perp  \frac{q\cdot h}{q\cdot P}
{\cal H}^3_-
+
\widetilde{\Delta}_\rho^\perp  \frac{q\cdot e}{q\cdot P}
{\cal E}^3_-
\, ,
\end{eqnarray}
where $\Delta_\rho^\perp \equiv \Delta_\rho + \xi P_\rho$,
$\widetilde\Delta_\rho^\perp \equiv i \epsilon_{\rho\Delta P q}/{P\cdot q}$,
and the Dirac bilinears are conventionally defined  by
\begin{eqnarray}
&& h_{\rho}
= \bar U (P_2, S_2) \gamma_\rho U (P_1, S_1) \, ,
\hspace{1.2cm}
e_{\rho} =
\bar U (P_2, S_2) i \sigma_{\rho\sigma} \frac{\Delta_\sigma}{2 M} U (P_1, S_1)
\, , \nonumber\\
&&\widetilde{h}_\rho
= \bar U (P_2, S_2) \gamma_\rho \gamma_5  U (P_1, S_1) \, ,
\qquad
\widetilde e_{\rho}
= \frac{\Delta_\rho}{2 M} \bar U (P_2, S_2) \gamma_5  U(P_1, S_1)
\, ,
\label{DiracBilinears}
\end{eqnarray}
with $U$ being the nucleon bispinor normalized as $\bar U (P_1) U (P_1) = 2 M$.

The CFFs introduced in Eqs.\ (\ref{Def-V1},\ref{Def-A1}) are given by a
convolution of perturbatively calculable coefficient functions $C^{(\pm)}$
and a set of twist-two and -three GPDs via
\begin{eqnarray}
\label{DefTw3}
&&\left\{
{\cal H}
,
{\cal E}
,
{\cal H}^3_+
,
{\cal E}^3_+
,
\widetilde {\cal H}^3_-
,
\widetilde {\cal E}^3_-
\right\}
(\xi)
= \int_{- 1}^{1} \! dx \, C^{(-)} (\xi, x)
\left\{
H
,
E
,
H^3_+
,
E^3_+
,
\widetilde H^3_-
,
\widetilde E^3_-
\right\}
(x, \eta)_{|\eta=-\xi}
\, , \\
&&\left\{
\widetilde {\cal H}
,
\widetilde {\cal E}
,
\widetilde {\cal H}^3_+
,
\widetilde {\cal E}^3_+
,
{\cal H}^3_-
,
{\cal E}^3_-
\right\}
(\xi)
= \int_{- 1}^{1} \! dx \, C^{(+)} (\xi, x)
\left\{
\widetilde H
,
\widetilde E
,
\widetilde H^3_+
,
\widetilde E^3_+
,
H^3_-
,
E^3_-
\right\}
(x, \eta)_{|\eta=-\xi}
\, ,
\nonumber
\end{eqnarray}
where $1/\cQ^2$-power suppressed effects have been neglected, i.e., $\eta=-\xi$.
The GPDs $\{H,\dots,\widetilde{E}^3_- \}$ and $\{\widetilde{H},\dots,{E}^3_- \}$
are related to the
off-forward matrix elements of the vector and axial quark operators, see Eqs.\
(\ref{vectorGPD}) and (\ref{axialGPD}), respectively. We have implied that there
is a summation on the right-hand side of the above equations over the quark
species, so that they have to be understood as follows
\begin{equation}
\label{Flavour}
C^{(\mp)} F \to \sum_{i = u, d, s} C^{(\mp)}_i F_i \, ,
\end{equation}
with $C^{(\mp)}$ having perturbative expansion
\begin{eqnarray*}
C^{(\mp)} = C_{(0)}^{(\mp)} + \frac{\alpha_s}{2\pi} C_{(1)}^{(\mp)} + {\cal
O}(\alpha_s^2)\, .
\end{eqnarray*}
To reduce the number of equations we have introduced unifying conventions
$F$ and $F^3_{\pm}$ for all twist-two and -three functions, which run over
$H$, $E$, $\widetilde H$, and $\widetilde E$ species as in Eq.\
(\ref{DefTw3}). At LO in the QCD coupling constant the coefficient functions
read for the even ($-$) and odd ($+$) parity sectors
\begin{equation}
\label{CoeffFunction}
\xi\, {C_{(0)i}^{(\mp)}} \left( \xi, x \right)
= \frac{Q^2_i}{1 - x/\xi - i 0}
\mp
\frac{Q^2_i}{1 + x/\xi - i 0} \, .
\end{equation}

According to \cite{BelMul00b}, all twist-three GPDs are decomposed into the
so-called WW term $F^{WW}_\pm$ (later recalculated in Ref.\
\cite{KivPol00}) and a function $F^{qGq}_\pm $ that contains new dynamical
information arising from antiquark-gluon-quark correlations:
\begin{eqnarray}
F^3_\pm = F^{WW}_\pm + F^{qGq}_\pm \, .
\end{eqnarray}

The WW parts are expressed solely in terms of the twist-two functions $F =
\{ H, E, \widetilde H, \widetilde E\}$ and have the following form
\begin{eqnarray}
\label{Def-Ftw3}
&& F^{WW}_+ (x, \xi)
= \int_{-1}^{1} \frac{d y}{\xi} \,
W_+ \left( \frac{x}{\xi}, \frac{y}{\xi} \right)
\left(
y \frac{\stackrel{\leftarrow}{\partial}}{\partial y}
-
\xi \frac{\stackrel{\rightarrow}{\partial}}{\partial \xi}
\right)
F (y, \xi)
- \frac{1}{\xi} F (x, \xi)
-
\frac{4 M^2 \, F^\perp_+ (x, \xi)}
{(1 - \xi^2)(\Delta^2 - \Delta^2_{\rm min})}
\, ,
\nonumber\\
&& F^{WW}_- (x, \xi)
= - \int_{- 1}^{1} \frac{d y}{\xi} \,
W_- \left( \frac{x}{\xi}, \frac{y}{\xi} \right)
\left(
y \frac{\stackrel{\leftarrow}{\partial}}{\partial y}
-
\xi \frac{\stackrel{\rightarrow}{\partial}}{\partial \xi}
\right)
F (y, \xi) -
\frac{4 M^2 \, F^\perp_- (x, \xi)}
{(1 - \xi^2)(\Delta^2 - \Delta^2_{\rm min})}
\, ,
\end{eqnarray}
for the `$+$' and `$-$' component of the twist-three GPDs in the
WW-approximation, respectively. The minimum value of the momentum transfer
squared $\Delta_{\rm min}^2$ is given below in Eq.\ (\ref{Def-tmin}).
 The functions $F^\perp_\pm$ specifically
appear  for the spin-$1/2$ targets, cf.\ Refs.\
\cite{BelMul00b,KivPolSchTer00,RadWei00,BelKirMulSch00c,AniTer01,Anietal01}
for the pion target, and read
\begin{eqnarray}
H^\perp_\pm (x, \xi)
\!\!\!&=&\!\!\! \mp \frac{\Delta^2}{4 M^2} \int_{- 1}^{1} d y
\left\{ \xi W_\pm \left( \frac{x}{\xi}, \frac{y}{\xi} \right)
\left( H + E \right) (y, \xi)
- W_\mp \left( \frac{x}{\xi}, \frac{y}{\xi} \right)
\widetilde H (y, \xi) \right\} ,
\nonumber\\
E^\perp_\pm (x, \xi)
\!\!\!&=&\!\!\! \pm \int_{- 1}^{1} d y
\left\{ \xi W_\pm \left( \frac{x}{\xi}, \frac{y}{\xi} \right)
\left( H + E \right) (y, \xi)
- W_\mp \left( \frac{x}{\xi}, \frac{y}{\xi} \right)
\widetilde H (y, \xi) \right\},
\nonumber\\
\widetilde
H^\perp_\pm (x, \xi)
\!\!\!&=&\!\!\! \pm \int_{- 1}^{1} d y
\left\{ \xi \left( 1 - \frac{\Delta^2}{4 M^2} \right)
W_\pm \left( \frac{x}{\xi}, \frac{y}{\xi} \right)
\widetilde H (y, \xi)
+ \frac{\Delta^2}{4 M^2} W_\mp \left( \frac{x}{\xi}, \frac{y}{\xi} \right)
\left( H + E \right) (y, \xi) \right\} ,
\nonumber\\
\widetilde E^\perp_\pm (x, \xi)
\!\!\!&=&\!\!\! \pm \frac{1}{\xi} \int_{- 1}^{1} d y
\left\{ W_\pm \left( \frac{x}{\xi}, \frac{y}{\xi} \right)
\widetilde H (y, \xi)
- \xi W_\mp \left( \frac{x}{\xi}, \frac{y}{\xi} \right)
\left( H + E \right) (y, \xi) \right\} .
\end{eqnarray}
The $W$-kernels introduced in these formulae are defined by
\begin{equation}
W_{\pm} \left( \frac{x}{\xi}, \frac{y}{\xi} \right)
= \frac{1}{2 \xi}
\left\{
W \left( \frac{x}{\xi}, \frac{y}{\xi} \right)
\pm
W \left( - \frac{x}{\xi}, - \frac{y}{\xi} \right)
\right\},
\qquad
W(x, y)
= \frac{\theta (1 + x) - \theta (x - y)}{1 + y} \, .
\end{equation}

The antiquark-gluon-quark contributions,
\begin{eqnarray*}
\label{Def-FqGq}
F^{qGq}_\pm
\!\!\!&=&\!\!\!
- \int_{-1}^1 \frac{dy}{\xi}
\int_{-1}^1 du \frac{1-u}{2}
\left\{
W \left( -\frac{x}{\xi}, - \frac{y}{\xi} \right)
\frac{\stackrel{\leftarrow}{\partial^2}}{\partial y^2}
S_F^+(y, u, - \xi)
\pm
W \left( \frac{x}{\xi}, \frac{y}{\xi} \right)
\frac{\stackrel{\leftarrow}{\partial^2}}{\partial y^2}
S_F^-(y,-u,-\xi)
\right\} \, ,
\end{eqnarray*}
can be read off from the parametrization of the corresponding operators
and result in eight independent functions
\begin{eqnarray}
\label{Def-qGqOpe}
S_\rho^{\pm}
= \Delta_\rho^\perp  \frac{q\cdot h}{q\cdot P}
S_H^{\pm}
+
\Delta_\rho^\perp \frac{q\cdot e}{q\cdot P}
S_E^{\pm}
\pm
\widetilde{\Delta}_\rho^\perp  \frac{q\cdot \widetilde{h}}{q\cdot P}
S_{\widetilde H}^{\pm}
\pm
\widetilde{\Delta}_\rho^\perp  \frac{q\cdot \widetilde{e}}{q\cdot P}
S_{\widetilde E}^{\pm} \, ,
\end{eqnarray}
(see Ref.\ \cite{BelMul00b} for details).

It turns out that only the difference ${\cal F}_+^3 - {\cal F}_-^3$ enters
the DVCS amplitude
\cite{KivPolSchTer00,RadWei00,BelKirMulSch00c,BelKirMulSch01b}, where, again
similar to the previous definitions, the CFF ${\cal F}$ in Eq.\
(\ref{DefTw3}) stands for ${\cal H}$, ${\cal E}$, $\widetilde {\cal H}$, and
$\widetilde {\cal E}$ species. Therefore, only four new GPDs remain at
twist-three level. In the WW-approximation, i.e., neglecting the
antiquark-gluon-quark correlation, all the twist-three CFFs are entirely
determined by the four twist-two GPDs $H$, $E$, $\widetilde H$, and
$\widetilde E$.

%%%%%%%%%%%%%%%%%%%%%%%%%%%%%%%%%%%%%%%%%%%%%%%%%%%%%%%%%%%%%%%%%%%%%
\section{DVCS amplitude with gluon transversity}
\label{GluonTransversity}
%%%%%%%%%%%%%%%%%%%%%%%%%%%%%%%%%%%%%%%%%%%%%%%%%%%%%%%%%%%%%%%%%%%%%

Apart from the conventional vector and axial composite quark operators,
whose matrix elements define the GPDs discussed in the previous section,
there is a specific gluon operator that requires the helicity of the
hadron to be flipped by two units. It can show up in the scattering
off a spin-$1/2$ target due to a non-zero angular momentum of partons in
the nucleon. The latter is a result of the non-vanishing $t$-channel momentum
transfer $\Delta$ in the present settings. The corresponding operator
forms the $(1,1)$ representation of the Lorentz group and reads in terms
of the gluon field strength tensors
\begin{eqnarray}
\label{TransOper}
{^G\!{\cal O}^T_{\mu\nu}}
(\kappa_1,\kappa_2)\!
=
G_{+ \rho} (\kappa_2 n)
\tau^\perp_{\mu\nu;\rho\sigma}
G_{\sigma +} (\kappa_1 n)
\, ,
\end{eqnarray}
with the omitted gauge link. Here the totally symmetric and traceless tensor is
\begin{equation}
\tau^\perp_{\mu\nu;\rho\sigma}
= \frac{1}{2}
\left(
g^\perp_{\mu\rho} g^\perp_{\nu\sigma} + g^\perp_{\mu\sigma}
g^\perp_{\nu\rho} - g^\perp_{\mu\nu} g^\perp_{\rho\sigma}
\right) \, .
\end{equation}
The two-dimensional metric $g^\perp_{\mu\nu} = g_{\mu\nu} - n_\mu n^\ast_\nu
- n^\ast_\mu n_\nu$ is defined in terms of two light-like vectors $n$ and
$n^\ast$, such that $n^2 = n^{\ast 2} = 0$ and $n \cdot n^\ast = 1$. The
off-forward matrix element of this operator is parametrized via four GPDs
\cite{Die01} (see \cite{JiHoo98,BelMul00d} for an earlier discussion)
\begin{eqnarray}
\label{gluonGPDT}
G^T_{\mu\nu} (x, \xi, \Delta^2)
\!\!\!&\equiv&\!\!\!
4 (P \cdot n)^{-1} \int \frac{d\kappa}{2\pi} {\rm e}^{i x \kappa (P \cdot n)}
\langle P_2 |
{^G\!{\cal O}^T_{\mu\nu}} (\kappa, -\kappa)
| P_1 \rangle
\\
&=&\!\!\!
\frac{\tau^\perp_{\mu\nu;\alpha\beta}}{2 M} \Delta_\alpha
\bar U (P_2)
\Bigg\{
H_T (x, \xi, \Delta^2)  \frac{q_\gamma}{P\cdot q} \, i \sigma_{\gamma\beta}
+
\widetilde H_T (x, \xi, \Delta^2) \frac{\Delta_\beta}{2 M^2}
\nonumber\\
&&\qquad
+
E_T (x, \xi, \Delta^2) \frac{1}{2 M}
\left(
\frac{\gamma \cdot q}{P\cdot q} \Delta_\beta - \eta \, \gamma_\beta
\right)
-
\widetilde E_T (x, \xi, \Delta^2)
\frac{\gamma_\beta}{2 M}
\Bigg\}
U (P_1) \, ,
\nonumber
\end{eqnarray}
The traceless symmetric projector $\tau^\perp$ in Eq.\ (\ref{gluonGPDT})
possesses the properties
\begin{eqnarray*}
\tau^\perp_{\mu\nu;\rho\sigma} \tau^\perp_{\mu\nu;\rho'\sigma'}
=
\tau^\perp_{\rho\sigma;\rho'\sigma'}
\, , \qquad
\tau^\perp_{\mu\nu;\rho\sigma} = \tau^\perp_{\rho\sigma;\mu\nu}
\, , \qquad
\tau^\perp_{\mu\mu;\rho\sigma} = 0
\, , \qquad
\tau^\perp_{\mu\nu;\mu\nu} = 2
\, .
\end{eqnarray*}
A simple calculation of one-loop diagrams \cite{JiHoo98,BelMul00d} gives us
the following result for the real final-state photon DVCS amplitude
\begin{equation}
T_{\mu\nu} = \frac{\alpha_s}{2 \pi} T_F \sum_{i = u,d,s}
\int_{- 1}^{1}
d x \  C^{(+)}_{(0)i} (x, \xi) \, G^T_{\mu\nu} (x, \xi, \Delta^2)
\, ,
\end{equation}
with $T_F = 1/2$ and the coefficient function $C^{(+)}_{i(0)}$ defined in
Eq.\ (\ref{CoeffFunction}). Substituting Eq.\ (\ref{gluonGPDT}) into the above
expression gives the amplitude, which we will use in our computation of the
cross section. We define the CFFs similarly to Eqs.\ (\ref{DefTw3}) via
\begin{eqnarray}
\label{Def-ConTra}
&&\left\{
{\cal H}_T
,
{\cal E}_T
,
\widetilde {\cal H}_T
,
\widetilde {\cal E}_T
\right\}
(\xi)
=
\frac{\alpha_s}{2 \pi} T_F \sum_{i = u,d,s}
\int_{- 1}^{1} \! d x \, C^{(+)}_{(0)i} (\xi, x)
\left\{
H_T
,
E_T
,
\widetilde H_T
,
\widetilde E_T
\right\}
(x, \xi)
\, .
\end{eqnarray}
Below we will use unifying conventions for these CFFs, i.e.,
${\cal F}_T=\{{\cal H}_T,\dots, \widetilde{\cal E}_T\}$.

%%%%%%%%%%%%%%%%%%%%%%%%%%%%%%%%%%%%%%%%%%%%%%%%%%%%%%%%%%%%%%%%%%%%%
\section{Angular dependence of the cross section}
\label{Sec-AziAngDep}
%%%%%%%%%%%%%%%%%%%%%%%%%%%%%%%%%%%%%%%%%%%%%%%%%%%%%%%%%%%%%%%%%%%%%

\begin{figure}[t]
\vspace{-2.5cm}
\begin{center}
\mbox{
\begin{picture}(0,250)(250,0)
\put(70,0){\insertfig{13}{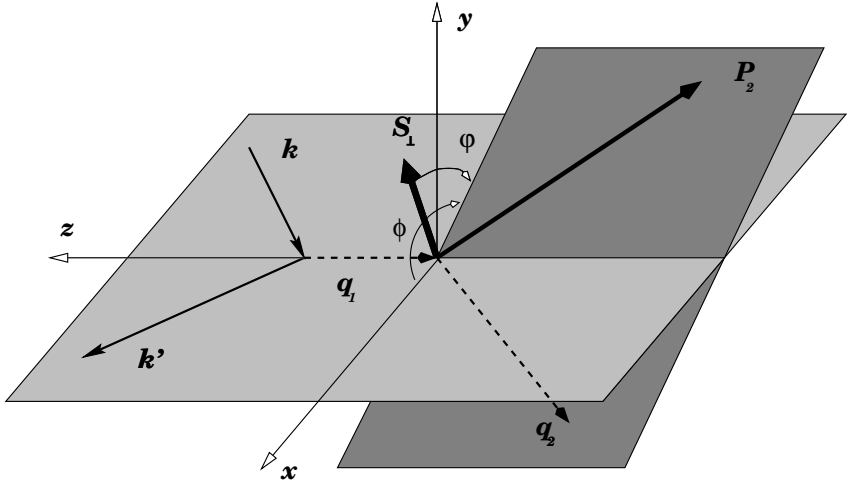}}
\end{picture}
}
\end{center}
\caption{\label{Fig-Kin}
The kinematics of the leptoproduction in the target rest frame. The
$z$-direction is chosen counter-along the three-momentum of the incoming
virtual photon. The lepton three-momenta form the lepton scattering plane,
while the recoiled proton and outgoing real photon define the hadron
scattering plane. In this reference system the azimuthal angle of the
scattered lepton is $\phi_l = 0$, while the azimuthal angle between the
lepton plane and the recoiled proton momentum is $\phi_N = \phi$. When the
hadron is transversely polarized (in this reference frame) $S_\perp = (0,
\cos {\mit\Phi}, \sin {\mit\Phi}, 0)$, the angle between the polarization
vector and the scattered hadron is denoted as $\varphi = {\mit\Phi} -
\phi_N$. }
\end{figure}

Now we are in a position to turn to physical observables, which give direct
access to GPDs in a measurement of the five-fold cross section for the
process $e (k) h (P_1) \to e (k^\prime) h (P_2) \gamma (q_2)$,
\begin{eqnarray}
\label{WQ}
\frac{d\sigma}{d\Bx dy d|\Delta^2| d\phi d\varphi}
=
\frac{\alpha^3  \Bx y } {16 \, \pi^2 \,  {\cal Q}^2 \sqrt{1 + \epsilon^2}}
\left| \frac{\cal T}{e^3} \right|^2 \, .
\end{eqnarray}
This cross section depends on the Bjorken variable $\Bx$, the squared
momentum transfer $\Delta^2 = (P_2 - P_1)^2$, the lepton energy fraction $y
= P_1\cdot q_1/P_1\cdot k$, with $q_1 = k - k'$, and, in general, two
azimuthal angles. We use throughout our presentation the convention
\begin{eqnarray*}
\epsilon \equiv 2 \Bx \frac{M}{{\cal Q}}
\, .
\end{eqnarray*}
In Eq.\ (\ref{WQ}), $\phi = \phi_N - \phi_l$ is the angle between the lepton
and hadron scattering planes and $\varphi = {\mit\Phi} - \phi_N$ is the
difference of the azimuthal angle $\mit\Phi$ of the transverse part of the
nucleon polarization vector $S$, i.e., $S_\perp = (0, \cos{\mit\Phi},
\sin{\mit\Phi}, 0)$, and the azimuthal angle $\phi_N$ of the recoiled
hadron. Our frame is rotated with respect to the laboratory one in such a
way that the virtual photon four-momentum has no transverse components, see
Fig.\ \ref{Fig-Kin}. We fix our kinematics by choosing the $z$-component of
the virtual photon momentum to be negative and the positive $x$-component of
the incoming lepton: $k = (E, E \sin\theta_l, 0, E \cos\theta_l )$, $q_1 =
(q_1^0, 0, 0,-|q_1^3|)$. Other vectors are $P_1 = (M, 0, 0, 0)$ and $P_2 =
(E_2, |\mbox{\boldmath$P$}_2| \cos\phi \sin\theta_N, |
\mbox{\boldmath$P$}_2| \sin\phi \sin\theta_N, |\mbox{\boldmath$P$}_2|
\cos\theta_N)$. The longitudinal part of the polarization vector is $S_{\rm
LP} = (0, 0, 0, \Lambda)$.

%%%%%%%%%%%%%%%%%%%%%%%%%%%%%%%%%%%%%%%%%%%%%%%%%%%%%%%%%%%%%%%%%%%%%
%                          Figure
%%%%%%%%%%%%%%%%%%%%%%%%%%%%%%%%%%%%%%%%%%%%%%%%%%%%%%%%%%%%%%%%%%%%%
%%%%%%%%%%%%%%%%%%%%%%%%%%%%%%%%%%%%%%%%%%%%%%%%%%%%%%%%%%%%%%%%%%%%%

The amplitude ${\cal T}$ is the sum of the DVCS ${\cal T}_{\rm DVCS}$ and
Bethe-Heitler (BH) ${\cal T}_{\rm BH}$ amplitudes. The latter one is real (to
the lowest order in the QED fine structure constant) and is parametrized in
terms of electromagnetic form factors, which we assume to be known from other
measurements. The azimuthal angular dependence of each of the three terms in
\begin{equation}
{\cal T}^2
= |{\cal T}_{\rm BH}|^2 + |{\cal T}_{\rm DVCS}|^2 + {\cal I}
\, ,
\end{equation}
with the interference term
\begin{equation}
{\cal I}
= {\cal T}_{\rm DVCS} {\cal T}_{\rm BH}^\ast
+ {\cal T}_{\rm DVCS}^\ast {\cal T}_{\rm BH}
\, ,
\end{equation}
arises from the contraction of leptonic and handronic tensors (see also
\cite{GouDiePirRal97}). In our frame these contractions yield finite sums of
Fourier harmonics, whose maximal frequencies are defined by the the rank-$m$
of the corresponding leptonic tensor in the incoming lepton momentum $k_\mu$.
Note, however, that in the polarized part of the leptonic tensors, proportional
to $\lambda$ times the $\epsilon$-tensor, one has one four-vector $k_\mu$ less
than in the unpolarized part. Thus, the highest harmonic which is proportional
to $\lambda$ will be $\cos/\sin([m-1]\phi)$ instead of $\cos/\sin(m\phi)$. The
parity and time reversal invariance provide further constraints on the Fourier
coefficients.

The BH term $|{\cal T}_{\rm BH}|^2$, squared DVCS amplitude
$|{\cal T}_{\rm DVCS}|^2$, and  interference term ${\cal I}$ read
\begin{eqnarray}
\label{Par-BH}
&&|{\cal T}_{\rm BH}|^2
= \frac{e^6}
{\Bx^2 y^2 (1 + \epsilon^2)^2 \Delta^2\, {\cal P}_1 (\phi) {\cal P}_2 (\phi)}
\left\{
c^{\rm BH}_0
+  \sum_{n = 1}^2
c^{\rm BH}_n \, \cos{(n\phi)} + s^{\rm BH}_1 \, \sin{(\phi)}
\right\} \, ,
\\
\label{AmplitudesSquared}
&& |{\cal T}_{\rm DVCS}|^2
=
\frac{e^6}{y^2 {\cal Q}^2}\left\{
c^{\rm DVCS}_0
+ \sum_{n=1}^2
\left[
c^{\rm DVCS}_n \cos (n\phi) + s^{\rm DVCS}_n \sin (n \phi)
\right]
\right\} \, ,
\\
\label{InterferenceTerm}
&&{\cal I}
= \frac{\pm e^6}{\Bx y^3 \Delta^2 {\cal P}_1 (\phi) {\cal P}_2 (\phi)}
\left\{
c_0^{\cal I}
+ \sum_{n = 1}^3
\left[
c_n^{\cal I} \cos(n \phi) +  s_n^{\cal I} \sin(n \phi)
\right]
\right\} \, ,
\end{eqnarray}
where the $+$ ($-$) sign in the interference stands for the negatively
(positively) charged lepton beam\footnote{Note that only in the case of massless
leptons, a partial cancellation of propagators occurs in the squared BH term
\cite{BerDiePir01}.}. The results for the Fourier coefficients, presented below,
show that the generation of new harmonics in the azimuthal angular dependence is
terminated at the twist-three level. The coefficients $c^{\cal I}_1, s^{\cal I}_1$
as well as $c^{\rm DVCS}_0$ arise at the twist-two level, and their dependence on
GPDs has been elaborated in Refs.\ \cite{GouDiePirRal97,BelMulNieSch00}. The
rest provides an additional angular dependence and is given in terms of
twist-two, i.e., $c^{\cal I}_0$, and twist-three, i.e., $c^{\rm DVCS}_1$,
$s^{\rm DVCS}_1$, $c^{\cal I}_2$, and $s^{\cal I}_2$, GPDs. The harmonics
proportional to $\cos{(3\phi)}$ [$\cos{(2\phi)}$] or $\sin{(3\phi)}$
[$\sin{(2\phi)}$] in the interference [squared DVCS] term stem from the
twist-two double helicity-flip gluonic GPDs alone. They are not
contaminated by twist-two quark amplitudes, however, will be affected by
twist-four power corrections \cite{KivMan01}. We neglect in our consequent
considerations the effects of dynamical higher-twist (larger than three)
contributions. They will give power-suppressed corrections to the Fourier
coefficients, we discussed.

There is an important difference between the interference term and the
squared DVCS amplitude. The former has a contaminating $\phi$-dependence due
to the lepton BH propagators,
\begin{equation}
\label{ExaBHpro}
{\cal Q}^2 {\cal P}_1 \equiv (k - q_2)^2 = {\cal Q}^2 + 2k\cdot \Delta
\, , \qquad
{\cal Q}^2 {\cal P}_2 \equiv (k - \Delta)^2 = - 2 k \cdot \Delta + \Delta^2
\, ,
\end{equation}
where
\begin{equation}
\label{kDelta}
k \cdot \Delta
= - \frac{{\cal Q}^2}{2y (1 + \epsilon^2)}
\Bigg\{
1 + 2 K \cos{\phi} - \frac{\Delta^2}{{\cal Q}^2}
\left( 1 - \Bx (2-y) + \frac{y \epsilon^2}{2} \right)
+  \frac{y \epsilon^2}{2}
\Bigg\} \, .
\end{equation}
The $1/{\cal Q}$-power suppressed kinematical factor $K$ appearing here
also shows up in the Fourier series (\ref{Par-BH}-\ref{InterferenceTerm}),
\begin{equation}
K^2 = -\frac{\Delta^2}{{\cal Q}^2} (1 - \Bx)
\left( 1 - y - \frac{y^2\epsilon^2}{4} \right)
\left( 1 - \frac{\Delta^2_{\rm min}}{\Delta^2} \right)
\left\{
\sqrt{1 + \epsilon^2}
+ \frac{4\Bx (1 - \Bx) + \epsilon^2}{4(1 - \Bx)}
\frac{\Delta^2 - \Delta^2_{\rm min}}{{\cal Q}^2}
\right\} \, ,
\end{equation}
(with the plus sign taken for the square root in Eq.\ (\ref{kDelta})). It
vanishes at the kinematical boundary $\Delta^2 = \Delta_{\rm min}^2$,
determined by the minimal value
\begin{eqnarray}
\label{Def-tmin}
-\Delta_{\rm min}^2
=  {\cal Q}^2
\frac{2(1 - \Bx) \left(1 - \sqrt{1 + \epsilon^2}\right) + \epsilon^2}
{4\Bx (1 - \Bx) + \epsilon^2}
\approx \frac{M^2 \Bx^2}{1 - \Bx + \Bx M^2/{\cQ}^2}
\, ,
\end{eqnarray}
as well as at
\begin{eqnarray*}
y(x,\cQ^2)
=
y_{\rm max}
\equiv
2 \frac{\sqrt{1 + \epsilon^2} - 1}{\epsilon^2}
\approx
1 - \frac{M^2 \Bx^2}{\cQ^2} \, .
\end{eqnarray*}
The square of the transverse momentum transfer is given by
\begin{eqnarray*}
\Delta_\perp^2 \approx (1 - \xi^2) (\Delta^2 - \Delta_{\rm min}^2) \, ,
\end{eqnarray*}
up to corrections suppressed by the hard-photon virtuality.

According to Eqs.\ (\ref{ExaBHpro}) and (\ref{kDelta}), we introduce the
following parametrization:
\begin{equation}
\label{Par-BH-Pro}
{\cal P}_1
= - \frac{1}{y (1 + \epsilon^2)} \left\{ J + 2 K \cos(\phi) \right\}
\, , \qquad
{\cal P}_2
=
1 + \frac{\Delta^2}{\cQ^2} +
\frac{1}{y (1 + \epsilon^2)}\left\{J   + 2 K \cos(\phi)
\right\}
\, ,
\end{equation}
where
\begin{eqnarray*}
J =
\left( 1 - y - \frac{y \epsilon^2}{2} \right)
\left( 1 + \frac{\Delta^2}{\cQ^2} \right)
-
(1 - x) (2 - y) \frac{\Delta^2}{\cQ^2} \, .
\end{eqnarray*}
As we see, the denominator of the $u$-channel lepton propagator, i.e.,
${\cal P}_1$, can be of order $1/\cQ^2$ at large $y$. In the Bjorken limit
it behaves like $(1 - y)$. Moreover, if the outgoing photon is collinear to
the incoming lepton, it vanishes. Of course, the photon then lies in the
lepton scattering plane, i.e., $\phi_\gamma = \phi + \pi = 0$, and both
polar angles coincide with each other. The latter condition is fulfilled if
\begin{eqnarray*}
y
=
y_{\rm col}
\equiv
\frac{\cQ^2 + \Delta^2}{\cQ^2 + x \Delta^2}
\approx
1 + (1 - \Bx) \frac{\Delta^2}{\cQ^2} \, ,
\end{eqnarray*}
(with $y_{\rm col} \le y_{\rm max}$) and this leads to the equality
\begin{eqnarray}
\label{Par-BH-Pro-ycol}
J_{|y = y_{\rm col}}
\!\!\!&=&\!\!\!
2 K_{|y = y_{\rm col}}
\\
\!\!\!&=&\!\!\!
- \frac{2 (1 - \Bx)\Delta^2}{{\cal Q}^2 + x\Delta^2}
\left( 1 - \frac{\Delta^2_{\rm min}}{\Delta^2} \right)
\left\{
\sqrt{1 + \epsilon^2}
+ \frac{4\Bx (1-\Bx) + \epsilon^2}{4(1-\Bx)}
\frac{\Delta^2 -\Delta^2_{\rm min}}{{\cal Q}^2}
\right\} \, .
\nonumber
\end{eqnarray}
Obviously, for large $y$ the squared BH (\ref{Par-BH}) and interference
(\ref{InterferenceTerm}) terms are enhanced with respect to the squared DVCS
one (\ref{AmplitudesSquared}). Furthermore, the expansion of ${\cal P}_1$ in
$\cQ$ is not justified, and, thus, the Fourier analysis of experimental data
must be modified. For small $y$ it is legitimate to expand ${\cal P}_1$ and
${\cal P}_2$ in power series with respect to $1/\cQ$. This generates higher
harmonics suppressed by powers of $K$.

The computation of the Fourier coefficients in Eq.\
(\ref{Par-BH}-\ref{InterferenceTerm}) is tedious, however, relatively
straightforward (see Refs.\ \cite{Ji98,GouDiePirRal97,BelMulNieSch00}).
After an extensive algebra we come to the Fourier harmonics summarized in
the following sections. We will employ the conventional definition of the
lepton helicity, i.e., $\lambda = +1$ if the spin is aligned with the
direction of the lepton three-momentum. To have a compact notation, we write
the cross section for a polarized target as
\begin{eqnarray}
\label{Def-DecCroSec}
d\sigma
=
d\sigma_{\rm unp}
+
\cos(\theta)\, d\sigma_{\rm LP}(\Lambda)
+
\sin(\theta)\, d\sigma_{\rm TP}(\varphi)
\, ,
\end{eqnarray}
where the polar angle $\theta$ appears in the decomposition of the spin
vector $S = \cos(\theta) S_{\rm LP}(\Lambda) + \sin(\theta) S_\perp(\Phi)$.
The same decomposition will be used for the Fourier coefficients.

%%%%%%%%%%%%%%%%%%%%%%%%%%%%%%%%%%%%%%%%%%%%%%%%%%%%%%%%%%%%%%%%%%%%%
\subsection{Bethe-Heitler amplitude squared}
\label{BHcrosssection}
%%%%%%%%%%%%%%%%%%%%%%%%%%%%%%%%%%%%%%%%%%%%%%%%%%%%%%%%%%%%%%%%%%%%%

This part of the leptoproduction cross section is expressed solely in
terms of $F_1(\Delta^2)$ and $F_2(\Delta^2)$, the known Dirac and Pauli
form factors of the nucleon. For diverse target polarizations the Fourier
coefficients are:
\begin{itemize}
\item Unpolarized target:
\end{itemize}
\begin{eqnarray}
\label{Def-FC-BH-unp0}
c^{\rm BH}_{0,{\rm unp}}
\!\!\!&=&\!\!\!
8 K^2
\left\{
\left( 2 + 3 \epsilon^2 \right)
\frac{{\cal Q}^2}{\Delta^2}
\left( F_1^2 - \frac{\Delta^2}{4 M^2} F_2^2 \right)
+ 2 \Bx^2 \left( F_1 + F_2 \right)^2
\right\}
\\
&+&\!\!\! (2 - y)^2
\Bigg\{
\left( 2 + \epsilon^2 \right)
\Bigg[
\frac{4 \Bx^2 M^2}{\Delta^2}
\left( 1 + \frac{\Delta^2}{{\cal Q}^2} \right)^2
+ 4 (1 - \Bx)
\left( 1 + \Bx  \frac{\Delta^2}{{\cal Q}^2} \right)
\Bigg]
\left( F_1^2 - \frac{\Delta^2}{4 M^2} F_2^2 \right)
\nonumber\\
&+&\!\!\!
4 \Bx^2
\Bigg[
\Bx + \left(1 - \Bx + \frac{\epsilon^2}{2} \right)
\left(1 -  \frac{\Delta^2}{Q^2} \right)^2
- \Bx (1 - 2 \Bx) \frac{\Delta^4}{Q^4}
\Bigg]
\left( F_1 + F_2 \right)^2
\Bigg\}
\nonumber\\
&+&\!\!\! 8 \left( 1 + \epsilon^2 \right)
\left(1 - y - \frac{\epsilon^2 y^2}{4} \right)
\Bigg\{
2 \epsilon^2
\left( 1 - \frac{\Delta^2}{4 M^2} \right)
\left( F_1^2 - \frac{\Delta^2}{4 M^2} F_2^2 \right)
\nonumber\\
&&\qquad\qquad\qquad\qquad\qquad\qquad\qquad\qquad\qquad
- \Bx^2
\left( 1 -  \frac{\Delta^2}{Q^2} \right)^2
\left( F_1 + F_2 \right)^2
\Bigg\} \, ,
\nonumber\\
\label{Def-FC-BH-unp1}
c^{\rm BH}_{1,{\rm unp}}
\!\!\!&=&\!\!\!
8 K (2 - y)
\Bigg\{
\left(
\frac{4\Bx^2 M^2}{\Delta^2} - 2\Bx - \epsilon^2
\right)
\left(
F_1^2 - \frac{\Delta^2}{4 M^2} F_2^2
\right)
\\
&&\qquad\qquad\qquad\qquad\qquad\qquad\qquad\qquad\qquad
+ \, 2 \, \Bx^2
\left( 1 - (1 - 2\Bx) \frac{\Delta^2}{{\cal Q}^2} \right)
\left( F_1 + F_2 \right)^2
\Bigg\},
\nonumber\\
\label{Def-FC-BH-unp2}
c^{\rm BH}_{2,{\rm unp}} \!\!\!&=&\!\!\!
8 \Bx^2 K^2  \left\{ \frac{4 M^2}{\Delta^2}
\left(F_1^2 - \frac{\Delta^2}{4 M^2} F_2^2\right) + 2 \left(F_1 +
F_2\right)^2 \right\} \, .
\end{eqnarray}

\begin{itemize}
\item Longitudinally polarized target:
\end{itemize}
\begin{eqnarray}
c^{\rm BH}_{0,{\rm LP}}
\!\!\!&=&\!\!\!
8 \lambda \Lambda \Bx (2 - y) y
\frac{\sqrt{1 + \epsilon^2}}{1 - \frac{\Delta^2}{4 M^2}}
\left( F_1 +F _2 \right)
\Bigg\{
\frac{1}{2}
\left[ \frac{\Bx}{2} \left( 1 - \frac{\Delta^2}{{\cal Q}^2} \right)
- \frac{\Delta^2}{4 M^2}
\right]
\Bigg[ 2 - \Bx
\\
&-&\!\!\!
2(1 - \Bx)^2  \frac{\Delta^2}{{\cal Q}^2}
+ \epsilon^2
\left( 1 - \frac{\Delta^2}{{\cal Q}^2} \right)
- \Bx (1 - 2\Bx) \frac{\Delta^4}{{\cal Q}^4}
\Bigg]
\left( F_1 + F_2 \right)
\nonumber\\
&+&\!\!\!
\left(
1 - (1 - \Bx)\frac{\Delta^2}{{\cal Q}^2}
\right)
\left[
\frac{\Bx^2 M^2}{\Delta^2}
\left( 1 + \frac{\Delta^2}{{\cal Q}^2} \right)^2
+
(1 - \Bx)
\left( 1 + \Bx \frac{\Delta^2}{{\cal Q}^2} \right)
\right]
\left( F_1 + \frac{\Delta^2}{4 M^2} F_2 \right)
\Bigg\} \, ,
\nonumber\\
c^{\rm BH}_{1,{\rm LP}}
\!\!\!&=&\!\!\!
- 8 \lambda \Lambda \Bx y K
\frac{\sqrt{1 + \epsilon^2}}{1 - \frac{\Delta^2}{4 M^2}}
\left( F_1 + F_2 \right)
\Bigg\{
\left[
\frac{\Delta^2}{2 M^2}
-
\Bx \left(1 - \frac{\Delta^2}{{\cal Q}^2} \right)
\right]
\left( 1 - \Bx + \Bx \frac{\Delta^2}{{\cal Q}^2} \right)
\left( F_1 + F_2 \right)
\nonumber\\
&+&\!\!\!
\left[
1 + \Bx - (3 - 2 \Bx)
\left( 1 + \Bx \frac{\Delta^2}{{\cal Q}^2} \right)
-
\frac{4 \Bx^2 M^2}{\Delta^2}
\left( 1 + \frac{\Delta^4}{{\cal Q}^4} \right)
\right]
\left( F_1 + \frac{\Delta^2}{4 M^2} F_2 \right)
\Bigg\} \, .
\end{eqnarray}

\begin{itemize}
\item Transversely polarized target:
\end{itemize}
\begin{eqnarray}
c^{\rm BH}_{0,{\rm TP}}
\!\!\!&=&\!\!\!
- 8 \lambda \cos{(\varphi)} (2 - y) y \frac{{\cal Q}}{M}
\frac{\sqrt{1 + \epsilon^2} K}{\sqrt{1 - y - \frac{\epsilon^2 y^2}{4}}}
\left( F_1 + F_2 \right)
\Bigg\{
\frac{\Bx^3 M^2}{{\cal Q}^2}
\left( 1 - \frac{\Delta^2}{{\cal Q}^2} \right)
\left( F_1 + F_2 \right)
\\
&+&\!\!\!
\left(1 - (1 - \Bx) \frac{\Delta^2}{{\cal Q}^2} \right)
\left[
\frac{\Bx^2 M^2}{\Delta^2} \left(1 - \frac{\Delta^2}{{\cal Q}^2}\right) F_1
+
\frac{\Bx}{2} F_2
\right]
\Bigg\},
\nonumber\\
c^{\rm BH}_{1,{\rm TP}}
\!\!\!&=&\!\!\!
- 16 \lambda\cos{(\varphi)} \Bx y
\sqrt{1 - y - \frac{\epsilon^2 y^2}{4}}\frac{M}{{\cal Q}}\sqrt{1 + \epsilon^2}
\left( F_1 + F_2 \right)
\Bigg\{
\frac{2 K^2 {\cal Q}^2}{\Delta^2 \left( 1 - y - \frac{\epsilon^2 y^2}{4} \right)}
\Bigg[
\Bx \left( 1 - \frac{\Delta^2}{{\cal Q}^2} \right) F_1
\nonumber\\
&+&\!\!\!
\frac{\Delta^2}{4 M^2} F_2 \Bigg]
+
(1 + \epsilon^2) \Bx \left( 1 - \frac{\Delta^2}{{\cal Q}^2} \right)
\left( F_1 + \frac{\Delta^2}{4 M^2} F_2 \right)
\Bigg\} \, ,
\\
s^{\rm BH}_{1,{\rm TP}}
\!\!\!&=&\!\!\!
16 \lambda \sin{(\varphi)} y \Bx^2 \sqrt{1 - y -  \frac{\epsilon^2 y^2}{4}}
\frac{M}{{\cal Q}}
\sqrt{(1 + \epsilon^2)^3} \left( 1 - \frac{\Delta^2}{{\cal Q}^2} \right)
\left( F_1 + F_2\right)
\left( F_1 + \frac{\Delta^2}{4 M^2} F_2 \right) \, .
\end{eqnarray}

%%%%%%%%%%%%%%%%%%%%%%%%%%%%%%%%%%%%%%%%%%%%%%%%%%%%%%%%%%%%%%%%%%%%%
\subsection{DVCS amplitude squared}
\label{SubSec-AziAngDep-DVCS}
%%%%%%%%%%%%%%%%%%%%%%%%%%%%%%%%%%%%%%%%%%%%%%%%%%%%%%%%%%%%%%%%%%%%%

$|{\cal T}_{\rm DVCS}|^2$ is bilinear in the CFFs, and its coefficients read
in terms of ${\cal C}^{\rm DVCS}$ functions, which are specified in section
\ref{App-CoeFun}:
\begin{itemize}
\item Unpolarized target:
\end{itemize}
\begin{eqnarray}
\label{Res-Mom-DVCS-UP}
c^{\rm DVCS}_{0,{\rm unp}}
\!\!\!&=&\!\!\!
2 ( 2 - 2 y + y^2 )
{\cal C}^{\rm DVCS}_{\rm unp}
\left(
{\cal F},{\cal F}^\ast\right) ,
\\
\label{Res-Mom-DVCS-UP-tw3}
\left\{
{c^{\rm DVCS}_{1,{\rm unp}} \atop s^{\rm DVCS}_{1,{\rm unp}}}
\right\}
\!\!\! &=&\!\!\!
\frac{8 K}{2 - \Bx} \left\{ {2-y \atop -\lambda y } \right\}
\left\{{ \Re{\rm e} \atop \Im{\rm m} }\right\} \, {\cal C}^{\rm DVCS}_{\rm unp}
\left(
{\cal F}^{\rm eff},{\cal F}^\ast\right) \, ,
\\
c^{\rm DVCS}_{2,{\rm unp}}
\!\!\!&=&\!\!\!
- \frac{4 {\cal Q}^2 K^2}{M^2 (2 - \Bx)}
\Re{\rm e}\,
{\cal C}^{\rm DVCS}_{T,{\rm unp}} \left( {\cal F}_T, {\cal F}^\ast \right).
\end{eqnarray}

\begin{itemize}
\item Longitudinally polarized target:
\end{itemize}
\begin{eqnarray}
\label{Res-Mom-DVCS-LP}
c^{\rm DVCS}_{0,{\rm LP}}
\!\!\!&=&\!\!\!
2 \lambda \Lambda y(2-y)
{\cal C}^{\rm DVCS}_{\rm LP}
\left(
{\cal F},{\cal F}^\ast\right) ,
\\
\left\{ c^{\rm DVCS}_{1,{\rm LP}} \atop   s^{\rm DVCS}_{1,{\rm LP}} \right\}
\!\!\! &=&\!\!\!
- \frac{8 \Lambda K}{2 - \Bx}
\left\{{-\lambda y \atop 2 - y}\right\}
\left\{{\Re{\rm e} \atop \Im{\rm m} }\right\} {\cal C}^{\rm DVCS}_{\rm LP}
\left( {\cal F}^{\rm eff},{\cal F}^\ast \right) \, ,
\\
s^{\rm DVCS}_{2,{\rm LP}}
\!\!\!&=&\!\!\! - \frac{4 \Lambda {\cal Q}^2 K^2}{M^2 (2 - \Bx)}
\Im{\rm m}\, {\cal C}^{\rm DVCS}_{T,{\rm LP}}
\left({\cal F}_T,{\cal F}^\ast\right) \, .
\end{eqnarray}

\begin{itemize}
\item Transversely polarized target:
\end{itemize}
\begin{eqnarray}
\label{Res-Mom-DVCS-TP}
c^{\rm DVCS}_{0,{\rm TP}}
\!\!\!&=&\!\!\! - \frac{{\cal Q} K}{M\sqrt{1-y}}
\left[
- \lambda y (2 - y) \cos (\varphi) {\cal C}^{\rm DVCS}_{\rm TP+}
+
(2 - 2 y + y^2) \sin (\varphi)
\Im{\rm m}\, {\cal C}^{\rm DVCS}_{\rm TP-}
\right]
\left( {\cal F}, {\cal F}^\ast \right) \, ,
\nonumber\\
\\
\label{Res-Mom-DVCS-TP-Tw3}
\left\{
{ c^{\rm DVCS}_{1,{\rm TP}} \atop s^{\rm DVCS}_{1,{\rm TP}} }
\right\}
\!\!\! &=&\!\!\! -
\frac{4 {\cal Q} K^2}{M (2 - \Bx) \sqrt{1 - y}}
\\
&\times&\!\!\!
\left[
\cos (\varphi)
\left\{
{ - \lambda y \atop 2 - y}
\right\}
\left\{
{ \Re{\rm e} \atop \Im{\rm m} }
\right\} {\cal C}^{\rm DVCS}_{\rm TP +}
+
\sin (\varphi)
\left\{
{ 2 - y \atop \lambda y }
\right\}
\left\{
{ \Im{\rm m} \atop \Re{\rm e } }
\right\} {\cal C}^{\rm DVCS}_{\rm TP-}
\right]
\left( {\cal F}^{\rm eff}, {\cal F}^\ast \right) \, ,
\nonumber\\
\left\{
{ c^{\rm DVCS}_{2,{\rm TP}} \atop s^{\rm DVCS}_{2,{\rm TP}} }
\right\}
\!\!\!&=&\!\!\! - \frac{4 {\cal Q} \sqrt{1 - y} K}{M (2 - \Bx)}
\Im{\rm m}
\left\{
{ \sin(\varphi)\, {\cal C}^{\rm DVCS}_{T,{\rm TP}-}
\atop
\cos(\varphi)\, {\cal C}^{\rm DVCS}_{T,{\rm TP}+ } }
\right\}
\left( {\cal F}_T, {\cal F}^\ast \right).
\end{eqnarray}

The  harmonic $c^{\rm DVCS}_{0}$ is given
in terms of twist-two CFFs ${\cal F}= \{{\cal H}, {\cal E},
\widetilde{\cal H}, \widetilde{\cal E}  \}$, defined in
Eq.\ (\ref{DefTw3}).
The twist-three coefficients $c^{\rm DVCS}_{1}$ and $s^{\rm
DVCS}_{1}$ arise from the interference of twist-two CFFs with
`effective' twist-three ones,
\begin{eqnarray}
\label{Tw3Eff}
{\cal F}^{\rm eff} \equiv - 2 \xi
\left(
\frac{1}{1 + \xi} {\cal F} + {\cal F}^3_+ - {\cal F}^3_-
\right) \, ,
\end{eqnarray}
where ${\cal F}^3_\pm$ are defined in Eqs.\ (\ref{Def-Ftw3}-\ref{Def-FqGq}).
These Fourier harmonics have the same functional dependence on CFFs
as the leading twist-two ones \cite{BelMulNieSch00}. However, this
is not the case for the Fourier coefficients $c^{\rm DVCS}_{2}$ and $s^{\rm
DVCS}_{2}$,  induced by the gluon transversity CFFs (\ref{Def-ConTra}).

%%%%%%%%%%%%%%%%%%%%%%%%%%%%%%%%%%%%%%%%%%%%%%%%%%%%%%%%%%%%%%%%%%%%%
\subsection{Interference of Bethe-Heitler and DVCS amplitudes}
\label{SubSec-AziAngDep-INT}
%%%%%%%%%%%%%%%%%%%%%%%%%%%%%%%%%%%%%%%%%%%%%%%%%%%%%%%%%%%%%%%%%%%%%

For the phenomenology of GPDs, ${\cal I}$ is the most interesting quantity
since it is linear in CFFs. This simplifies their disentanglement from
experimental measurements. The Fourier harmonics have the form:
\begin{itemize}
\item Unpolarized target:
\end{itemize}
\begin{eqnarray}
c^{\cal I}_{0,\rm{unp}}
\!\!\!&=&\!\!\!
- 8  (2 - y)
\Re{\rm e}
\Bigg\{
\frac{(2 - y)^2}{1-y} K^2 {\cal C}^{\cal I}_{\rm unp} \left({\cal F}\right)
+
\frac{\Delta^2}{{\cal Q}^2}  (1-y)(2 - \Bx)
\left(
{\cal C}^{\cal I}_{\rm unp} + \Delta {\cal C}^{\cal I}_{\rm unp}
\right)
\left( {\cal F} \right)
\Bigg\} \, ,
\nonumber\\
\label{Res-IntTer-unp-c0}\\
\label{Res-IntTer-unp}
\left\{{c^{\cal I}_{1, \rm unp} \atop s^{\cal I}_{1, \rm unp}}\right\}
&\!\!\!=\!\!\!&
8 K
\left\{  {-(2 - 2y + y^2) \atop \lambda y (2-y)} \right\}
\left\{{\Re{\rm e} \atop \Im{\rm m} } \right\}
{\cal C}^{\cal I}_{\rm unp}\left({\cal F} \right) \, ,
\\
\label{Res-IntTer-unp-tw3}
\left\{ c^{\cal I}_{2, \rm unp} \atop  s^{\cal I}_{2, \rm unp} \right\}
&\!\!\!=\!\!\!&
\frac{16 K^2}{2 - \Bx}  \left\{ { -(2 - y) \atop \lambda y} \right\}
\left\{{\Re{\rm e} \atop \Im{\rm m} } \right\}
{\cal C}^{\cal I}_{\rm unp}
\left({\cal F}^{\rm eff}\right) \, ,
\\
\label{Res-IntTer-Tra-unp}
c^{\cal I}_{3, \rm unp}
\!\!\!&=&\!\!\! - \frac{8 {\cal Q}^2 K^3}{M^2 (2 - \Bx)^2}
\Re{\rm e}\, {\cal C}^{\cal I}_{T,{\rm unp}}
\left( {\cal F}_T \right) \, .
\end{eqnarray}

\begin{itemize}
\item Longitudinally polarized target:
\end{itemize}
\begin{eqnarray}
c^{\cal I}_{0, \rm{LP}}
\!\!\!&=&\!\!\!
-8 \lambda \Lambda y
\Re{\rm e}\Bigg\{
\left(\frac{(2 - y)^2}{1-y} +2 \right) K^2
{\cal C}^{\cal I}_{\rm LP} \left({\cal F}\right)
+\frac{\Delta^2}{{\cal Q}^2} (1-y)(2 - \Bx)
\left( {\cal C}^{\cal I}_{\rm LP}
+\Delta {\cal C}^{\cal I}_{\rm LP}\right)\left({\cal F}\right)
\Bigg\} \, ,
\nonumber\\
\label{Res-IntTer-LP-c0}\\
\label{Res-IntTer-LP}
\left\{{c^{\cal I}_{1, \rm LP} \atop s^{\cal I}_{1, \rm LP}}\right\}
&\!\!\!=\!\!\!&
8\Lambda K
\left\{  {-\lambda y (2 - y) \atop 2 - 2y + y^2 } \right\}
\left\{{\Re{\rm e} \atop \Im{\rm m} } \right\}
{\cal C}^{\cal I}_{\rm LP}\left({\cal F} \right) \, ,
\\
\left\{ c^{\cal I}_{2, \rm LP} \atop  s^{\cal I}_{2, \rm LP} \right\}
&\!\!\!=\!\!\!&
\frac{16 \Lambda K^2}{2 - \Bx}  \left\{ { -\lambda y \atop  2 - y} \right\}
\left\{{\Re{\rm e} \atop \Im{\rm m} } \right\}
{\cal C}^{\cal I}_{\rm LP}
\left(
{\cal F}^{\rm eff}\right) \, ,
\\
\label{Res-IntTer-Tra-pol}
s^{\cal I}_{3, \rm LP}
\!\!\!&=&\!\!\!  \frac{8\Lambda {\cal Q}^2 K^3}{M^2 (2 - \Bx)^2}
\Im{\rm m}\, {\cal C}^{\cal I}_{T,{\rm LP}} \left( {\cal F}_T \right) \, .
\end{eqnarray}

\begin{itemize}
\item Transversely polarized target:
\end{itemize}
\begin{eqnarray}
\label{Res-IntTer-TP-c0}
c^{\cal I}_{0,\rm TP}
\!\!\!&=&\!\!\!
\frac{8 M \sqrt{1-y}K}{\cal Q}\Bigg[
-\lambda y \cos(\varphi)\,
\Re{\rm e}
 \Bigg\{
\left(\frac{(2-y)^2}{1-y} +2\right)
{\cal C}^{\cal I}_{{\rm TP}+}\left({\cal F} \right) +
\Delta{\cal C}^{\cal I}_{{\rm TP}+}\left({\cal F} \right)
\Bigg\}
\\
&&\qquad\qquad\qquad
+ (2 - y)\sin(\varphi)\, \Im{\rm m}
\Bigg\{
\frac{(2-y)^2}{1-y} {\cal C}^{\cal I}_{{\rm TP}-}\left({\cal F} \right) +
\Delta{\cal C}^{\cal I}_{{\rm TP}-}\left({\cal F} \right)
\Bigg\}
\Bigg] \, ,
\nonumber\\
\label{Res-IntTer-TP}
\left\{
{c^{\cal I}_{1, \rm TP} \atop s^{\cal I}_{1, \rm TP}}
\right\}
\!\!\!&=&\!\!\!
\frac{8 M \sqrt{1-y}}{\cal Q}
\\
&\times&\!\!\!\left[
\cos(\varphi)
\left\{
{-\lambda y (2 - y) \atop 2 - 2y + y^2 }
\right\}
\left\{
{\Re{\rm e} \atop \Im{\rm m} }
\right\}
{\cal C}^{\cal I}_{{\rm TP}+}
+
\sin(\varphi)
\left\{
{2 - 2y + y^2  \atop \lambda y (2 - y) }
\right\}
\left\{
{\Im{\rm m} \atop \Re{\rm e} }
\right\}
{\cal C}^{\cal I}_{{\rm TP}-}
\right]
\left( {\cal F} \right) \, ,
\nonumber\\
\left\{
c^{\cal I}_{2, \rm TP} \atop  s^{\cal I}_{2, \rm TP}
\right\}
\!\!\!&=&\!\!\!
\frac{16 M \sqrt{1-y}K}{{\cal Q}(2 - \Bx)}
\\
&\times&\!\!\!
\left[
\cos(\varphi)
\left\{
{-\lambda y \atop 2 - y}
\right\}
\left\{
{\Re{\rm e} \atop \Im{\rm m}}
\right\}
{\cal C}^{\cal I}_{{\rm TP}+}
+
\sin(\varphi)
\left\{
{2 - y  \atop \lambda y}
\right\}
\left\{
{\Im{\rm m}\atop \Re{\rm e}}
\right\}
{\cal C}^{\cal I}_{{\rm TP}-}
\right]
\left( {\cal F}^{\rm eff} \right) \, ,
\nonumber\\
\label{Res-IntTer-Tra-Tp}
s^{\cal I}_{3, \rm TP}
\!\!\!&=&\!\!\!
\frac{8 {\cal Q} \sqrt{1 - y} K^2}{M (2 - \Bx)^2}
\cos(\varphi) \Im{\rm m}\, {\cal C}^{\cal I}_{T,{\rm TP}+}
\left( {\cal F}_T\right) \, ,
\\
c^{\cal I}_{3, \rm TP}
\!\!\!&=&\!\!\!
\frac{8{\cal Q} \sqrt{1 - y} K^2}{M (2 - \Bx)^2}
\sin(\varphi) \Im{\rm m}\, {\cal C}^{\cal I}_{T,{\rm TP}-}
\left({\cal F}_T\right) \, .
\end{eqnarray}

The higher twist-three harmonics, i.e., $ c^{\cal I}_{2}$ and $s^{\cal I}_{2}$
have again the same functional dependence as the twist-two ones. However, this
is not the case for $c^{\cal I}_{0}$, which depends only on twist-two CFFs
$\cal F$, and for  $ c^{\cal I}_{3}$ and $s^{\cal I}_{3}$, induced by ${\cal F}_T$
CFFs.

%%%%%%%%%%%%%%%%%%%%%%%%%%%%%%%%%%%%%%%%%%%%%%%%%%%%%%%%%%%%%%%%%%%%%
\subsection{Angular harmonics in terms of GPDs}
\label{App-CoeFun}
%%%%%%%%%%%%%%%%%%%%%%%%%%%%%%%%%%%%%%%%%%%%%%%%%%%%%%%%%%%%%%%%%%%%%

The Fourier coefficients displayed above are expressed in terms of the
coefficients ${\cal C}$. They depend on GPDs, integrated over the momentum
fraction, and are functions of the kinematical variables $\Bx$, $\Delta^2$,
and ${\cal Q}^2$. For the harmonics involving $H$, $E$, $\widetilde H$, and
$\widetilde E$-type GPDs they are:
\begin{itemize}
\item Squared DVCS amplitude:
\end{itemize}
\begin{eqnarray}
\label{Def-C-DVCS-unp}
{\cal C}^{\rm DVCS}_{\rm{unp}}
\left(
{\cal F},{\cal F}^\ast
\right)
\!\!\!&=&\!\!\!
\frac{1}{(2 - \Bx)^2}
\Bigg\{
4 (1 - \Bx)
\left(
{\cal H} {\cal H}^\ast
+
\widetilde{\cal H} \widetilde {\cal H}^\ast
\right)- \Bx^2
\bigg(
{\cal H} {\cal E}^\ast
+ {\cal E} {\cal H}^\ast
+ \widetilde{{\cal H}} \widetilde{{\cal E}}^\ast
+ \widetilde{{\cal E}} \widetilde{{\cal H}}^\ast
\bigg)
\nonumber\\
&&\qquad\qquad\;
-
\left( \Bx^2 + (2 - \Bx)^2 \frac{\Delta^2}{4M^2} \right)
{\cal E} {\cal E}^\ast
- \Bx^2 \frac{\Delta^2}{4M^2}
\widetilde{{\cal E}} \widetilde{{\cal E}}^\ast
\Bigg\},
\end{eqnarray}
\begin{eqnarray}
\label{Def-C-DVCS-LP}
{\cal C}^{\rm DVCS}_{\rm{LP}}
\left(
{\cal F},{\cal F}^\ast
\right)
\!\!\!&=&\!\!\!
\frac{1}{(2 - \Bx)^2}
\Bigg\{
4 (1 - \Bx)
\left(
{\cal H} \widetilde{\cal H}^\ast
+
\widetilde{\cal H} {\cal H}^\ast
\right)- \Bx^2
\bigg(
{\cal H} \widetilde{\cal E}^\ast
+ \widetilde{\cal E} {\cal H}^\ast
+ \widetilde{\cal H} {\cal E}^\ast
+ {\cal E} \widetilde{{\cal H}}^\ast
\bigg)
\nonumber\\
&&\qquad\qquad\;
-
\Bx \left( \frac{\Bx^2}{2} + (2 - \Bx) \frac{\Delta^2}{4M^2} \right)
\left(
{\cal E} \widetilde{{\cal E}}^\ast + \widetilde{\cal E} {\cal E}^\ast
\right)
\Bigg\},
\end{eqnarray}
\begin{eqnarray}
\label{Def-C-DVCS-TP}
{\cal C}^{\rm DVCS}_{\rm{TP}+}
\left(
{\cal F},{\cal F}^\ast
\right)
\!\!\!&=&\!\!\!
\frac{1}{(2 - \Bx)^2}
\Bigg\{
2 \Bx(
{\cal H}\widetilde{\cal E}^\ast
+ \widetilde{\cal E} {\cal H}^\ast)
- 2 (2 - \Bx) (\widetilde {\cal H} {\cal E}^\ast
+ \widetilde {\cal H}^\ast {\cal E})
+ \Bx^2 ({\cal E}  \widetilde{\cal E}^\ast
+ \widetilde{\cal E} {\cal E}^\ast )
\Bigg\},
\nonumber\\
{\cal C}^{\rm DVCS}_{\rm{TP}-}
\left(
{\cal F},{\cal F}^\ast
\right)
\!\!\!&=&\!\!\!
\frac{2}{(2 - \Bx)^2}
\Bigg\{
(2 - \Bx)({\cal H} {\cal E}^\ast - {\cal E} {\cal H}^\ast)
- \Bx ( \widetilde{\cal H} \widetilde{\cal E}^\ast
- \widetilde{\cal E} \widetilde{\cal H}^\ast )
\Bigg\} \, .
\end{eqnarray}

\begin{itemize}
\item Interference of Bethe-Heitler and DVCS amplitudes:
\end{itemize}
Here a part of the result at the twist-two level is expressed in terms of the
functions, which show up in the lowest twist approximation, and they read
\cite{BelMulNieSch00}
\begin{eqnarray}
\label{Def-C-Int-unp}
{\cal C}^{\cal I}_{\rm unp}
\!\!\!&=&\!\!\!
F_1 {\cal H} + \frac{\Bx}{2 - \Bx}
(F_1 + F_2) \widetilde {\cal H}
-
\frac{\Delta^2}{4M^2} F_2 {\cal E} \, ,
\\
\label{Def-C-Int-LP}
{\cal C}^{\cal I}_{\rm LP}
\!\!\!&=&\!\!\!
\frac{\Bx }{2 - \Bx}(F_1 + F_2) \left({\cal H} +
\frac{\Bx}{2} {\cal E} \right)
+ F_1 \widetilde{\cal H}
- \frac{\Bx}{2 - \Bx}
\left(
\frac{\Bx}{2} F_1 + \frac{\Delta^2}{4 M^2} F_2
\right)
\widetilde{\cal E}
\, , \\
\label{Def-C-Int-TP}
{\cal C}^{\cal I}_{{\rm TP}+}
\!\!\!&=&\!\!\!
(F_1 + F_2)
\left\{
\frac{\Bx^2}{2 - \Bx}
\left( {\cal H} + \frac{\Bx}{2} {\cal E} \right)
+  \frac{\Bx \Delta^2}{4 M^2}{\cal E}
\right\}
- \frac{\Bx^2}{2 - \Bx} F_1
\left( \widetilde{\cal H} + \frac{\Bx}{2} \widetilde{\cal E} \right)
\nonumber\\
&&
+ \frac{\Delta^2}{4M^2} \left\{4 \frac{1 - \Bx}{2 - \Bx} F_2
\widetilde{\cal H}
- \left( \Bx F_1 + \frac{\Bx^2}{2 - \Bx} F_2 \right)
\widetilde{\cal E} \right\}
\, , \\
{\cal C}^{\cal I}_{{\rm TP}-}
\!\!\!&=&\!\!\!
\frac{1}{2 - \Bx}
\left( \Bx^2 F_1 - (1 - \Bx) \frac{\Delta^2}{M^2} F_2  \right)
{\cal H}
+ \Bigg\{
\frac{\Delta^2}{4 M^2}
\Bigg(
(2 - \Bx) F_1 + \frac{\Bx^2}{2 - \Bx} F_2
\Bigg)
\nonumber\\
&&
+ \frac{\Bx^2}{2 - \Bx} F_1 \Bigg\} {\cal E}
- \frac{\Bx^2}{2 - \Bx} (F_1 + F_2)
\left(
\widetilde{\cal H} + \frac{\Delta^2}{4 M^2}\widetilde{\cal E} \right) \, .
\nonumber
\end{eqnarray}
The addenda arising in power-suppressed contributions are defined as
\begin{eqnarray}
\label{Def-C-IntAdd-unp}
\Delta {\cal C}^{\cal I}_{{\rm unp}}
\!\!\!&=&\!\!\!
- \frac{\Bx}{2-\Bx}  (F_1 + F_2)
\left\{
\frac{\Bx}{2 - \Bx} ({\cal H} + {\cal E})
+ \widetilde {\cal H}
\right\} \, ,
\\
\label{Def-C-IntAdd-LP}
\Delta {\cal C}^{\cal I}_{{\rm LP}}
\!\!\!&=&\!\!\! - \frac{\Bx}{2 - \Bx} (F_1 + F_2)
\left\{
{\cal H} + \frac{\Bx}{2}{\cal E} + \frac{\Bx}{2 - \Bx}
\left(
\widetilde{\cal H} + \frac{\Bx}{2}\widetilde{\cal E}
\right)
\right\} \, ,
\\
\label{Def-C-IntAdd-TP}
\Delta {\cal C}^{\cal I}_{{\rm TP+}}
\!\!\!&=&\!\!\!
- \frac{\Delta^2}{M^2}
\left\{
F_2\widetilde {\cal H} - \frac{\Bx}{2 - \Bx}
\left( F_1 + \frac{\Bx}{2} F_2 \right) \widetilde{\cal E}
\right\} \, ,
\\
\Delta {\cal C}^{\cal I}_{{\rm TP-}}
\!\!\!&=&\!\!\!
\frac{\Delta^2}{M^2}
\left( F_2 {\cal H} - F_1 {\cal E} \right) \, .
\end{eqnarray}

Let us now list the coefficients involving the gluon transversity:
\begin{itemize}
\item Squared DVCS amplitude:
\end{itemize}
\begin{eqnarray}
{\cal C}^{\rm DVCS}_{T,{\rm unp}}
\!\!\!&=&\!\!\!
\frac{1}{(2 -\Bx)^2}
\Bigg\{
{\cal H}_T
\left[
(2 - \Bx) {\cal E}^\ast - \Bx \widetilde{\cal E}^\ast
\right]
- 2 (2 - \Bx) \widetilde{\cal H}_T
\left[
{\cal H}^\ast
+
\frac{\Delta^2}{4 M^2} {\cal E}^\ast
\right]
\\
&&\qquad\qquad- {\cal E}_T
\left[
(2 - \Bx) {\cal H}^\ast
- \Bx \widetilde{\cal H}^\ast
\right]
+ \widetilde{\cal E}_T
\left[
\Bx ( {\cal H}^\ast + {\cal E}^\ast )
- (2 - \Bx) \widetilde{\cal H}^\ast
\right]
\Bigg\} \, ,
\nonumber\\
{\cal C}^{\rm DVCS}_{T,{\rm LP}}
\!\!\!&=&\!\!\!
\frac{1}{(2 - \Bx)^2}
\Bigg\{
{\cal H}_T
\left[
(2 - \Bx) {\cal E}^\ast - \Bx \widetilde{\cal E}^\ast
\right]
+ \widetilde{\cal H}_T
\left[
2 (2 - \Bx) \widetilde{\cal H}^\ast
-
\Bx \left( \Bx - \frac{\Delta^2}{2M^2} \right) \widetilde{\cal E}^\ast
\right]
\\
&&\qquad\qquad
- {\cal E}_T
\Bigg[
\Bx {\cal H}^\ast - (2 - \Bx) \widetilde{\cal H}^\ast
+ \frac{\Bx^2}{2}
\left( {\cal E}^\ast + \widetilde{\cal E}^\ast \right)
\Bigg]
\nonumber\\
&&\qquad\qquad
+ \widetilde{\cal E}_T
\left[
(2 - \Bx)
\left( {\cal H}^\ast + \frac{\Bx}{2} {\cal E}^\ast \right)
- \Bx
\left(
\widetilde{\cal H}^\ast + \frac{\Bx}{2} \widetilde{\cal E}^\ast
\right)
\right]
\Bigg\} \, ,
\nonumber\\
{\cal C}^{\rm DVCS}_{T,{\rm TP+}}
\!\!\!&=&\!\!\!
\frac{1}{(2 - \Bx)^2}
\Bigg\{
\left[
4 (1 - \Bx) {\cal H}_T - \Bx^2 {\cal E}_T + \Bx (2 - \Bx) \widetilde{\cal E}_T
\right]
\left(
{\cal H}^\ast + \widetilde{\cal H}^\ast
\right)
\\
&&\qquad\qquad
- \widetilde{\cal H}_T \left( \Bx^2 + (1 - \Bx) \frac{\Delta^2}{M^2} \right)
(2 \widetilde{\cal H}^\ast + \Bx \widetilde{\cal E}^\ast )
\nonumber\\
&&\qquad\qquad
- \Bx
\left[
\Bx {\cal H}_T
+ \left(
\frac{\Bx^2}{2} + (2 - \Bx) \frac{\Delta^2}{4 M^2}
\right) {\cal E}_T
\right]
( {\cal E}^\ast + \widetilde{\cal E}^\ast )
\nonumber\\
&&\qquad\qquad
+ \widetilde{\cal E}_T
\Bigg[ (2 - \Bx)
\left(
\frac{\Bx^2}{2} + (2 - \Bx) \frac{\Delta^2}{4 M^2} \right)
{\cal E}^\ast
- \Bx^2 \left( \frac{\Bx}{2} - \frac{\Delta^2}{4 M^2} \right)
\widetilde{\cal E}^\ast
\Bigg]
\Bigg\} \, ,
\nonumber\\
{\cal C}^{\rm DVCS}_{T,{\rm TP-}}
\!\!\!&=&\!\!\!
\frac{1}{(2 - \Bx)^2}
\Bigg\{
\left[
4 (1 - \Bx) {\cal H}_T - \Bx^2 {\cal E}_T + \Bx (2 - \Bx) \widetilde{\cal E}_T
\right]
\left( {\cal H}^\ast + \widetilde{\cal H}^\ast \right)
\\
&&\qquad\qquad
- 2 \widetilde{\cal H}_T
\left( \Bx^2 + (1 - \Bx) \frac{\Delta^2}{M^2} \right)
\left( {\cal H}^\ast + {\cal E}^\ast \right)
\nonumber\\
&&\qquad\qquad
-\Bx
\left(
\Bx {\cal H}_T - (2 - \Bx) \frac{\Delta^2}{4 M^2} \widetilde{\cal E}_T
\right)
\left( {\cal E}^\ast + \widetilde{\cal E}^\ast \right)
\nonumber\\
&&\qquad\qquad
- {\cal E}_T
\Bigg[
\left( \Bx^2 + (2 - \Bx)^2 \frac{\Delta^2}{4 M^2} \right)
{\cal E}^\ast
 +
\Bx^2 \frac{\Delta^2}{4 M^2} \widetilde{\cal E}^\ast
\Bigg]
\Bigg\} \, .
\nonumber
\end{eqnarray}

\begin{itemize}
\item Interference of Bethe-Heitler and DVCS amplitudes:
\end{itemize}
\begin{eqnarray}
{\cal C}^{\cal I}_{T,{\rm unp}}
\!\!\!&=&\!\!\!
- F_2 {\cal H}_T
+ 2 \left(F_1 + \frac{\Delta^2}{4 M^2} F_2\right) \widetilde{\cal H}_T
+ F_1 {\cal E}_T \, ,
\\
{\cal C}^{\cal I}_{T,{\rm LP}}
\!\!\!&=&\!\!\!
F_2
\left\{
{\cal H}_T +
\frac{\Bx}{2}
( 2 \widetilde{\cal H}_T + {\cal E}_T  + \widetilde{\cal E}_T )
\right\}
+ F_1 ( \Bx \widetilde{\cal H}_T   + \widetilde{\cal E}_T ) \, ,
\\
{\cal C}^{\cal I}_{T,{\rm TP}+}
\!\!\!&=&\!\!\!
(2 F_1 + \Bx F_2) {\cal H}_T
+ \Bx \left( \frac{x}{2} - \frac{\Delta^2}{4 M^2} \right)
\left[ 2 (F_1 + F_2) \widetilde{\cal H}_T + F_2 {\cal E}_T \right]
\\
&&+
\left\{
\Bx F_1
+
\left( \frac{\Bx^2}{2} + (2 - \Bx) \frac{\Delta^2}{4M^2} \right) F_2
\right\}
\widetilde{\cal E}_T \, ,
\nonumber\\
{\cal C}^{\cal I}_{T,{\rm TP}-}
\!\!\!&=&\!\!\! (2 F_1 + \Bx F_2) {\cal H}_T
- (2 - \Bx) \frac{\Delta^2}{4 M^2}
\left[
2(F_1 + F_2) \widetilde{\cal H}_T  +  F_2 {\cal E}_T
\right]
+ \Bx \left( F_1 + F_2 \frac{\Delta^2}{4 M^2} \right) \widetilde{\cal E}_T \, .
\nonumber\\
\end{eqnarray}
This set of formulae is the complete result for the real-photon
leptoproduction cross section in the twist-three approximation. Below,
presenting quantitative estimates,
we will not discuss the case of transversely polarized target, therefore,
the integration with respect to $\varphi$ gives $2 \pi$ on the
right-hand side of Eq.\ (\ref{WQ}).

%%%%%%%%%%%%%%%%%%%%%%%%%%%%%%%%%%%%%%%%%%%%%%%%%%%%%%%%%%%%%%%%%%%%%
\subsection{Structure of effective twist-three GPDs}
%%%%%%%%%%%%%%%%%%%%%%%%%%%%%%%%%%%%%%%%%%%%%%%%%%%%%%%%%%%%%%%%%%%%%

Note that the twist-three GPDs, having generic discontinuities at $|x|
= \xi$ \cite{KivPolSchTer00,RadWei00}, enter the CFFs
in a singularity-free combination (\ref{Tw3Eff}). Using the integrals of
the type
\begin{eqnarray*}
\int_{-1}^1 \frac{d x}{|\eta|}
\frac{1}{\xi - x - i 0} W\left(\frac{x}{\eta}, \frac{y}{\eta}\right)
= \frac{{\rm sign}(\eta)}{\eta + y} \ln\left(\frac{\eta + \xi}
{\xi - y - i 0} \right)
\end{eqnarray*}
in Eq.\ (\ref{Tw3Eff}),
the convolution of the coefficient functions $C^{(\mp)}_{(0)i}$ and the
$W$-kernels
results into an explicit form for the effective twist-three function,
\begin{eqnarray}
\label{Res-tw3eff}
{\cal F}^{\rm eff} (\xi)
\!\!\!&=&\!\!\!
\frac{2}{1 + \xi}{\cal F} (\xi)
+ 2 \xi
\frac{\partial}{\partial \xi}
\int_{- 1}^{1} d x \,
C^{3(\mp)} (\xi, x) F (x, \xi)
+
\frac{8 M^2 \xi}{( 1 - \xi^2 )( \Delta^2 - \Delta^2_{\rm min} )}
{\cal F}^\perp (\xi)
\\
&&\!\!\! - 2 \xi \int_{-1}^{1} d u \int_{-1}^{1} d x \,
C^{qGq} (\xi, x, u)
\left( S^+_{F} (-x, -u, -\xi) - S^-_{F} (x, u, -\xi) \right)
\Bigg\},
\nonumber
\end{eqnarray}
where ${\cal F}$ is defined in Eq.\ (\ref{DefTw3}). The summation over
quark flavors is implied, as in Eq.\ (\ref{Flavour}), and we have used
new notations for the coefficient functions weighted by the quark charge
\begin{eqnarray}
C_{(0)i}^{3(\mp)} (\xi, x)
\!\!\!&=&\!\!\!
\frac{Q^2_i}{\xi + x} \ln \frac{2 \xi}{\xi - x - i 0}
\mp
\frac{Q^2_i}{\xi - x} \ln \frac{2 \xi}{\xi + x - i 0} \, ,
\\
C_{(0)i}^{qGq} (\xi, x, u) \!\!\!&=&\!\!\! Q^2_i
\frac{{\partial^2}}{\partial x^2}
\frac{1 + u}{\xi + x}
\ln \left( \frac{2 \xi}{\xi - x - i 0} \right) \, .
\end{eqnarray}
The $-$ ($+$) sign stands here for the vector (axial) sectors, i.e., for
$\{{\cal H} (\widetilde {\cal H}), \ {\cal E} (\widetilde {\cal E})\}$.
Finally, the functions ${\cal F}^\perp$ are
\begin{eqnarray}
\label{Def-Fperb}
{\cal H}^\perp (\xi)
\!\!\!&=&\!\!\!\
- \frac{\Delta^2}{4 M^2}
\int_{- 1}^{1} d x
\left\{
\xi \, C^{3(-)} (x, \xi) \left( H + E \right) (x, \xi)
-
C^{3(+)} (x, \xi) \widetilde H (x, \xi)
\right\} \, ,
\\
{\cal E}^\perp (\xi)
\!\!\!&=&\!\!\!\
\int_{- 1}^{1} d x
\left\{
\xi \, C^{3(-)} (x, \xi) \left( H + E \right) (x, \xi)
-
C^{3(+)} (x, \xi) \widetilde H (x, \xi)
\right\}
\, ,
\nonumber\\
\widetilde {\cal H}^\perp (\xi)
\!\!\!&=&\!\!\!
\int_{- 1}^{1} d x
\left\{
\xi \left(1 - \frac{\Delta^2}{4 M^2} \right)
C^{3(+)} (x, \xi) \widetilde H (x, \xi)
+
\frac{\Delta^2}{4 M^2} C^{3(-)} (x, \xi)
\left( H + E \right) (x, \xi)
\right\}
\, ,
\nonumber\\
\widetilde {\cal E}^{\perp } (\xi)
\!\!\!&=&\!\!\!
\frac{1}{\xi}
\int_{- 1}^{1} d x
\left\{
C^{3(+)} (x, \xi) \widetilde H (x, \xi)
-
\xi \, C^{3(-)} (x, \xi) \left( H + E \right) (x, \xi)
\right\} \, .
\nonumber
\end{eqnarray}

We should note that the kinematical factor $(\Delta^2 - \Delta_{\rm
\min}^2)^{- 1}$ in Eq.\ (\ref{Res-tw3eff}) cancels out in the final
results for the angular harmonics calculated in sections
\ref{SubSec-AziAngDep-DVCS} and \ref{SubSec-AziAngDep-INT}. Therefore,
it does not lead to a singular behavior of the Fourier coefficients
in the cross section (\ref{AmplitudesSquared}). For all interference
as well as for unpolarized and longitudinally polarized squared DVCS terms
this cancellation already appears in the $\cal C$ functions given in
sections \ref{App-CoeFun}, see Ref.\ \cite{BelKirMulSch01b} for
explicit examples. However, in $c^{\rm DVCS}_{1,{\rm TP}}$ and
$s^{\rm DVCS}_{1,{\rm TP}}$ this kinematical singularity is annihilated
by the prefactor in Eq.\ (\ref{Res-Mom-DVCS-TP-Tw3}).

%%%%%%%%%%%%%%%%%%%%%%%%%%%%%%%%%%%%%%%%%%%%%%%%%%%%%%%%%%%%%%%%%%%%%
\section{Physical observables and access to GPDs}
\label{Sec-PhyObs}
%%%%%%%%%%%%%%%%%%%%%%%%%%%%%%%%%%%%%%%%%%%%%%%%%%%%%%%%%%%%%%%%%%%%%

As we have seen, the cross section for leptoproduction of the real photon
possesses a quite rich angular structure. The goal of experimental
measurements is to pin down the GPDs and this requires a clean
disentanglement of different components of the cross section (\ref{WQ}).
A general strategy, which follows from the analytical results, given above,
will be discussed in section \ref{SubSec-PhyObs-GPDs}: separation of
twist-two and -three sectors\footnote{We talk about the twist-two and
twist-three sectors, however, beyond the twist-three approximation each
of the associated observables gets corrected by higher-twist, i.e.,
power-suppressed contributions.}, measuring the CFFs, and access to the
GPDs. To go along this line, we introduce appropriate asymmetries in
section \ref{SubSec-PhyObs-Asy} and demonstrate how a Fourier transform
can distinguish between the interference and squared DVCS contributions.

%%%%%%%%%%%%%%%%%%%%%%%%%%%%%%%%%%%%%%%%%%%%%%%%%%%%%%%%%%%%%%%%%%%%%
\subsection{Unraveling GPDs}
\label{SubSec-PhyObs-GPDs}
%%%%%%%%%%%%%%%%%%%%%%%%%%%%%%%%%%%%%%%%%%%%%%%%%%%%%%%%%%%%%%%%%%%%%

\begin{table}[t]
\begin{center}
\begin{tabular}{|l|c||c|c|c|c|c|c|c|c|c|}
\hline
\multicolumn{2}{|c||}{sector}
&
\multicolumn{4}{c|}{harmonics in $\cal I$}
&
\multicolumn{1}{c|}{extraction}
&
\multicolumn{1}{c|}{$P$ of}
&
\multicolumn{2}{c|}{$\Delta_\perp^{l}$ behavior}
\\
\multicolumn{1}{|c|}{twist}
&
\multicolumn{1}{c||}{$\cal C$'s}
&
\multicolumn{1}{c|}{unp}
&
\multicolumn{1}{c|}{LP}
&
\multicolumn{1}{c|}{TP$_x$}
&
\multicolumn{1}{c|}{TP$_y$}
&
\multicolumn{1}{c|}{of CFFs}
&
\multicolumn{1}{c|}{$\cQ^{-P}$}
&
\multicolumn{1}{c|}{unp, LP}&\multicolumn{1}{c|}{TP}
\\
\hline\hline
two
&
$\Re{\rm e} {\cal C}({\cal F}),\ \Delta{\cal C}({\cal F}) $
&
$c_1$, $c_0$
&
$c_1$, $c_0$
&
$c_1$, $c_0$
&
$s_1$, -
&
over compl.
&
1,2
&
1,0
&
0,1
\\
\hline
&
$\Im{\rm m} {\cal C}({\cal F}),\ \Delta{\cal C}({\cal F})$
&
$s_1$, -
&
$s_1$, -
&
$s_1$, -
&
$c_1$, $c_0$
&
over compl.
&
1,2
&
1,0
&
0,1
\\
\hline\hline
three
&
$\Re{\rm e} {\cal C}({\cal F}^{\rm eff})$
&
$c_2$
&
$c_2$
&
$c_2$
&
$s_2$
&
complete
&
2
&
2
&
1
\\
\hline
&
$\Im{\rm m} {\cal C}( {\cal F}^{\rm eff})$
&
$s_2$
&
$s_2$
&
$s_2$
&
$c_2$
&
complete
&
2
&
2
&
1
\\
\hline\hline
two
&
$\Re{\rm e} {\cal C}_T({\cal F}_T)$
&
$c_3$
&
-
&
-
&
-
&
$1\times \Re{\rm e}$ of 4
&
1
&
3
&
2
\\
\hline
&
$\Im{\rm m} {\cal C}_T({\cal F}_T)$
&
-
&
$s_3$
&
$s_3$
&
$c_3$
&
$3\times \Im{\rm m}$ of 4
&
1
&
3
&
2
\\
\hline
\end{tabular}
\end{center}
\caption{\label{Tab-FCs-Int} Fourier coefficients $c_{i}^{\cal I}$
and $s_{i}^{\cal I}$ of the interference term defined in section
\ref{SubSec-AziAngDep-INT}, while the corresponding $\cal C$ coefficients
are given in section \ref{App-CoeFun}.}
\end{table}

\begin{table}[t]
\begin{center}
\begin{tabular}{|l|c||c|c|c|c|c|c|c|c|c|}
\hline
\multicolumn{2}{|c||}{interference of}
&
\multicolumn{4}{c|}{harmonics in $|{\cal T}^{\rm DVCS}|^2$}
&
\multicolumn{1}{c|}{extraction}
&
\multicolumn{1}{c|}{$P$ of}
&
\multicolumn{2}{c|}{$\Delta_\perp^{l}$ behavior}
\\
\multicolumn{1}{|c|}{twist}
&
\multicolumn{1}{c||}{$\cal C$'s}
&
\multicolumn{1}{c|}{unp}
&
\multicolumn{1}{c|}{LP}
&
\multicolumn{1}{c|}{TP$_x$}
&
\multicolumn{1}{c|}{TP$_y$}
&
\multicolumn{1}{c|}{of CFFs}
&
\multicolumn{1}{c|}{$\cQ^{-P}$}
&
\multicolumn{1}{c|}{unp, LP}
&
\multicolumn{1}{c|}{TP}
\\
\hline\hline
two \& two
&
$\Re{\rm e} {\cal C}({\cal F},{\cal F}^\ast)$
&
$c_0$
&
$c_0$
&
$c_0$
&
-
&
$3\times \Re{\rm e}/\Im{\rm m}$
&
2
&
0
&
1
\\
\hline
&
$\Im{\rm m} {\cal C}({\cal F},{\cal F}^\ast)$
&
-
&
-
&
-
&
$c_0$
&
$1\times \Im{\rm m}/\Re{\rm e}$
&
2
&
-
&
1
\\
\hline\hline
two \& three
&
$\Re{\rm e} {\cal C}({\cal F}^{\rm eff},{\cal F}^\ast)$
&
$c_1$
&
$c_1$
&
$c_1$
&
$s_1$
&
$4 \times \Re{\rm e}/\Im{\rm m}$
&
3
&
1
&
0
\\
\hline
&
$\Im{\rm m} {\cal C}({\cal F}^{\rm eff},{\cal F}^\ast)$
&
$s_1$
&
$s_1$
&
$s_1$
&
$c_1$
&
$4 \times\Im{\rm m}/\Re{\rm e}$
&
3
&
1
&
0
\\
\hline\hline
two \& two
&
$\Re{\rm e} {\cal C}_T({\cal F}_T,{\cal F}^\ast)$
&
$c_2$
&
-
&
-
&
-
&
$1\times \Re{\rm e}/\Im{\rm m}$
&
2
&
2
&
-
\\
\hline
&
$\Im{\rm m} {\cal C}_T({\cal F}_T,{\cal F}^\ast)$
&
-
&
$s_2$
&
$s_2$
&
$c_2$
&
$3\times \Im{\rm m}/\Re{\rm e}$
&
2
&
2
&
1
\\
\hline
\end{tabular}
\end{center}
\caption{\label{Tab-FCs-DVCS} Fourier coefficients $c_{i}^{\rm DVCS}$ and
$s_{i}^{\rm DVCS}$ of the squared DVCS amplitude $|{\cal T}^{\rm DVCS}|^2$
defined in section \ref{SubSec-AziAngDep-DVCS}, while the corresponding
$\cal C$ coefficients are given in section \ref{App-CoeFun}. }
\end{table}

So far we have introduced eight CFFs at the twist-two level, with four
of them from the gluonic transversity contribution. Another four new
CFFs arise at the twist-three level. These sectors can be separated due
to their characteristic azimuthal dependence as summarized in Table
\ref{Tab-FCs-Int} for the interference $\cal I$ and in Table
\ref{Tab-FCs-DVCS} for the squared DVCS amplitude, respectively.

Let us first examine the issue of the dominance of each of the three terms
(\ref{Par-BH}-\ref{InterferenceTerm}) in the leptoproduction cross section
(\ref{WQ}) in different kinematical regions. To do so we have to know the
functional dependence of the Fourier coefficients given in Eqs.\
(\ref{Par-BH}-\ref{InterferenceTerm}) on scaling variables and transferred
momenta. For instance, apart from an explicit $\Bx$-dependence of
multiplicative prefactors there is one hidden in CFFs. It will be argued in
section \ref{Sec-CFF} that the latter behave like $\Bx^{-1}$ in the
small-$\Bx$ region. Thus, for general kinematical settings we expect for the
scattering on the unpolarized target from Eqs.\ (\ref{Def-FC-BH-unp0}),
(\ref{Res-Mom-DVCS-UP}), and (\ref{Res-IntTer-unp}) that $c_0^{\rm BH}
\sim \Bx^2 c_0^{\rm DVCS} \sim \Bx c_1^{\cal I}/K$. Taking now the
kinematical prefactors in Eqs.\ (\ref{Par-BH}-\ref{InterferenceTerm}) into
account, and the behavior of the BH-propagators (\ref{Par-BH-Pro}), we
realize that the ratio of the DVCS to BH amplitude behaves like $\sqrt{-
(1-y) \Delta^2/ y^2 \cQ^2 }$. Obviously, for small (large) $y$ the DVCS (BH)
one dominates. As compared to the squared amplitudes, the interference term
has an additional factor $\sqrt{\Delta^2_\perp/\Delta^2}$. Note that the
beam spin-flip contributions provide always an additional damping by the
factor $y$. For the unpolarized or longitudinally polarized target, higher
harmonics in any of the three terms are suppressed by  powers of $K$.
However, in the case of gluonic transversity, this goes hand in hand with an
enhancement by $\cQ^2/M^2$. It is important that lower harmonics in the
interference term, i.e., $c_0^{\cal I}$, appear at the twist-three level.
Since, they are not proportional to the factor $K$ as compared to $c_1$,
which is, they can be rather important and sizable close to the kinematical
boundaries. Fortunately, they only depend on twist-two CFFs, thus, we listed
them in Table \ref{Tab-FCs-Int} in the twist-two sector. In the case of
transversely polarized target, we observe that both higher and lower
twist-three harmonics are suppressed by one power of $K$ in the interference
term.

The analytical structure and simple counting rules, we gave, provide a
guideline to separate the three different parts in the leptoproduction
cross section. In single spin-flip experiments, which give access to the
imaginary part of CFFs, the BH cross section drops out, while in unpolarized or
double spin-flip experiments it does not and has to be subtracted. This can
certainly be done for not too small values of $y$. The interference and
squared DVCS terms have different azimuthal angular dependencies due to the
presence of the BH-propagators in the former. In principle, this fact can be
used to separate them by a Fourier analysis. However, this method requires
from the experimental side very high-statistics data and from the
theoretical side a better understanding of twist-four contributions.

Due to the different charge parity of individual components, it is possible
to use the charge asymmetry
to separate the interference and squared DVCS terms. The interference term
is charge-odd and can be extracted in facilities that possess both
positively and negatively charged leptons \cite{KroSchGui96}, i.e.,
\begin{eqnarray}
d\sigma^+ - d\sigma^-
\propto
\frac{1}{\Bx y^3 {\cal P}_1 (\phi) {\cal P}_2 (\phi) \Delta^2}
\left\{
c_0^{\cal I}
+
\sum_{n = 1}^3
\left[
c_n^{\cal I} \cos(n \phi) +  s_n^{\cal I} \sin(n \phi)
\right]
\right\}
\, .
\end{eqnarray}
Its measurement and consequent extraction of separate harmonics provides the
real (unpolarized or double spin-flip experiments) and imaginary (single
spin-flip experiments) part of linear combinations of twist-two and -three
CFFs. An explicit projection procedure of these harmonics will be discussed
below in section \ref{SubSec-PhyObs-ChaAsy}. Moreover, the charge-even part
is given by the sum of the BH and DVCS cross section. The subtraction of the
former gives then the Fourier coefficients of the latter:
\begin{eqnarray}
d^+ \sigma + d ^-\sigma - 2d^{\rm BH} \sigma  \propto
\frac{1}{y^2 {\cal Q}^2}\left\{
c^{\rm DVCS}_0
+ \sum_{n = 1}^2
\left[
c^{\rm DVCS}_n \cos (n\phi) + s^{\rm DVCS}_n \sin (n \phi)
\right]
\right\}
\, .
\end{eqnarray}

In the case of the four CFFs ${\cal F} = \{ {\cal H}, {\cal E}, \widetilde
{\cal H}, \widetilde {\cal E} \}$ we have eight observables given by the
first harmonics $\cos(\phi)$ and $\sin(\phi)$ of the interference term,
which are accessible away from the kinematical boundaries in polarized beam
and target experiments. Thus, experiments with both longitudinally and
transversely polarized target can measure all eight Fourier coefficients
$c_{1,{\mit\Lambda}}^{\cal I}$ and $s_{1,{\mit\Lambda}}^{\cal I}$ and, thus,
also $\Re{\rm e}/\Im{\rm m}{\cal C}^{\cal I}_{\mit\Lambda}$ with
${\mit\Lambda} = \{ {\rm unp}, {\rm LP}, {\rm TP}_x, {\rm TP}_y \}$. Knowing
these ${\cal C}$ functions, we can invert them to obtain the CFFs:
\begin{eqnarray}
\label{Inv-CFF-H}
{\cal H}
\!\!\!&=&\!\!\!
\frac{2 - \Bx}{(1 - \Bx) D}
\Bigg\{
\left[
\left(
2 - \Bx + \frac{4 \Bx^2 M^2}{(2 - \Bx) \Delta^2}
\right)
F_1 + \frac{\Bx^2}{2 - \Bx} F_2
\right] {\cal C}^{\cal I}_{\rm unp}
\\
&&\hspace{2cm} - (F_1 + F_2)
\left[
\Bx {\cal C}^{\cal I}_{{\rm LP}}
+
\frac{2 \Bx^2  M^2}{(2 - \Bx)\Delta^2}
\left(
\Bx {\cal C}^{\cal I}_{{\rm LP}} - {\cal C}^{\cal I}_{{\rm TP}+}
\right)
\right]
+
F_2 {\cal C}^{\cal I}_{{\rm TP}-}
\Bigg\}
\, , \nonumber\\
\label{Inv-CFF-E}
{\cal E}
\!\!\!&=&\!\!\!
\frac{2 - \Bx}{(1 - \Bx) D}
\Bigg\{
\left[
4 \frac{1 - \Bx}{2 - \Bx}F_2
-
\frac{4 M^2 \Bx^2}{(2 - \Bx) \Delta^2} F_1
\right]
{\cal C}^{\cal I}_{\rm unp}
+
\frac{4 \Bx M^2}{(2 - \Bx) \Delta^2}
\left( F_1 + F_2 \right)
\\
&&\hspace{2cm}
\times
\left(
\Bx {\cal C}^{\cal I}_{{\rm LP}} - {\cal C}^{\cal I}_{{\rm TP}+}
\right)
+
\frac{4 M^2}{\Delta^2} F_1 {\cal C}^{\cal I}_{{\rm TP}-}
\Bigg\}
\, , \nonumber\\
\label{Inv-CFF-tH}
\widetilde{\cal H}
\!\!\!&=&\!\!\!
\frac{2 - \Bx}{(1 - \Bx) D}
\Bigg\{
(2 - \Bx) F_1 {\cal C}^{\cal I}_{{\rm LP}}
- \Bx (F_1 + F_2) {\cal C}^{\cal I}_{{\rm unp}}
+
\left[
\frac{2 \Bx M^2}{\Delta^2} F_1 + F_2
\right]
\nonumber\\
&&\hspace{2cm}
\times
\left(
\Bx {\cal C}^{\cal I}_{{\rm LP}} - {\cal C}^{\cal I}_{{\rm TP}+}
\right)
\Bigg\}
\, , \\
\label{Inv-CFF-tE}
\widetilde{\cal E}
\!\!\!&=&\!\!\!
\frac{2 - \Bx}{(1 - \Bx) D}
\Bigg\{
\frac{4 M^2}{\Delta^2} \left( F_1 + F_2 \right)
\left(
\Bx {\cal C}^{\cal I}_{{\rm unp}} + {\cal C}^{\cal I}_{{\rm TP}-}
\right)
+
\left[
4\frac{1 - \Bx}{\Bx} F_2 - \frac{4\Bx M^2}{\Delta^2} F_1
\right]
{\cal C}^{\cal I}_{\rm LP}
\\
&&\hspace{2cm}
- \frac{4 (2 - \Bx) M^2}{\Bx \Delta^2} F_1
{\cal C}^{\cal I}_{{\rm TP}+}
\Bigg\}
\, , \nonumber
\end{eqnarray}
where
\begin{eqnarray*}
D = 4 \left( F_1^2 - \frac{\Delta^2}{4 M^2} F_2^2 \right)
\left(
1 - \frac{\Delta^2_{\rm min}}{\Delta^2}
\right) \, .
\end{eqnarray*}
Consequently, the four Fourier coefficients $c_{0,\mit\Lambda}^{\cal I}$ as
well as the four twist-two DVCS coefficients $c_{0,\mit\Lambda}^{\rm DVCS}$
can serve as experimental consistency checks. Alternatively, they can be
used to extract CFFs. Thus, experiments with longitudinally polarized target
have the potential to extract the real part of all four CFFs as well as two
linear combinations of their imaginary parts from the interference term
alone. The missing two imaginary parts could then, in principle, be obtained
from the DVCS cross section, i.e., by measuring $c_{0,{\rm unp}}^{\rm DVCS}$
and $c_{0,{\rm LP}}^{\rm DVCS}$.

For a polarized beam and target with all polarization options, the
real and imaginary part of all four CFFs in the twist-three sector can be
extracted from the interference term alone by projection onto the
$\cos(2\phi)$ and $\sin(2\phi)$ harmonics and using Eqs.\
(\ref{Inv-CFF-H}-\ref{Inv-CFF-tE}). Alternatively, knowing the twist-two
sector and having only a longitudinally polarized target, one can employ in
addition the squared DVCS term, i.e., its $\cos(\phi)$ and $\sin(\phi)$
harmonics, to access the full twist-three sector.

For gluonic transversity, the $\cos(3\phi)$ and $\sin(3\phi)$ harmonics in
the interference term can only provide  one imaginary and three real parts of
certain linear combinations of ${\cal F}_T$. Missing information can, in
principle, be obtained from the $\cos(2\phi)$ and $\sin(2\phi)$ harmonics of
the squared DVCS term. Note that here again a polarized beam and target with
all polarizations is necessary. Moreover, the gluonic transversity is
suppressed by $\alpha_s/\pi$, so one expects a stronger contamination by
twist-four effects \cite{KivMan01}.

As we discussed, a combination of the charge asymmetry with different
nucleon/lepton polarizations and projection of the corresponding harmonics
provides, at least in principle, a way to explore the real and imaginary
part of all CFFs (cf.\ \cite{Ji98,GouDiePirRal97,BelMulNieSch00}). This
gives maximal access to all GPDs, which enter in a convolution with the
real or imaginary part of the coefficient functions, as we have established
above in Eqs.\ (\ref{DefTw3}) and (\ref{Def-ConTra}). Since these formulae
cannot be deconvoluted in practice \cite{Fre00}, one has to rely on models
with a set of free parameters, which has to be adjusted to experimental data
on CFFs. This issue will be discussed below for different available
experiments.

%%%%%%%%%%%%%%%%%%%%%%%%%%%%%%%%%%%%%%%%%%%%%%%%%%%%%%%%%%%%%%%%%%%%%
\subsection{Asymmetries}
\label{SubSec-PhyObs-Asy}
%%%%%%%%%%%%%%%%%%%%%%%%%%%%%%%%%%%%%%%%%%%%%%%%%%%%%%%%%%%%%%%%%%%%%

The measurements of the cross section (\ref{WQ}) in different setups, as
discussed in the preceding section, would directly lead to determination of
the CFFs. However, on the experimental side it is simpler to measure
asymmetries since they escape the normalization issue. Thus, we now discuss
the separation of twist-two and -three sectors in their terms. The charge
and lepton-spin asymmetries used in our previous studies, see, e.g.,
\cite{BelMulNieSch00}, were restricted to the leading twist approximation
for the amplitudes and cannot serve the purpose since they are inevitably
contaminated by power-suppressed effects. Due to our poor knowledge of
multi-particle correlations inside the nucleon, this affects the theoretical
predictions in an uncontrollable manner\footnote{Note that experimental
results from polarized deeply inelastic scattering give only a constraint
for the forward limit of the antiquark-gluon-quark content of $\widetilde
{\cal H}^{qGq}$. On the other hand, theoretical arguments that flavor small
multi-parton contributions rely on model assumptions \cite{BalPolWei98}. In
our opinion, one cannot exclude that these correlations might be important.
Thus, one should not rely on the WW-approximation and rather search for
their signature in data.}.

The charge asymmetry
\begin{equation}
\label{Def-CA-old}
A_{\rm C}
= \left.
\left(
\int_{-\pi/2}^{\pi/2} d \phi
\frac{d^+\sigma^{\rm unp} - d^-\sigma^{\rm unp}}{d \phi}
- \int_{\pi/2}^{3\pi/2} d\phi
\frac{d^+ \sigma^{\rm unp} - d^-\sigma^{\rm unp}}{d \phi}
\right)
\right/
\int_{0}^{2\pi} d\phi
\frac{d{^{-}\!\sigma}^{\rm unp} +  d{^{+}\!\sigma}^{\rm unp}}{d\phi}
\, ,
\end{equation}
for unpolarized settings contains the contribution of all harmonics due
to the presence of the non-negligible $\phi$-dependence of BH propagators:
\begin{equation}
A_{\rm C} \propto \sum_{n = 0}^3 I^c_{1,n} (2 K/J) c_n^{\cal I}
\, ,
\end{equation}
with
\begin{equation}
I^c_{1,n} (2 K/J) \propto
\left(
\int_{-\pi/2}^{\pi/2} d \phi
\frac{\cos(n \phi)}{{\cal P}_1(\phi) {\cal P}_2(\phi)}
-
\int_{\pi/2}^{3\pi/2} d \phi
\frac{\cos(n \phi)}{{\cal P}_1(\phi) {\cal P}_2(\phi)}
\right)
\, ,
\end{equation}
while the normalization is not affected by twist-three corrections. If the
final photon is collinear to the incoming (massless) lepton, ${\cal P}_1(\phi)$
is peaked at $\phi = \pi$. Thus, the ratio $I^c_{1,n}/I^c_{1,1}$ approaches
plus or minus $1$ and all harmonics contribute on equal footing. However, since
$K$ is then of  order $\Delta^2/\cQ^2$, only $c_0^{\cal I}$ and $c_1^{\cal I}$
give essential contributions at the same order in $\Delta^2/\cQ^2$. Thus,
for this asymmetry $c_0^{\cal I}$ may gives an essential effect for large
$y$, since the twist-two part becomes small. For $y \ll y_{\rm col}$, all
twist-three harmonics  are suppressed in addition by the $K$-factor.
From the expansion of $I^c_{1,n}$ in powers of $K$ we get
\begin{eqnarray}
A_{\rm C}
\propto
c_{1,{\rm unp}}^{\cal I}
-
\frac{1}{3} c_{3,{\rm unp}}^{\cal I}
-
\frac{2 (3 - 2 y)}{2 - y} \frac{K}{1 - y}
\left(
c_{0,{\rm unp}}^{\cal I} - \frac{1}{3} c_{2,{\rm unp}}^{\cal I}
\right)
+
{\cal O} \left( K^2, \epsilon \right)
\, .
\end{eqnarray}
If $K$ is not too small, e.g., $K \sim 1/3$ for $-\Delta^2/\cQ^2 \sim 1/10$
and $- \Delta^2 \gg - \Delta^2_{\rm min}$, as it is the case in the present
fixed target experiments, we realize that $c_{0,{\rm unp}}^{\cal I}$ is in
fact not numerically suppressed in comparison to $c_{1,{\rm unp}}^{\cal I}$.
So we would expect a contamination of the leading twist-two prediction by a
$\sqrt{-\Delta^2/\cQ^2}$ suppressed term, which, however, contains only
twist-two CFFs. Even if this $1/\cQ$-contamination is small, $A_{\rm C}$
still fails to extracts solely the $c^{\cal I}_{1, {\rm unp}}$ coefficient
of the interference term and gets additive correction from $c^{\cal I}_{3,
{\rm unp}}$, which stems from the gluon transversity (and higher, more or
equal to four, twist effects). Since the gluon contribution is suppressed by
a power of the strong coupling $\alpha_s$, it may affect the twist-two
coefficient $c^{\cal I}_{1, {\rm unp}}$ in a modest way, though. Therefore,
as a first order approximation the definition (\ref{Def-CA-old}) can be used
for an order of magnitude estimate of the effects.

The (definite charge) beam-spin asymmetry on an unpolarized target
\begin{equation}
\label{Def-SSA-old}
A_{\rm SL}
=
\left.
\left(
\int_0^\pi
d \phi
\frac{
d \sigma^\uparrow - d \sigma^\downarrow
}{
d\phi
}
- \int_\pi^{2\pi}
d \phi
\frac{
d \sigma^\uparrow - d \sigma^\downarrow
}{
d \phi
}
\right)
\right/
\int_0^{2\pi} d \phi
\frac{
d \sigma^\uparrow + d \sigma^\downarrow
}{
d \phi
}
\, ,
\end{equation}
does not separate the interference term alone. So it does contain contributions
from the squared DVCS amplitude:
\begin{equation}
A_{\rm SL}
\propto
\pm \sum_{n = 1}^2 I^s_{1,n}(2 K/J) s_{n,{\rm unp}}^{\cal I}
-
\frac{\Delta^2}{y \cQ^2} \Bx s_{1,{\rm unp}}^{\rm DVCS}
\, ,
\end{equation}
with
\begin{equation}
I^s_{1,n}(2 K/J)
= - \frac{1}{y^2}
\left(
\int_{0}^{\pi} d \phi
\frac{\sin(n \phi)}{{\cal P}_1(\phi) {\cal P}_2(\phi) }
-
\int_{\pi}^{2\pi} d \phi
\frac{\sin(n \phi)}{{\cal P}_1(\phi) {\cal P}_2(\phi)}
\right)
\, . \nonumber
\end{equation}
In the collinear limit, the asymmetry vanishes. While for $y \to 0$, it is
determined by the twist-three coefficient $s_{1,{\rm unp}}^{\rm DVCS}$. The
expansion with respect to $K$ reads:
\begin{eqnarray}
\label{ASLexp}
A_{\rm SL}
\propto
s_{1,{\rm unp}}^{\cal I}
-
\frac{2 (3 - 2 y)}{3(2 - y)} \frac{K}{1 - y} s_{2,{\rm unp}}^{\cal I}
-
\frac{(1-y)(2-y)\Delta^2}{y \cQ^2} \Bx s_{1,{\rm unp}}^{\rm DVCS}
+
{\cal O} \left( K^2, \epsilon \right) \, .
\end{eqnarray}
The $1/{\cal Q}$-suppressed effect stemming from the BH propagators induces
a contamination by the second harmonic suppressed by $K/(1-y)$. Depending on
the kinematics and the size of multi-particle contributions, this
contamination together with $s_{1,{\rm unp}}^{\rm DVCS}$ may not allow clean
access to the twist-two GPDs even from high-precision measurements of this
asymmetry. Note that for general reasons, $s_{3,{\rm unp}}^{\cal I}$
($s_{2,{\rm unp}}^{\rm DVCS}$) is absent in the unpolarized interference
(squared DVCS) term. The normalization of $A_{\rm SL}$ is affected by
$1/\cQ$-effects in the interference term, however, mainly due to the
coefficient $c^{\cal I}_{0,{\rm unp}}$.

%%%%%%%%%%%%%%%%%%%%%%%%%%%%%%%%%%%%%%%%%%%%%%%%%%%%%%%%%%%%%%%%%%%%%
\subsubsection{Facilities with positively and negatively charged lepton beams}
\label{SubSec-PhyObs-ChaAsy}
%%%%%%%%%%%%%%%%%%%%%%%%%%%%%%%%%%%%%%%%%%%%%%%%%%%%%%%%%%%%%%%%%%%%%

As we have already mentioned, when lepton beams of both charges are
available, this provides a clean separation of twist-two and -three GPDs. In
these settings one can discuss charge-odd and -even parts of the cross
section (\ref{WQ}), which extract the interference, and squared DVCS and BH
amplitudes, respectively. The integrated charge-even part of the cross
section does not contain any twist-three contributions --- the interference
term cancels there, while after azimuthal averaging the $c_0$ coefficient of
the squared DVCS and all harmonics of squared BH amplitudes survive. So we
will use it as a unique normalization of the asymmetries discussed below.
Namely, independent from the target polarization we introduce
\begin{equation}
\label{Def-Nor-Asy}
{\cal N}^{- 1}_{+-}
\equiv
\int_0^{2\pi} d \phi \frac{d^+ \sigma^{\rm unp} + d^-\sigma^{\rm unp}}{d\phi}
=
2\int_0^{2\pi} d\phi \frac{d^{\rm BH} \sigma^{\rm unp} +
d^{\rm DVCS}\sigma^{\rm unp}}{d\phi}
\, .
\end{equation}

\begin{itemize}
\item Charge-odd part:
\end{itemize}
In this case, we end up with the interference term alone. However, because of
the $\phi$-dependence of the BH propagators we have to include an additional
weight factor and use the measure
\begin{eqnarray}
d w =  2 \pi
\frac{
{\cal P}_1 (\phi) {\cal P}_2 (\phi) d \phi
}{
\int_0^{2\pi} {\cal P}_1 (\phi') {\cal P}_2 (\phi') d \phi'
}
\, ,
\end{eqnarray}
for the azimuthal integration in order to compensate for the strong
$\phi$-dependence of the product of lepton propagators. The measure $d w$ has
the properties
\begin{eqnarray*}
\int_0^{2\pi} d w = 2 \pi
\, , \qquad
d w (\phi) = d w (- \phi) = d w (\phi + 2 \pi) \, .
\end{eqnarray*}
Now we can exactly separate the Fourier coefficients in Eq.\
(\ref{InterferenceTerm}). Note that $d w$ has its minimum at $\phi = \pi$,
when the outgoing photon lies in the lepton scattering plane. In the case
when its momentum is collinear to the lepton beam, this minimum approaches
zero in the massless limit.

To project out different harmonics, on can either (i) do the azimuthal
averaging with appropriate weights, namely,
\begin{equation}
\cos (n \phi) d w
\, , \qquad\qquad
\sin (n \phi) d w
\, ,
\end{equation}
where $n = 0, \dots, 3$, or, (ii) use the fact that there is only a finite
number of terms in the Fourier sum of the cross section and integrate over
different partitions of the azimuthal sphere. For $n>0$ this has an advantage of
having a
numerical enhancement by the factor of $4/\pi$ with respect to the first method.
Let us present the charge odd asymmetries (CoA), which distinguish the cosine,
$\cos (n \phi)$, and sine, $\sin (n \phi)$, harmonics.

The $\cos$-harmonics, i.e., $c^{\cal I}_n$ coefficients, are projected out
by means of the integrals
\begin{eqnarray}
\label{Def-CA-Mom-c-0dd}
\CoA^{\mit\Lambda}_{c(0)}
\!\!\!&=&\!\!\!
{\cal N}_{+-}
\int_0^{2\pi} d w
\frac{
d^+ \sigma^{\mit\Lambda} - d^- \sigma^{\mit\Lambda}
}{
d\phi
}
\, , \\
\CoA^{\mit\Lambda}_{c(1)}
\!\!\!&=&\!\!\!
{\cal N}_{+-}
\left(
\int_{-\pi/2}^{\pi/2} d w
\frac{
d^+ \sigma^{\mit\Lambda} - d^- \sigma^{\mit\Lambda}
}{
d\phi
}
-
\int_{\pi/2}^{3\pi/2} d w
\frac{
d^+ \sigma^{\mit\Lambda} - d^- \sigma^{\mit\Lambda}
}{
d\phi
}
\right)
+ \frac{1}{3} \CoA^{\mit\Lambda}_{c(3)}
\, , \\
\CoA^{\mit\Lambda}_{c(n)}
\!\!\!&=&\!\!\!
{\cal N}_{+-}
\sum_{k = 1}^{2 n} (-1)^{k + 1}
\int_{(2 k - 3) \pi / (2n)}^{( 2 k - 1)\pi / (2n)} d w
\frac{
d^+ \sigma^{\mit\Lambda} - d^- \sigma^{\mit\Lambda}
}{
d\phi
} \, ,
\end{eqnarray}
with $n$ running over $n = 2, 3$ in the last equation. Here we adopt the analogous
decomposition of asymmetries as for the cross section in Eq.\ ({\ref{Def-DecCroSec})
so that ${\mit\Lambda} = \{ {\rm unp}, {\rm LP}, {\rm TP} \}$. The projection can be
achieved by an appropriate flip of the target polarization vector. For a transversely
polarized target, asymmetries are given in terms of two different combinations of
CFFs. They are separable by the projection of the first odd and even harmonics in
$\varphi$, while the average $\int_0^{2\pi} d \varphi\, {\CoA}^{\rm TP} = 0$ vanishes.

Next, the $\sin$-harmonics, $s^{\cal I}_n$, can analogously be separated  with
the help of the formulae
\begin{eqnarray}
\label{Def-CA-Mom-s-Odd}
{\CoA}^{\mit\Lambda}_{s(1)}
\!\!\!&=&\!\!\!
{\cal N}_{+-}
\left(
\int_{0}^{\pi} d w
\frac{
d^+ \sigma^{\mit\Lambda} - d^- \sigma^{\mit\Lambda}
}{
d \phi
}
-
\int_{\pi}^{2\pi} d w
\frac{
d^+ \sigma^{\mit\Lambda} - d^- \sigma^{\mit\Lambda}
}{
d \phi
}
\right)
- \frac{1}{3} {\CoA}^{\mit\Lambda}_{s(3)}
\, , \\
\label{CoAsn}
{\CoA}^{\mit\Lambda}_{s(n)}
\!\!\!&=&\!\!\!
{\cal N}_{+-}
\sum_{k = 1}^{2n} (-1)^{k + 1}
\int_{(k - 1)\pi / n}^{k \pi / n}
d w
\frac{
d^+ \sigma^{\mit\Lambda} - d^- \sigma^{\mit\Lambda}
}{
d \phi
} \, ,
\end{eqnarray}
where $n = 2, 3$.

\begin{itemize}
\item Charge-even part:
\end{itemize}
Furthermore, we define the azimuthal asymmetries of the charge-even part.
To do this in the cleanest way, we subtract at first the BH term.
This might be possible in practice in the analysis of experimental data
since the BH cross section is known exactly (up to electromagnetic
radiative corrections), see section \ref{BHcrosssection}, with the nucleon
form factors measured elsewhere. This gives us the squared DVCS amplitude
\begin{eqnarray*}
2 d^{\rm DVCS} \sigma = d^+ \sigma + d ^-\sigma - 2d^{\rm BH} \sigma \, .
\end{eqnarray*}
Then an appropriate azimuthal averaging, now with the conventional measure
$d \phi$, separates the $\cos$-harmonics, $c^{\rm DVCS}_n$, via
\begin{eqnarray}
\label{Def-CA-Mom-c-Even}
{\CeA}^{\mit\Lambda}_{c(0)}
\!\!\!&=&\!\!\!
2 {\cal N}_{+-}
\int_0^{2\pi} d \phi
\frac{
d^{\rm DVCS} \sigma^{\mit\Lambda}
}{
d\phi
}
\, , \\
\CeA^{\mit\Lambda}_{c(n)}
\!\!\!&=&\!\!\!
2 {\cal N}_{+-}
\sum_{k = 1}^{2n} (-1)^{k + 1}
\int_{(2 k - 3) \pi / (2n)}^{(2 k - 1) \pi / (2n)} d \phi
\frac{
d^{\rm DVCS} \sigma^{\mit\Lambda}
}{
d\phi
}
\, ,
\end{eqnarray}
with $n = 1, 2$,
and $\sin$-dependent coefficients by means of
\begin{eqnarray}
\label{Def-CA-Mom-s-Even}
\CeA^{\mit\Lambda}_{s(n)}
= 2 {\cal N}_{+-}
\sum_{k = 1}^{2n} (-1)^{k + 1}
\int_{(k - 1)\pi / n}^{k \pi / n}
d \phi
\frac{
d^{\rm DVCS} \sigma^{\mit\Lambda}
}{
d\phi
} \, ,
\end{eqnarray}
with $n = 1, 2$.

To conclude, an experimental facility having electron and positron beams
is an ideal place to study GPDs and, thus, to extract the fundamental
information about the spin structure of the nucleon.

%%%%%%%%%%%%%%%%%%%%%%%%%%%%%%%%%%%%%%%%%%%%%%%%%%%%%%%%%%%%%%%%%%%%%
\subsubsection{Facilities with single-charge lepton beam}
%%%%%%%%%%%%%%%%%%%%%%%%%%%%%%%%%%%%%%%%%%%%%%%%%%%%%%%%%%%%%%%%%%%%%

Now let us come to the situation when only one kind of the lepton beam is
available. Here the study of single (lepton or nucleon) spin asymmetries
allows to remove the background BH cross section. Note, however, that when
both the beam and target are polarized, and one studies double-spin
asymmetries, one gets the contamination from the BH harmonics too. In the
single (lepton or hadron) spin experiments, still the twist-two coefficient
$s_1^{\cal I}$ is contaminated by power-suppressed effects, since both the
interference and squared DVCS terms contribute. The best one can do in these
circumstances is to cancel completely the twist-three part of the
interference term in the numerator. However, still one will have the
presence of the power-suppressed DVCS cross section. For instance, for the
single lepton-spin experiment one enables to probe $s^{\cal I}_{1,{\rm
unp}}$ plus $(1-y)\Delta^2/y{\cal Q}^2$ corrections from $|{\cal T}_{\rm
DVCS}|^2$ and this can be done with the formula
\begin{eqnarray}
\label{Def-SSA-Mom}
\SSA_{1} \!\!\!&=&\!\!\!
\Bigg(
\int_{0}^\pi d w
\frac{ d \sigma^\uparrow - d \sigma^\downarrow
}{
d\phi
}
-
\int_\pi^{2\pi} d w
\frac{
d \sigma^\uparrow - d \sigma^\downarrow
}{
d\phi
}
\\
&&\qquad\qquad\qquad\qquad
- \frac{1}{3}
\sum_{k = 1}^{6} (-1)^{k + 1}
\int_{(k - 1)\pi / 6}^{k \pi / 6}
d w
\frac{
d \sigma^\uparrow - d \sigma^\downarrow
}{
d \phi
}
\left.\Bigg) \right/
\int_0^{2\pi} d \phi
\frac{
d \sigma^\uparrow + d \sigma^\downarrow
}{
d \phi
}
\, .
\nonumber
\end{eqnarray}
Analogous extraction of the twist-two coefficient $s^{\cal I}_{1,{\rm LP}}$
is available for the nucleon-spin asymmetry (with the unpolarized lepton beam).
The projection of the same components can be achieved by weighting the
integral with $\sin(\phi) d w$.

For smaller value of $y$ the contamination from the squared DVCS term may
be large. Let us demonstrate that a separation of the interference and
squared DVCS terms can be achieved by a Fourier transform. The multiplication
of the cross section with $d w / d \phi$ induces new harmonics in the squared
DVCS term. Projection of all three harmonics, i.e., measuring also
\begin{equation}
\SSA_{n}
=
\Bigg(
\sum_{k = 1}^{2n} (-1)^{k + 1}
\int_{(k - 1)\pi / n}^{k \pi / n}
d w
\frac{
d \sigma^\uparrow - d \sigma^\downarrow
}{
d \phi
}
\left.\Bigg) \right/
\int_0^{2\pi} d \phi
\frac{
d \sigma^\uparrow + d \sigma^\downarrow
}{
d \phi
}
\, ,
\end{equation}
(where $n = 2, 3$) provides the desired Fourier coefficients:
\begin{eqnarray}
s_{1,{\rm unp}}^{\cal I}  \propto \left(\SSA_{1} -
\frac{w_{s,11}}{w_{s,13}} \SSA_{3} \right) ,\quad
s_{2,{\rm unp}}^{\cal I} \propto \left(\SSA_{2} -
\frac{w_{s,12}}{w_{s,13}} \SSA_{3} \right),\quad
s_{1,{\rm unp}}^{\rm DVCS} \propto \SSA_{3},
\end{eqnarray}
with $w_{s,km}= \int_{0}^{2\pi} d w \sin(k \phi) \sin(m \phi)$. Thus the
extraction of twist-three harmonics is, in principle, possible, however, it
requires high precision data.

Such a modified Fourier analysis might be employed to separate the
coefficients of the interference and squared DVCS term also in unpolarized
and double spin-flip experiments. One has to be aware that the weighted
cross section contains now four odd and five even harmonics and that the
BH cross section must be removed.

%%%%%%%%%%%%%%%%%%%%%%%%%%%%%%%%%%%%%%%%%%%%%%%%%%%%%%%%%%%%%%%%%%%%%
\section{Properties of Compton form factors}
\label{Sec-CFF}
%%%%%%%%%%%%%%%%%%%%%%%%%%%%%%%%%%%%%%%%%%%%%%%%%%%%%%%%%%%%%%%%%%%%%

To compare twist-two and -three contributions quantitatively, we use
essentially the same ansatz for the GPDs as in \cite{BelMulNieSch00}. GPDs
are constrained by sum rules and reduction formulae to the forward
kinematics, which give us a guideline on how to model them. One may hope
that these constraints, once implemented into a parametrization, provide a
realistic order of magnitude estimate for cross sections and asymmetries.
However, we should emphasize that our evaluations (as well as by other
authors) are strongly affected by the model ambiguities involved.
Experimental data will necessarily constrain GPDs via theoretical formulae.
Therefore, it is important to understand the (i) theoretical uncertainties
and (ii) influence of the model parameters on predictions. While the first
issue is relatively straightforward to handle, provided one has enough
knowledge on higher-order and -twist corrections, the second issue relies,
more or less, on the experience that has to be collected by applying
different GPD models.

%%%%%%%%%%%%%%%%%%%%%%%%%%%%%%%%%%%%%%%%%%%%%%%%%%%%%%%%%%%%%%%%%%%%%
\subsection{Ansatz for GPDs}
\label{SubSec-NumEst-Ans}
%%%%%%%%%%%%%%%%%%%%%%%%%%%%%%%%%%%%%%%%%%%%%%%%%%%%%%%%%%%%%%%%%%%%%

To give a quantitative estimate of the CFFs we have to model, or parametrize,
GPDs. Let us discuss this issue for the twist-two sector, where we set $\eta =
- \xi$. Note that the time-reversal invariance and hermiticity imply that the
twist-two GPDs are real valued functions symmetric in $\eta$.

%%%%%%%%%%%%%%%%%%%%%%%%%%%%%%%%%%%%%%%%%%%%%%%%%%%%%%%%%%%%%%%%%%%%%
\subsubsection{General remarks}
\label{GeneralRemarks}
%%%%%%%%%%%%%%%%%%%%%%%%%%%%%%%%%%%%%%%%%%%%%%%%%%%%%%%%%%%%%%%%%%%%%

For the reader's convenience let us remind the definition of the
twist-two GPDs in terms of light-ray operators and their reduction to
the conventional parton densities. We will use this information for the
modeling of the former. The twist-two quark GPDs are given by
\begin{eqnarray}
\label{vectorGPD}
\langle P_2 |
\bar \psi (- \kappa n)
\gamma \cdot n \,
\psi (\kappa n)
| P_1 \rangle
\!\!\!&=&\!\!\!
\int_{- 1}^{1} d x
{\rm e}^{- i x \kappa ( P \cdot n )}
\left\{
h \cdot n \, H (x, \xi, \Delta^2)
+
e \cdot n \, E (x, \xi, \Delta^2)
\right\}
\, ,
\\
\label{axialGPD}
\langle P_2 |
\bar \psi (- \kappa n)
\gamma \cdot n \, \gamma_5
\psi (\kappa n)
| P_1 \rangle
\!\!\!&=&\!\!\!
\int_{- 1}^{1} d x
{\rm e}^{- i x \kappa (P \cdot n)}
\left\{
\tilde h \cdot n \, \widetilde H (x, \xi, \Delta^2)
+
\tilde e \cdot n \, \widetilde E (x, \xi, \Delta^2)
\right\}
\, ,
\end{eqnarray}
for the vector and axial channels, respectively. The Dirac bilinears are
introduced in Eq.\ (\ref{DiracBilinears}). We have contracted the free
Lorentz index with the light-like vector $n_\mu$ in order to render the
operators twist-two. For brevity we dropped the path ordered exponential
between the quark fields, which makes the nonlocal operators gauge
invariant, and did not indicate their scale dependence. This dependence
arise from the renormalization procedure of the light-ray operators and is
governed by a renormalization group equation
\cite{GeyRobBorHor85,BukFroKurLip85}.

On the other hand unpolarized and polarized quark $(f, \Delta f)$ and
antiquark $(\bar f, \Delta \bar f)$ densities are defined similarly by
equations
\begin{eqnarray}
\langle p |
\bar \psi (0)
\gamma \cdot n
\psi (\kappa n)
| p \rangle
\!\!\!&=&\!\!\!
2 p \cdot n
\int_{0}^{1} d x
\left\{
f (x) {\rm e}^{- i x \kappa (p \cdot n)}
-
\bar f (x) {\rm e}^{i x \kappa (p \cdot n)}
\right\} \, ,
\\
\langle p |
\bar \psi (0)
\gamma \cdot n \, \gamma_5
\psi (\kappa n)
| p \rangle
\!\!\!&=&\!\!\!
2 s \cdot n
\int_{0}^{1} d x
\left\{
\Delta f (x) {\rm e}^{- i x \kappa (p \cdot n)}
+
\Delta \bar f (x) {\rm e}^{i x \kappa (p \cdot n)}
\right\} \, ,
\end{eqnarray}
with the nucleon polarization vector $s_\mu = \frac{1}{2} \bar U (p)
\gamma_\mu \gamma_5 U (p)$. Here, the limiting procedure from the
off-forward kinematics is defined as $\Delta \to 0$, so that $P_1 = P_2
\equiv p$, and one gets a restriction on the GPDs $H$ and $\widetilde H$:
\begin{equation}
\label{ForwardLimit}
H (x, 0, 0) = f (x) \theta (x) - \bar f (- x) \theta (-x) \, ,
\qquad
\widetilde H (x, 0, 0)
=  \Delta f (x) \theta (x) + \Delta \bar f (- x) \theta (-x) \, .
\end{equation}
The first moments of the twist-two GPDs are equal to the corresponding parton
form factors in the nucleon
\begin{eqnarray}
\label{SumRule}
&&\int_{- 1}^{1} d x \, H (x, \xi, \Delta^2) = F_1 (\Delta^2) \, ,
\qquad
\int_{- 1}^{1} d x \, E (x, \xi, \Delta^2) = F_2 (\Delta^2) \, ,
\\
&&\int_{- 1}^{1} d x \, \widetilde H (x, \xi, \Delta^2) = G_1 (\Delta^2) \, ,
\qquad
\int_{- 1}^{1} d x \, \widetilde E (x, \xi, \Delta^2) = G_P (\Delta^2) \, ,
\nonumber
\end{eqnarray}
i.e., Dirac, Pauli, axial, and pseudoscalar form factors, respectively.

For the gluonic GPDs we have the parametrization
\begin{eqnarray}
\langle P_2 |
n_\alpha G_{\alpha \mu} (- \kappa n)
G_{\mu \beta} (\kappa n) n_\beta
| P_1 \rangle
\!\!\!&=&\!\!\!
\frac{1}{4} \left( P \cdot n \right)
\int_{- 1}^{1} d x \
{\rm e}^{- i x \kappa ( P \cdot n )}
\nonumber\\
&&\qquad\times\left\{
h \cdot n \, H_G (x, \xi, \Delta^2)
+
e \cdot n \, E_G (x, \xi, \Delta^2)
\right\} \, ,
\\
\langle P_2 |
n_\alpha G_{\alpha \mu} (- \kappa n)
i \widetilde G_{\mu \beta} (\kappa n) n_\beta
| P_1 \rangle
\!\!\!&=&\!\!\!
\frac{1}{4} \left( P \cdot n \right)
\int_{- 1}^{1} d x \
{\rm e}^{- i x \kappa ( P \cdot n )}
\nonumber\\
&&\qquad\times
\left\{
\tilde h \cdot n \, \widetilde H_G (x, \xi, \Delta^2)
+
\tilde e \cdot n \, \widetilde E_G (x, \xi, \Delta^2)
\right\}
\, ,
\end{eqnarray}
for the even- and odd-parity sectors, respectively. To make the contact with
the forward parton densities, let us recall the definitions of the gluonic
distributions,
\begin{eqnarray}
\langle p |
n_\alpha G_{\alpha \mu} (0)
G_{\mu \beta} (\kappa n) n_\beta
| p \rangle
\!\!\!&=&\!\!\!
\left( p \cdot n \right)^2
\int_{0}^{1} d x \
x g (x)
\left\{
{\rm e}^{- i x \kappa p_+}
+
{\rm e}^{i x \kappa p_+}
\right\} \, ,
\\
\langle p |
n_\alpha G_{\alpha \mu} (0)
i \widetilde G_{\mu \beta} (\kappa n) n_\beta
| p \rangle
\!\!\!&=&\!\!\!
\left( s \cdot n \right)
\left( p \cdot n \right)
\int_{0}^{1} d x \
x \Delta g (x)
\left\{
{\rm e}^{- i x \kappa p_+}
-
{\rm e}^{i x \kappa p_+}
\right\}
\, .
\end{eqnarray}
Therefore, we must have
\begin{equation}
H_G (x, 0, 0) = x g (x) \theta (x) - x g (- x) \theta (-x) \, ,
\qquad
\widetilde H_G (x, 0, 0)
=  x \Delta g (x) \theta (x) + x \Delta g (- x) \theta (-x) \, .
\end{equation}
The first moment of these GPDs is related to gluonic form factors.
The vector form factor is measurable in diffractive meson production
\cite{Breetal99} and was theoretically estimated in Ref.\ \cite{BraGorManSch93}.

In different regions of the phase space GPDs share common properties with
conventional inclusive parton densities for $|x| > \xi$, and exclusive
distribution amplitudes for $|x| < \xi$. Moreover, their experimental
exploration will resolve the target from different perspectives and will
enable to obtain its real-space image \cite{Bur00,RalPir01}.

To model the GPDs we will use the representation in terms of the so-called
double distributions (DDs) \cite{MulRobGeyDitHor94,Rad96}. In this
representation one treats both $s$- and $t$-channel momentum flows
independently with the corresponding fractions $y$ and $z$, respectively.
For the matrix element of the composite operator constructed out of scalar
field operators, $\phi$, we would have
\begin{eqnarray}
\label{ScalarDD}
\langle P_2 |
\phi (- \kappa n) \phi (\kappa n)
| P_1 \rangle
\!\!\!&=&\!\!\!
\int_{-1}^{1} d y \int_{- 1 + |y|}^{1 - |y|} d z
{\rm e}^{- i y \kappa (P \cdot n) - i z \kappa (\Delta \cdot n)}
f (y, z, \Delta^2)
\nonumber\\
\!\!\!&=&\!\!\!
\int_{- 1}^{1} d x {\rm e}^{- i x \kappa (P \cdot n)}
q (x, \xi, \Delta^2) \, .
\end{eqnarray}
The function introduced on the first line of Eq.\ (\ref{ScalarDD}) is the
aforementioned double distribution $f (y, z, \Delta^2)$, and $q (x, \xi,
\Delta^2)$ is the corresponding GPD. The expression of $q (x, \xi,
\Delta^2)$ in terms of $f (y, z, \Delta^2)$ is given via
\begin{equation}
\label{nontodouble}
q (x, \xi, \Delta^2)
= \int_{- 1}^1 d y \int_{- 1 + |y|}^{1 - |y|} d z \,
\delta ( y + \xi z - x ) f ( y, z, \Delta^2 ) \, ,
\end{equation}
and the inversion is known as the Radon transformation \cite{BelKirMulSch00c,Ter01}.

However, in the case of spin-$1/2$ quarks there is an additional vector
index associated with the Dirac matrix inserted in between the field
operators. This leads to the appearance of extra Lorentz structures in the
decomposition of the matrix element, which once discarded, lead to a violation
of the polynomiality condition for the GPDs. In the historically first
definition, the relevant structure was erroneously neglected and one used
the same Eq.\ (\ref{nontodouble}) even for the nucleon matrix elements of
QCD operators. This problem has been resolved by taking into account an
extra term, the so-called D-term \cite{PolWei99}. It is only concentrated in
the exclusive region, i.e., $|x/\xi| \le 1$. Such contributions arise from
isolated mesonic-like states\footnote{From the partonic interpretation of
GPDs, it is clear that the exclusive region is always governed by
mesonic-like states. In general, there is a cross talk between the exclusive
and  inclusive regions, e.g., this ensures the polynomiality condition for moments
\cite{DieFelJakKro00BroDieHwa00}. If this is not the case, we call them
isolated mesonic-like states.}. However, there is yet another possibility
to resolve the polynomiality problem \cite{BelKirMulSch00c}. Unfortunately, in
this case one has to deal with a more singular double distribution at $y = 0$,
which requires, in general, a regularization that is invisible in the forward
case. We will accept for the time being the former solution, which allows us
to have a comparison with numerical results of Ref.\ \cite{KivPolVan00}.

%%%%%%%%%%%%%%%%%%%%%%%%%%%%%%%%%%%%%%%%%%%%%%%%%%%%%%%%%%%%%%%%%%%%%
\subsubsection{Models}
\label{Models}
%%%%%%%%%%%%%%%%%%%%%%%%%%%%%%%%%%%%%%%%%%%%%%%%%%%%%%%%%%%%%%%%%%%%%

As we have said above, we assume a two-component form for the GPDs.
Namely, for the unpolarized case one has for each $i$-type quark
\begin{eqnarray}
\label{H-GPD}
H_i (x, \xi, \Delta^2)
= q_i (x, \xi, \Delta^2)
+ \frac{1}{N_f}
\theta (1 - |x/\xi|) D (x/\xi, \Delta^2)
\, , \\
\label{E-GPD}
E_i (x, \xi, \Delta^2)
=
r_i (x, \xi, \Delta^2)
- \frac{1}{N_f} \theta (1 - |x/\xi|) D (x/\xi, \Delta^2)
\, ,
\end{eqnarray}
with the D-term contributing equally to all $N_f$ active quark species.
Similar equations hold for the gluonic GPDs with $i = G$ and a gluonic
D-term, $N_f^{-1} D (x/\xi) \to D_G (x/\xi)$. Note that the D-term drops out
in the sum $H + E$ and, thus, has to be an antisymmetric function of $x/\xi$.
For three active quark flavors we set $N_f = 3$. In the following we take
an oversimplified factorized ansatz of the $\Delta^2$-dependence from the
other two scaling variables $x$ and $\xi$ for all functions. Note that due
to the antisymmetry of the D-term it does not enter into the sum rule
(\ref{SumRule}) and, therefore, its $\Delta^2$-dependence is not constrained
by it. As a model we take
\begin{equation}
\label{DeltaDsea}
D (z, \Delta^2)
=
\left(
1 - \frac{\Delta^2}{m^2_{\rm D}}
\right)^{-3} D (x/\xi)\, ,
\end{equation}
where the $\Delta^2$-dependence is characterized by the cutoff mass
$m_{\rm D}$, considered as a free parameter.

Obviously, $q^i$ is a sum of valence- and sea-quark contributions.
According to this we make a decomposition in these components and
extract the momentum transfer dependence into valence and sea
form factors, namely,
\begin{equation}
q_i (x, \xi, \Delta^2)
=
F^{i, \rm val}_1 (\Delta^2) \, q^{\rm val}_i (x, \xi)
+
F^{\rm sea}_1 (\Delta^2) \, q^{\rm sea}_i (x, \xi) \, ,
\end{equation}
for $i = u, d, s$, and $q^{\rm val}_s (x, \xi) = 0$ for the $s$-quarks.

At this point let us say that the $(x, \xi)$-dependence of the function
$r (x, \xi, \Delta^2)$ is not constrained at all. However, since its
first moment is given by the Pauli form factor, we extract it from
$r (x, \xi, \Delta^2)$ and naively set the remainder of $r$ equal
to $q (x, \xi)$, i.e.,
\begin{equation}
r_i (x, \xi, \Delta^2)
=
F^{i, \rm val}_2 (\Delta^2) \, q^{\rm val}_i (x, \xi)
+
F^{\rm sea}_2 (\Delta^2) \, q^{\rm sea}_i (x, \xi) \, .
\end{equation}

The valence $u$- and $d$-quark form factors can be extracted from the proton
and neutron form factors via the formulae
\begin{equation}
2 F^{u, \rm val}_{1,2} (\Delta^2)
= 2 F_{1,2}^p (\Delta^2) + F_{1,2}^n (\Delta^2) \, ,
\qquad
F^{d, \rm val}_{1,2} (\Delta^2)
= F_{1,2}^p (\Delta^2) + 2 F_{1,2}^n (\Delta^2) \, .
\end{equation}
The latter are known fairly well from experimental measurements and can
be pa\-ra\-met\-rized by dipole formulae in the small-$\Delta^2$ region
\begin{equation}
\label{dipoleFF}
G_E^p (\Delta^2)
= \frac{1}{1 + \kappa_p} G_M^p (\Delta^2)
= \frac{1}{\kappa_n} G_M^n (\Delta^2)
= \left( 1 - \frac{\Delta^2}{m_V^2} \right)^{- 2} ,
\qquad
G_E^n (\Delta^2) = 0 \, .
\end{equation}
Note that we have set the neutron electric form factor equal to zero since
at small momentum transfer it vanishes as a first power of $\Delta^2$ with
$-1/6$ times the neutron charge radius $r_n^2$ as a coefficient, which is quite
small, $r_n^2 \approx - 0.113\ {\rm fm}^2$ . Here the proton and neutron
magnetic moments are $\kappa_p = 1.793$ and $\kappa_n = - 1.913$, respectively.
We have used Sachs electric and magnetic form factors related to Dirac and
Pauli ones by
\begin{eqnarray}
G_E^i (\Delta^2)
\!\!\!&=&\!\!\! F_1^i (\Delta^2) + \frac{\Delta^2}{4 M^2} F_2^i (\Delta^2)
\, , \\
G_M^i (\Delta^2)
\!\!\!&=&\!\!\! F_1^i (\Delta^2) + F_2^i (\Delta^2) \, ,
\end{eqnarray}
respectively. They are characterized by the cutoff mass $m_V = 0.84\, {\rm GeV}$,
see, e.g., \cite{Oku90}. The sea-quark form factors are guided by the counting
rules and read in terms of Sachs form factors
\begin{equation}
\label{SQFF}
G_E^{\rm sea }(\Delta^2)
= \frac{1}{1 + \kappa_{\rm sea}} G_M^{\rm sea} (\Delta^2)
= \left( 1 - \frac{\Delta^2}{m_{\rm sea}^2} \right)^{- 3},
\end{equation}
with yet another mass cutoff $m_{\rm sea}$. The slope of $F_1^{\rm sea}$
is given by $B_{\rm sea} = \partial F_1^{\rm sea} / \partial \Delta^2
|_{\Delta^2 = 0} = 3 / m^2_{\rm sea} - \kappa_{\rm sea}/4M^2$. The two free
parameters $B_{\rm sea}$ and $\kappa_{\rm sea}$ will be specified below. Note
that $\kappa_{\rm sea}$ enters in the sum rule that gives the fraction of the
orbital angular momentum carried by quarks \cite{Ji96}. With our ad hoc
assumption $r^i = q^i$ this sum rule reads
\begin{eqnarray}
\label{SumRul-Jq}
J^q \!\!\!&=&\!\!\!
\lim_{\Delta\to 0}\sum_{i = u, d, s}
\frac{1}{2}\int_{-1}^1 dx\;
x \left\{
H^i(x, \xi, \Delta^2, \cQ^2)
+
E^i(x, \xi, \Delta^2, \cQ^2)
\right\}
\\
\!\!\!&=&\!\!\! \frac{1}{2}
\left\{
\left( 1 + \kappa_p + \kappa_n/2 \right) P^{u_{\rm val}}
+
\left( 1+\kappa_p + 2 \kappa_n \right) P^{d_{\rm val}}
+
\left( 1 + \kappa_{\rm sea} \right) P^{\rm sea}
\right\} \, ,
\nonumber
\end{eqnarray}
where the momentum fractions $P^i$ carried by quarks can be deduced from
deeply inelastic data alone.

The D-term does not affect the forward limit
of GPDs. Therefore, we use Eq.\ (\ref{nontodouble}) and model
the quark $f_i (y, z)$  according to Ref.\
\cite{Rad99} as a product of a profile function $\pi$ with the
conventional type-$i$ quark $f_i (y)$ and antiquark $\bar f_i$ densities.
Namely, for valence and sea contributions we have
\begin{eqnarray}
&&f^{\rm val}_i (y, z)
= f^{\rm val}_i (y) \theta (y) \pi (|y|, z; b_{\rm val}) \, ,
\\
&&f^{\rm sea}_i (y, z) =
\left\{
\bar f_i (y) \theta (y) - \bar f_i (- y) \theta (- y)
\right\} \pi (|y|, z; b_{\rm sea}) \, ,
\end{eqnarray}
with the usual definition for the valence density $f^{\rm val}_i (y)
\equiv f_i (y) - \bar f_i (y)$. Here the profile is assumed to be universal
for all valence- and sea-quark species and reads
\begin{equation}
\label{DD-Ansatz}
\pi (y, z; b) =
\frac{{\mit\Gamma} \left( b + \ft32 \right)}
{\sqrt{\pi} {\mit\Gamma} (b + 1)}
\frac{\left[ (1 - y)^2 - z^2 \right]^b}{(1 - y)^{2 b + 1}}
\, .
\end{equation}
This ansatz ensures that the reduction formula (\ref{ForwardLimit}) holds true.
The parameter $b$ encodes the skewedness effect, i.e., a larger value of $b$
suppresses the $\xi$-dependence, and $q^i (x, \xi, \cQ^2)$ reduces to the parton
density $f^i ( x, \cQ^2 )$ in the limit $b \to \infty$. This limit will be called
the forward parton distribution (FPD) model \cite{GuiVan98}.
The idea driving the form of the
ansatz (\ref{DD-Ansatz}) is that the profile function $\pi^i (y = 0, z)$ mimics
a mesonic-like two-parton state characterized by the longitudinal momentum
fraction $z$. In the case of valence quarks one may choose the asymptotic
distribution amplitude, which implies $b_{\rm val} = 1$. In the case of sea
quarks we consider $b_{\rm sea}$ as a free parameter, which will be discussed
below. It will be adjusted by the comparison of our predictions to the H1
small-$\Bx$ data.

We similarly assume  a factorized ansatz for the gluonic DD
\begin{equation}
f_G (y, z, \Delta^2) = F_G (\Delta^2) f_G (y, z) \, ,
\end{equation}
where
\begin{equation}
f_G (y, z)
=
y \left\{ g (y) \theta (y) - g (- y) \theta (- y) \right\}
\pi (|y|, z; b_G) \, .
\end{equation}

The quark D-term was given in \cite{KivPolVan00}, making use of the results
of the chiral quark soliton model \cite{Petetal97} at the very low
normalization point $\mu^2 \approx 0.36 \ {\rm GeV}^2$, in the form of an
expansion in terms of Gegenbauer polynomials, $D(z) \approx - 4.0 \,
C_1^{3/2} (z) - 1.2 \, C_3^{3/2} (z) - 0.4 \, C_5^{3/2}(z)$. The expansion
coefficients have been fixed under the assumption that the polynomiality can
only be restored with the help of the D-term and fit to the chiral soliton
model evaluation. Compared to Ref.\ \cite{GoePolVan01}, we can adjust its
fall-off with $\Delta^2$ in Eq.\ (\ref{DeltaDsea}) by the cutoff mass
$m_{\rm D}$. Note that the evolution upwards to our input scale $\cQ^2_0 =
4\ \GeV^2$ from $\mu^2\approx 0.36\ \GeV^2$ reduces the absolute value of
the Gegenbauer polynomial coefficients by about 25\% or even more, provided
that the gluonic D-term is neglected. This reduction is larger (smaller) for
positive (negative) expansion coefficients of the gluonic D-term. However,
to study the significance of such a term for DVCS observables, we ignore any
evolution effects.

The twist-two CFF ${\cal H} (\xi, \Delta^2)$ at LO in the QCD coupling
constant reads in terms of the model (\ref{nontodouble}) with the D-term
included, i.e., Eq.\ (\ref{H-GPD}),
\begin{eqnarray}
&&\!\!\!\!\!\!\!\!{\cal H} (\xi, \Delta^2)
= \frac{1}{\xi} \int_{- 1}^{1} \frac{d x}{1 - x/\xi - i 0}
\Bigg\{
\frac{2}{N_f}
\sum_{i = u, d, s} Q_i^2 \,
\left(
1 - \frac{\Delta^2}{m^2_{\rm D}}
\right)^{-3} \,
\theta (1 - |x/\xi|)
D (x/\xi)
\\
&&\!\!\!\!+
\int_{- 1}^{1} d y \int_{- 1 + |y|}^{1 - |y|} d z \,
\delta (y + \xi z - x)
\sum_{i = u, d, s} Q_i^2
\bigg\{
F_1^{i, \rm val} (\Delta^2)
\left(
f^{\rm val}_i (y) \theta (y) - f^{\rm val}_i (- y) \theta (-y)
\right) \pi (|y|, z; b_{\rm val})
\nonumber\\
&&\qquad\qquad\qquad\qquad\qquad\qquad\qquad\qquad
+ \,
2 \, F_1^{\rm sea} (\Delta^2)
\left(
\bar f_i (y) \theta (y) - \bar f_i (- y) \theta (- y)
\right)\pi (|y|, z; b_{\rm sea})
\bigg\}
\Bigg\}
\, , \nonumber
\end{eqnarray}
where $N_f = 3$. Obviously, the replacements $D (x/\xi) \to -D(x/\xi)$
and $F_1 \to F_2$ lead to the definition of $\cal E$.

In the following we take the MRS A' \cite{MarRobSti95} parametrizations
at $\cQ_0^2 = 4\ {\rm GeV}^2$. This input scale
ensures that evolution effects are small and are, therefore, not considered.
We will neglect SU(2) breaking in the sea and its charm component,
i.e., $\bar{u} = \bar{d} = \bar{s}/2$.
We use these forward parton densities in our LO analysis of DVCS and we are
aware that this may result in a deviation of our predictions from the forward
structure functions, measured experimentally in deeply inelastic scattering,
\begin{eqnarray}
\label{Red-for-F1}
F_2 (\Bx, \cQ^2) \approx 2\Bx F_1 (\Bx, \cQ^2)  = \frac{\Bx}{\pi} \Im{\rm m}
{\cal H}^{\rm LO} (\xi = \Bx, \eta = 0, \Delta^2 = 0, \cQ^2) \, ,
\end{eqnarray}
up to 20\%. Here we employed the Callan-Gross relation, valid in LO of
perturbation theory. Recall that for a general two-photon process in the
light-cone dominated region with a virtual outgoing $\gamma$-quantum,
the CFF is given by ${\cal H} (\xi, \eta, \Delta^2, \cQ^2)$. We will
demonstrate below that the error involved in this approximation is small in
comparison to the uncertainties induced by other sources.

Note that the fraction of the orbital angular momentum, given by the sum rule
(\ref{SumRul-Jq}) with the  assumed MRS parametrization, results into
\begin{eqnarray}
J^q \approx 0.3 + 0.09 \kappa_{\rm sea}
\, , \qquad
\mbox{for}
\qquad
\cQ^2 = 4\ \GeV^2.
\end{eqnarray}
Estimates based on lattice calculations \cite{MatDonLiuManMul00} and QCD sum
rules \cite{BalJi97} provide $J^q = 0.3 \pm 0.14$ and $J^q = 0.15\pm 0.13$
at $\mu^2 \approx 1\ \GeV^2$, respectively. So we conclude that for consistency
of these predictions, $\kappa_{\rm sea}$ may vary in a wide range
$-3 < \kappa_{\rm sea} < 2$.

By the same token as above, the polarized quark GPD $\widetilde H$ is modeled
according to
\begin{equation}
\widetilde H_i (x, \xi, \Delta^2)
=
G^{i, \rm val}_1 (\Delta^2) \, \Delta q^{\rm val}_i (x, \xi)
+
G^{\rm sea}_1 (\Delta^2) \, \Delta q^{\rm sea}_i (x, \xi) \, ,
\end{equation}
for $i = u, d, s$, and $\Delta q^{\rm val}_s (x, \xi) = 0$ for the $s$-quarks.
Note that there is no D-term in this function \cite{BelMul01a}. The polarized
valence- and sea-quark contributions are
\begin{eqnarray}
&&\Delta f^{\rm val}_i (y, z)
= \Delta f^{\rm val}_i (y) \theta (y) \pi (|y|, z; b_{\rm val}) \, ,
\\
&&\Delta f^{\rm sea}_i (y, z) =
\left\{
\Delta \bar f_i (y) \theta (y) + \Delta \bar f_i (- y) \theta (- y)
\right\} \pi (|y|, z; b_{\rm sea}) \, ,
\end{eqnarray}
with the usual definition for the polarized valence-quark density
$\Delta f^{\rm val}_i (y) \equiv \Delta f_i (y) - \Delta \bar f_i (y)$.
Obviously, $\Delta f^{\rm val}_s = 0$. In the forward limit this
formula leads to Eq.\ (\ref{ForwardLimit}). For the forward densities we
use the GS A \cite{GehSti96} parametrization with an SU(3) symmetric sea,
again at the input scale $\cQ^2 = 4 \mbox{\ GeV}^2$.

The axial form factors are
\begin{equation}
G^{i, \rm val}_1 (\Delta^2) = \left( 1 - \frac{\Delta^2}{m_A^2} \right)^{- 2}
\, ,
\qquad
G^{\rm sea}_1 (\Delta^2) = \left( 1 - \frac{B_A}{3} \Delta^2\right)^{- 3},
\end{equation}
where $i$ runs over $u$- and $d$-quark species and the scale is $m_A = 0.9\,
{\rm GeV}$ \cite{Oku90}. The slope $B_A$ is considered as a free parameter,
which should be fixed from experimental data.

The polarized CFF $\widetilde {\cal H} (\xi, \Delta^2)$ in
terms of the model (\ref{nontodouble}) has the form
\begin{eqnarray}
\widetilde {\cal H} (\xi, \Delta^2)
\!\!\!&=&\!\!\! \frac{1}{\xi} \int_{- 1}^{1} \frac{d x}{1 - x/\xi - i 0}
\int_{- 1}^{1} d y \int_{- 1 + |y|}^{1 - |y|} d z \,
\delta (y + \xi z - x)
\\
&&\qquad\qquad\qquad\times
\sum_{i = u, d, s} Q_i^2
\bigg\{
G_1^{i, \rm val} (\Delta^2)
\left(
\Delta f^{\rm val}_i (y) \theta (y) + \Delta f^{\rm val}_i (- y) \theta (-y)
\right) \pi (|y|, z; b_{\rm val})
\nonumber\\
&&\qquad\qquad\qquad\qquad\qquad+ 2 \, G_1^{\rm sea} (\Delta^2)
\left(
\Delta \bar f_i (y) \theta (y) + \Delta \bar f_i (- y) \theta (- y)
\right)\pi (|y|, z; b_{\rm sea})
\bigg\}
\Bigg\}
\, .
\nonumber
\end{eqnarray}

For $\widetilde E$ we accept the pion pole-dominated ansatz of Ref.\
\cite{PenPolGoe00}
\begin{equation}
\label{Etilde}
\widetilde E^{u} (x, \xi, \Delta^2)
=
- \widetilde E^{d} (x, \xi, \Delta^2)
= \frac{1}{2} F_\pi (\Delta^2)
\frac{\theta \left( \xi > |x| \right)}{2\xi}
\phi_\pi\!\!\left(\frac{x + \xi}{2\xi}\right) ,
\end{equation}
with $F_\pi (\Delta^2)$ taken in our estimates in the form given in Eq.\
(39) of Ref.\ \cite{PenPolGoe00},
\begin{eqnarray}
F_\pi (\Delta^2)
\approx
\frac{4 g_A^{(3)} M^2}{m_\pi^2 - \Delta^2}
\left(
1 + \frac{1.7 \, (\Delta^2 - m_\pi^2)/\GeV^{2}}{(1 - \Delta^2/2\, \GeV^2)^2}
\right)
\, ,
\end{eqnarray}
valid for $|\Delta^2| \ll M^2$. The pion mass is $m_\pi \approx 0.14 \ \mbox{GeV}$.
For the pion distribution amplitude $\phi_\pi$ we take for simplicity its
asymptotic form $\phi_\pi (u) = \phi^{\rm asy}(u) \equiv 6 u (1 - u)$,
$0 \leq u \leq 1$.

Finally, the parity-odd gluonic GPD, $\widetilde H_G$, is modeled via the reduction
to DD, and the latter taken in the form
\begin{equation}
\Delta f_G (y, z)
=
y \left\{
\Delta g (y) \theta (y) + \Delta g (- y) \theta (- y)
\right\} \pi (|y|, z; b_G) \, .
\end{equation}

%%%%%%%%%%%%%%%%%%%%%%%%%%%%%%%%%%%%%%%%%%%%%%%%%%%%%%%%%%%%%%%%%%%%%
\subsection{Features of Compton form factors}
\label{SubSec-CFF-Feature}
%%%%%%%%%%%%%%%%%%%%%%%%%%%%%%%%%%%%%%%%%%%%%%%%%%%%%%%%%%%%%%%%%%%%%

Now, before we come in section \ref{Sec-NumEst} to numerical estimates for
the cross sections and asymmetries, it is worthwhile to study the properties
of CFFs in the twist-two and -three approximations: the magnitude and shape
of CFFs, the ratio of their real to imaginary part, the relation between
twist-two and -three CFFs in the WW-approximation, and the effect of NLO
corrections. Within our approximation it is consistent to take $\xi=
\Bx/(2 - \Bx)$. Since we use a factorized form of GPDs with $\Delta^2$-dependence
separated from the scaling variables, we can safely set $\Delta^2 = 0$ in the
most of estimates made in this section. These calculations are done with
$\cQ^2 = \cQ^2_0 = 4\ \GeV^2$.

%%%%%%%%%%%%%%%%%%%%%%%%%%%%%%%%%%%%%%%%%%%%%%%%%%%%%%%%%%%%%%%%%%%%%
\subsubsection{Twist-three versus twist-two effects}
%%%%%%%%%%%%%%%%%%%%%%%%%%%%%%%%%%%%%%%%%%%%%%%%%%%%%%%%%%%%%%%%%%%%%

As observed above, the real and imaginary parts of the partonic CFFs are
convex or concave functions of $\xi$ (see the archive version of
\cite{BelKirMulSch00c}). In Fig.\ \ref{Fig-CFF-Vec} we show the CFFs for
the valence up-quarks (a,b) and up-antiquarks (c,d) based on the MRS A'
parametrization without the D-term and with $b_{\rm val}=b_{\rm sea} = 1$.
Here we rescaled
the CFFs with $\xi^{\alpha_i}$, and used $\alpha^{\rm val}_u = 0.441$ and
$\alpha^{\rm sea}_u = 1.17$, which determine the small momentum fraction
behavior of the corresponding forward parton densities \cite{MarRobSti95}.
The D-term affects only the real part of the CFFs by a shift, e.g.,
${\cal H}^{{\rm sea}}_u$ and $C_u^{3(-)} \otimes H^{\rm sea}_u$
are shifted by a constant value of $\approx -1.66$ and $\approx -0.22$,
respectively. Then, the entire CFF ${\cal H}(\xi) = \sum_{i = u,d,s}
{\cal H}_i (\xi)$ will have a significant change: it starts now
with a negative value at large $\xi$ and turns over to a positive one at
$\Bx\approx 0.2$. Obviously, the sign of $\Re{\rm e}{\cal H}$ in the
valence-quark region\footnote{It implies that $\Bx \approx 1/3$, which we
understand in the following as the region $0.1 < \Bx < 0.6$ .} is determined
by the cancellation between the positive sea-quark and negative valence-quark
effects, both for double distribution ansatz alone and with the addition of
the D-term. We will demonstrate below that a clear signature for the D-term
can be washed away at $\Delta^2 \neq 0$ since the $\Delta^2$-dependence of
the sea quarks and the D-term is not constrained by the sum rules
(\ref{SumRule}) as we have discussed in section \ref{Models} after Eq.\
(\ref{E-GPD}).

%%%%%%%%%%%%%%%%%%%%%%%%%%%%%%%%%%%%%%%%%%%%%%%%%%%%%%%%%%%%%%%%%%%%%
%                          Figure
%%%%%%%%%%%%%%%%%%%%%%%%%%%%%%%%%%%%%%%%%%%%%%%%%%%%%%%%%%%%%%%%%%%%%
\begin{figure}[t]
\vspace{0cm}
\begin{center}
\mbox{
\begin{picture}(0,300)(300,0)
\put(65,150){\insertfig{8}{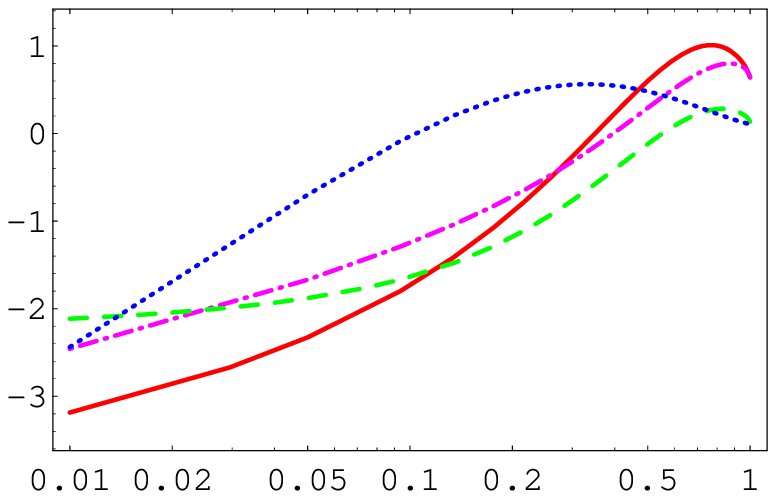}}
\put(95,275){$\Re{\rm e}{\cal H}_u^{\rm val}$}
\put(265,175){(a)}
\put(330,175){(b)}
\put(480,275){$\Im{\rm m}{\cal H}_u^{\rm val}$}
\put(305,150){\insertfig{8}{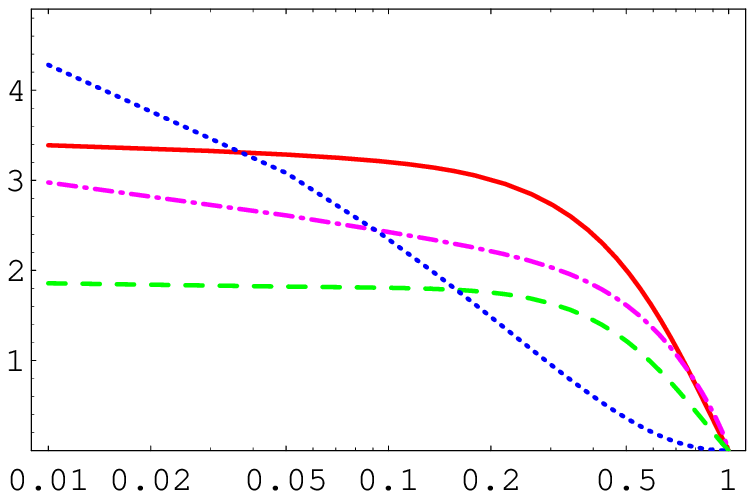}}
\put(55,0){\insertfig{8.2}{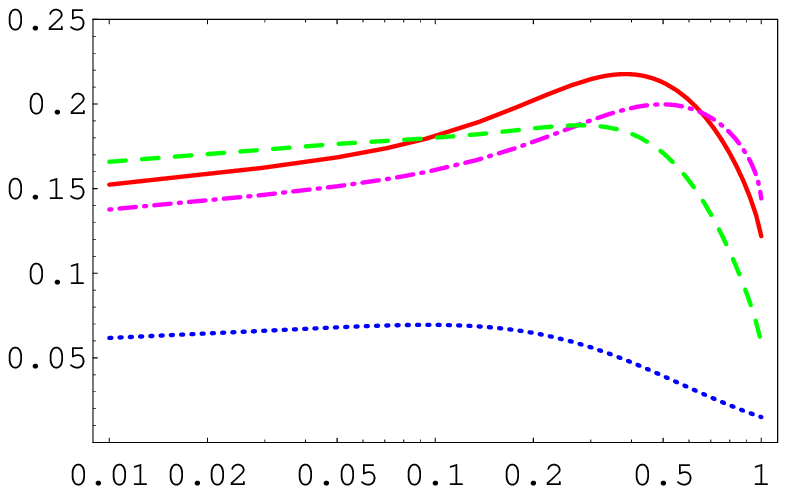}}
\put(300,0){\insertfig{8}{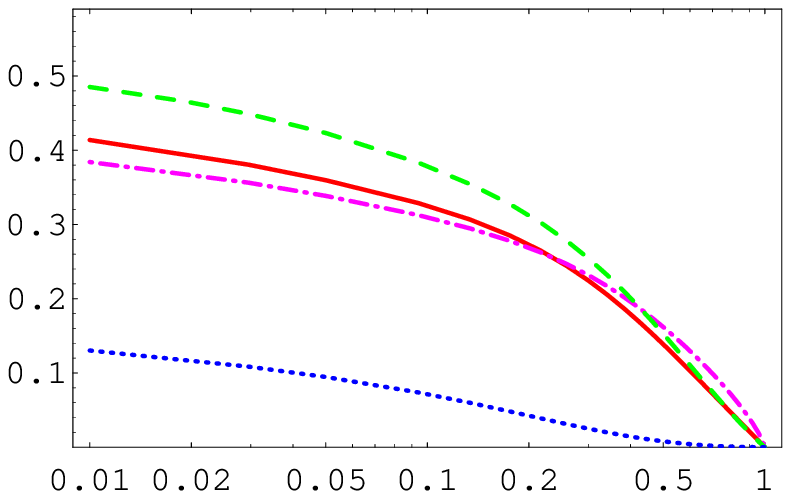}}
\put(95,123){$\Re{\rm e}{\cal H}_u^{\rm sea}$}
\put(265,123){(c)}
\put(330,123){(d)}
\put(480,123){$\Im{\rm m}{\cal H}_u^{\rm sea}$}
\put(290,-10){$\Bx$}
\put(530,-10){$\Bx$}
\end{picture}
}
\end{center}
\caption {\label{Fig-CFF-Vec}
The real (a,c) and imaginary (b,d) part of the u-quark CFFs
$\xi^{\alpha} {\cal H}$ in LO (solid) and NLO
(dash-dotted). It displays also the twist-three functions in the WW
approximation $\xi^{\alpha} {\cal H}^{\rm eff-WW}$ with
${\cal H}^{\perp} = 0$ (dashed), and the additive contribution
$\xi^{\alpha} C^{3(-)} \otimes H$ (dotted), which enter the CFF
${\cal H}^{\rm eff}$ multiplied by a kinematical factor singular at
$\Delta^2 = \Delta_{\rm min}^2$. The valence and sea $u$-quark
contributions are shown in (a,b) and (c,d), respectively, as a function
of $\Bx$ with $\Bx \ge 0.05$. The GPDs are taken at $\Delta^2=0$ and
$\cQ^2= 4\ \GeV^2$ with $b_{\rm val}=b_{\rm sea}=1$. }
\end{figure}
%%%%%%%%%%%%%%%%%%%%%%%%%%%%%%%%%%%%%%%%%%%%%%%%%%%%%%%%%%%%%%%%%%%%%

The shapes of the CFFs can be understood in the following way. The phases
$\phi_i(\xi)= {\rm Arg}\; {\cal H}_i(\xi)$ and the
absolute values $|{ \cal H}_i(\xi)| $ are monotonically
increasing functions of the scaling variable $\xi$. It is interesting to
note that in the WW-approximation the phase difference $\Delta\phi_i =
\phi_i (\xi) - \phi_i^{\rm eff-WW} (\xi)$ is not very large.
Neglecting the D-term and ${\cal H}_i^{\perp}$, we find that
$| \Delta\phi^{\rm val}_u | < 0.4 $ and $| \Delta\phi^{\rm sea}_u | < 0.16 $
for all values of $\xi$. Moreover, these differences will vanish in the
limit $\xi \to 0$. This property is obviously significant for observables
that arise from the interference of twist-two and twist-three CFFs. Note
also that the absolute value of the effective twist-three CFFs in the
WW-approximation is of the same order as the twist-two ones. Thus, there is
no numerical enhancement of the twist-three effects in this approximation.
The relative size of twist-three contributions as compared to the twist-two
harmonics is governed solely by  kinematical factors appearing in the
Fourier coefficients computed in section \ref{Sec-AziAngDep}.

%%%%%%%%%%%%%%%%%%%%%%%%%%%%%%%%%%%%%%%%%%%%%%%%%%%%%%%%%%%%%%%%%%%%%
\subsubsection{Small-$\Bx$ behavior}
%%%%%%%%%%%%%%%%%%%%%%%%%%%%%%%%%%%%%%%%%%%%%%%%%%%%%%%%%%%%%%%%%%%%%

Let us now study the properties of the CFFs for small $\Bx$ in LO
approximation. At intermediate momentum transfer this kinematics was
discussed within the BFKL approach in Ref.\ \cite{BalKuc00}. In Fig.\
\ref{Fig-CFF-Vec} we observe a scaling power-law behavior of the CFFs for
small $\xi$, which has been already established in Refs.\
\cite{ShuGolMarRys99,HebTeu01} and the archive version of
\cite{BelKirMulSch00c} for the FPD model. The small-$\xi$ behavior of the
CFFs is governed by the small momentum fraction asymptotics of the parton
densities. If the latter behave like $x^{-\alpha}$ with $\alpha > 0$, then
the real and imaginary part of $\cal H$ goes like $\xi^{-\alpha}$, and the
ratio of the real to imaginary part is given by $\tan\left((\alpha - 1 )
\pi/2 \right) = - \cot\left(\alpha \pi/2\right) $. The small $x$-behavior
of polarized parton densities, i.e., $y^{-\widetilde\alpha}$, induces an
analogous growth of $\widetilde{\cal H}$, however, the ratio of the real to
imaginary part is now $\tan\left( \widetilde\alpha \pi/2\right)$.

Such a behavior can be understood in a model-independent way. The essential
assumption is that the double distributions, which belong to the class of
mathematical distributions, develop an expansion in the vicinity of the
point $y = 0$ of the form
\begin{eqnarray}
f_i (y, z) = y^{- \alpha_i} \left\{ d_i (0, z) + \dots  \right\}.
\end{eqnarray}
Note that a $\delta$-like singularity at $y = 0$ gives a contribution
that is entirely concentrated in the exclusive region. We refer to it as
an isolated mesonic-like state. In the forward limit of GPDs, such a
contribution appears as a $\delta$-like singularity and, thus, it will
affect sum rules for parton densities. So we may conclude that such
singularities are absent in the spin non-flip GPDs\footnote{Unfortunately,
this is not necessarily true since they can vanish in the forward limit with
$\Delta^2$, too.} $H$ and $\widetilde H$, however, they may be present
in the spin-flip ones. Furthermore, we assume that $f (y, z)$ and, consequently,
also $d (0,z)$ vanish fast enough when they approach the boundary of their
support. By `fast enough' we mean that $|d (0, z)| < |1 \mp z|^{{\rm Max}
(0, \alpha - 1) + \varepsilon}$ with $\varepsilon > 0$ for $z \to \pm 1$. We
only discuss the region $0 \le y$ in the following. The results of
antiquark contributions for negative $y$ can be easily deduced from the
sea quarks at $y > 0$ employing the (anti)symmetry property of
corresponding GPDs. We also assume that $\alpha_i < 2$ in the vector
and $\widetilde \alpha_i < 1$ in the axial-vector sector to ensure the
existence of certain integrals.

Since GPDs are reduced to the parton densities $f_i (y)$ according to
Eqs.\ (\ref{ForwardLimit}), we see that at small momentum fraction $y$, the
parton densities are given by $y^{- \alpha_i} \int_{-1}^1 d z \; d_i (0,
z)$ with the exponential $\alpha_i$, which determines their small-$y$
behavior. Furthermore, a straightforward calculation shows that the
imaginary and real parts of the twist-two CFFs at LO behave like:
\begin{eqnarray}
\label{Res-ReImVec}
\Im{\rm m} {\cal H}_i (\xi, \Delta^2, \cQ^2)
\!\!\!&=&\!\!\! Q_i^2 \pi \xi^{-\alpha_i}
\int_{-1}^{1} d z \; \frac{d_i (0, z, \Delta^2, \cQ^2)}{(1 - z)^{\alpha_i}}
\, , \\
\Re{\rm e} {\cal H}_i (\xi, \Delta^2, \cQ^2)
\!\!\!&=&\!\!\! \tan \left( (\alpha_i - 1) \frac{\pi}{2} \right)
\Im{\rm m} {\cal H}_i (\xi, \Delta^2, \cQ^2) \, ,
\nonumber
\end{eqnarray}
where the parameter $\alpha_i$ may depend on $\Delta^2$ and $\cQ^2$. The first
equation agrees with the earlier considerations in Ref.\ \cite{ShuGolMarRys99}
and the second one was discussed in Ref.\ \cite{HebTeu01}.
An analogous formula holds true for the spin-flip contributions ${\cal E}_i$.
Note that the D-term is not important in this limit, however, the small-$\xi$
behavior can, in principle, be altered by other terms concentrated at $y = 0$,
i.e., $\frac{d^n}{dy^n} \delta(y)$. So far such a behavior is not excluded by sum
rules.

Let us also give the ratio of the imaginary part of the CFFs to  parton
densities,
\begin{eqnarray}
\label{RatioImForw}
\frac{
\Im{\rm m} {\cal H}_i (\xi, \Delta^2, \cQ^2)
}{
q_i (\xi, \cQ^2)
}
= Q_i^2
\xi^{- \alpha^0_i/\alpha_i}
\,
\frac{
\pi \int_{-1}^{1} d z \; (1 - z)^{- \alpha_i} d_i (0, z, \Delta^2, \cQ^2)
}
{
\int_{-1}^{1} d z \; d_i (0, z, 0, \cQ^2)
} \, ,
\end{eqnarray}
where $\alpha_i^0 \equiv \alpha_i (\Delta^2 = 0)$. For the GPD ansatz, we
are using with $b_{\rm sea}=1$, the leading behavior comes from the sea
quarks with $\alpha_{\rm sea} = 1.17$, which provides their ratio
(\ref{RatioImForw}) of order $\approx 1.75 \pi Q_i^2 F_1^{\rm sea} (\Delta^2)$
for each sea quark species.

For the axial-vector channel, analogous relations can be derived, however,
with a difference that the ratio of the real to imaginary part is now
tangent rather than minus cotangent,
\begin{eqnarray}
\label{Res-ReImAxiVec}
\Im{\rm m} \widetilde{{\cal H}}_i (\xi, \Delta^2, \cQ^2)
\!\!\!&=&\!\!\! Q_i^2 \pi \xi^{- \widetilde\alpha_i}
\int_{-1}^{1} d z \;
\frac{\widetilde{d}_i (0, z, \Delta^2, \cQ^2)}{(1 - z)^{\widetilde\alpha_i}}
\, , \\
\Re{\rm e} \widetilde{\cal H}_i (\xi, \Delta^2, \cQ^2)
\!\!\!&=&\!\!\! \tan \left( \widetilde\alpha_i \frac{\pi}{2} \right)
\Im{\rm m} \widetilde{\cal H}_i (\xi, \Delta^2, \cQ^2) \, .
\nonumber
\end{eqnarray}
The ratio of the imaginary part of $\widetilde {\cal H}_i$ to the polarized
parton density $\Delta q_i$ is determined by
\begin{eqnarray}
\frac{
\Im{\rm m} \widetilde{\cal H}_i (\xi, \Delta^2, \cQ^2)
}{
\Delta q_i (\xi, \cQ^2)
} = Q_i^2
\xi^{- \widetilde\alpha^0_i/\widetilde\alpha_i}
\,
\frac{
\pi\int_{-1}^{1} d z \;
(1-z)^{- \widetilde\alpha_i}
\widetilde d_i (0, z, \Delta^2, \cQ^2)
}
{
\int_{-1}^{1} d z \; \widetilde d_i (0, z, 0, \cQ^2)
} \, ,
\end{eqnarray}
with $\widetilde\alpha^0_i = \widetilde\alpha_i (\Delta^2 = 0)$. However,
in contrast to the vector channel, the small-$\Bx$ behavior of the polarized
parton distributions has not been explored in experiments. Usually, one
assumes a similar behavior of the polarized valence- and sea-quark distributions
with $\widetilde\alpha_i \approx 0.5$. Thus, the leading behavior of
$\widetilde {\cal H}$ is determined by all parton species and in the forward
limit it is suppressed with respect to ${\cal H}$ by an extra $\sqrt{\xi}$ or
similar.

In the case of twist-three contributions, the power behavior remains the
same, however, the normalization of the real and imaginary part
will be changed. It is sufficient to discuss the convolution with the
$C^{3(\pm)}_{(0)}$-kernels. The behavior of the effective twist-three
contribution in the WW-approximation of Eq.\ (\ref{Res-tw3eff}), can then be
simply predicted. For the imaginary part we find a modification of the
integrand by the hypergeometric function:
\begin{eqnarray}
\label{Res-ImGenTw3}
\Im{\rm m} \left\{C^{3(\mp)}_{(0)i}\otimes F_i \right\}
= Q_i^2
\frac{\pi}{\alpha_i}
\xi^{- \alpha_i}
\int_{-1}^{1} d z \;
\frac{ d_i (0, z, \Delta^2, \cQ^2)}{(1 - z)^{\alpha_i}}
\;
{_2F_1}
\left(
{1, \alpha_i \atop 1 + \alpha_i }
\Bigg|
- \frac{1 + z}{1 - z}
\right) \, .
\end{eqnarray}
The ratio of the real to imaginary part remains the same as in the twist-two
case, i.e., $-\cot(\pi \alpha/2)$ and $\tan(\pi \widetilde\alpha/2)$ for the
convolution with $C^{3(-)}$ and $C^{3(+)}$ kernels [see Eqs.\
(\ref{Res-ReImVec}) and (\ref{Res-ReImAxiVec})], respectively.

%%%%%%%%%%%%%%%%%%%%%%%%%%%%%%%%%%%%%%%%%%%%%%%%%%%%%%%%%%%%%%%%%%%%%
%                          Figure
%%%%%%%%%%%%%%%%%%%%%%%%%%%%%%%%%%%%%%%%%%%%%%%%%%%%%%%%%%%%%%%%%%%%%
\begin{figure}[t]
\vspace{-0.2cm}
\begin{center}
\mbox{
\begin{picture}(-100,150)(300,0)
\put(0,69){\rotate{[\%]}}
\put(0,0){\insertfig{17}{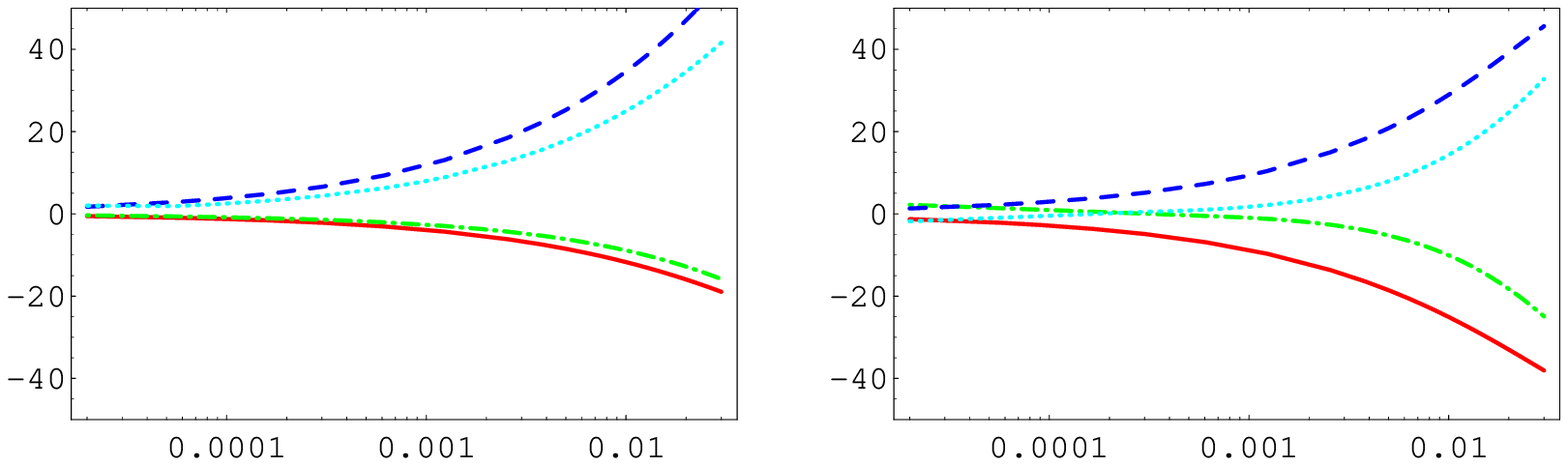}}
\put(40,32){(a)}
\put(40,127){$u$-sea}
\put(280,32){(b)}
\put(280,127){$u$-val}
\put(222,0){$\Bx$}
\put(465,0){$\Bx$}
\end{picture}
}
\end{center}
\caption{\label{Fig-Asy-Bx} The relative deviation of the sea $u$-quark CFF
from its asymptotic value is displayed for $\Im{\rm m} {\cal H}$
and $\Re{\rm e} {\cal H}/\Im{\rm m} {\cal H}$ as the
solid and dashed line, respectively, in (a). The same quantities are plotted
for ${\cal H}^{\rm eff-WW}$ with ${\cal H}^\perp =0$ as dash-dotted and
dotted lines, respectively. The right panel (b) shows the same as in (a) but
for the valence $u$-quark CFF $\widetilde {\cal H}$. The GPD-parameters
are the same as in Fig.\ \ref{Fig-CFF-Vec}.}
\end{figure}
%%%%%%%%%%%%%%%%%%%%%%%%%%%%%%%%%%%%%%%%%%%%%%%%%%%%%%%%%%%%%%%%%%%%%

From Fig.\ \ref{Fig-Asy-Bx} we observe that the relative deviation
of the dominant partonic CFFs ${\cal H}^{\rm sea}_u$ and
$\widetilde{\cal H}^{\rm val}_u$ from their asymptotic values given
in Eqs.\ (\ref{Res-ReImVec}) and (\ref{Res-ReImAxiVec}), respectively,
is of order of 20\% for $\Bx = 10^{-2}$ and decreases to 10\% level,
and even further at $\Bx \le 10^{-3}$. Note, however, that a faster
convergence of the effective twist-three CFFs to its asymptotic form
(\ref{Res-ImGenTw3}) in comparison to the twist-two sector is caused
by an essential cancellation occurred in the WW-approximation. For
instance, the convolution $\left\{C^{3(\mp)}_i\otimes F_i \right\}$ alone
has a large deviation of order of 40\% for $\Bx = 10^{-2}$.

Finally, let us add a word on the isolated mesonic-like states, which we
mentioned at the beginning of this section. It is easy to show that these
$\delta$-like singularities, i.e., $\frac{d^n}{dy^n}\delta(y)$, convoluted
with the hard scattering part yield contributions proportional to
$\xi^{-n-1}$. Since the hard scattering part possesses definite symmetry
properties, the exponent $n$ is odd (even) for $\cal E$ ($\widetilde{\cal
E}$).

%%%%%%%%%%%%%%%%%%%%%%%%%%%%%%%%%%%%%%%%%%%%%%%%%%%%%%%%%%%%%%%%%%%%%
\subsubsection{Radiative corrections}
\label{SubSec-CFF-NLO}
%%%%%%%%%%%%%%%%%%%%%%%%%%%%%%%%%%%%%%%%%%%%%%%%%%%%%%%%%%%%%%%%%%%%%

Now we address the theoretical uncertainties due to radiative corrections,
which have been analytically worked out in a number of papers for the
twist-two sector at NLO
\cite{BelMul97a,JiOsb98,ManPilSteVanWei97,BelMul98,BelFreMul00}, and
numerical studies have been performed in Refs.\
\cite{BelMulNieSch99,FreMcd01a,FreMcd01b}. For the quark sector alone the
NLO amplitudes are shown as dash-dotted curves in Fig.\ \ref{Fig-CFF-Vec}.
The one-loop corrections turn out to be moderate for each quark specie.
However, model-dependent considerations that involve gluonic contributions
showed that NLO effects can be quite sizable both in the valence
\cite{BelMulNieSch99,FreMcd01b} and small-$\Bx$ \cite{FreMcd01b} region. In
the following we have a closer look to this issue and propose a solution to
the problem of large NLO effects.

The NLO corrections are evaluated with the same set of GPDs as earlier. We
disregard the evolution of the coupling and set it equal to $\alpha_s/\pi =
0.1$. The D-term and pion-pole contributions will separately be considered
below. If the real part of a CFF is close to zero at some $\xi$, e.g., due to
cancellations between different quark species, the relative radiative
corrections are blowing up. Thus, it is more convenient to
discuss their magnitude in terms of the absolute values $|{\cal F}| = \sqrt{
(\Re{\rm e}{\cal F})^2 + (\Im{\rm m}{\cal F})^2 }$ and phases ${\rm
Arg} \, {\cal F}$.

%%%%%%%%%%%%%%%%%%%%%%%%%%%%%%%%%%%%%%%%%%%%%%%%%%%%%%%%%%%%%%%%%%%%%
%                          Figure
%%%%%%%%%%%%%%%%%%%%%%%%%%%%%%%%%%%%%%%%%%%%%%%%%%%%%%%%%%%%%%%%%%%%%
\begin{figure}[t]
\vspace{0cm}
\begin{center}
\mbox{
\begin{picture}(0,300)(300,0)
\put(55,215){\rotate{$\cal R$ [\%]}}
\put(55,70){\rotate{$\cal R$ [\%]}}
%\put(75,295){\%}
%\put(302,295){\%}
\put(65,150){\insertfig{8}{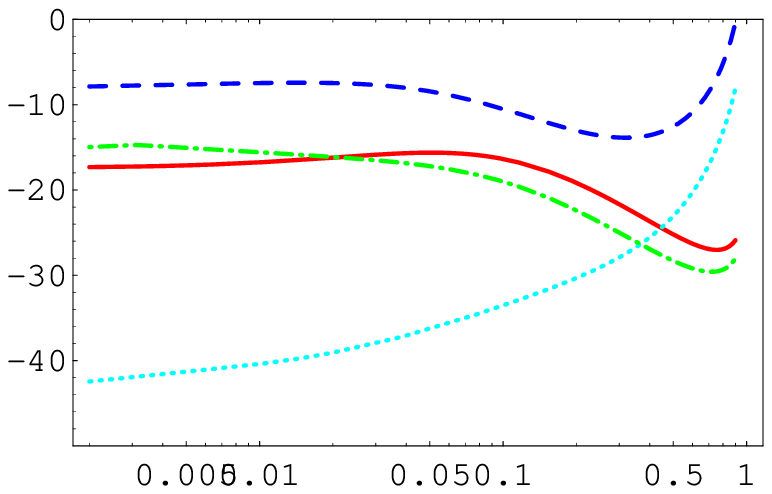}}
\put(305,150){\insertfig{8}{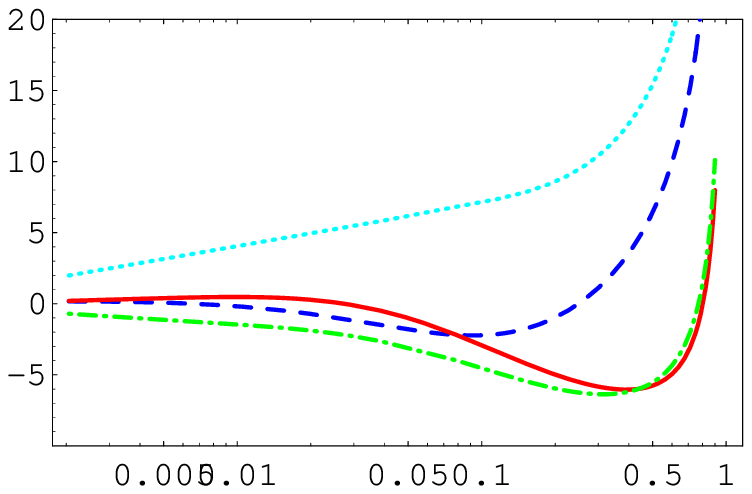}}
\put(260,180){(a)}
\put(340,180){(b)}
\put(65,0){\insertfig{8}{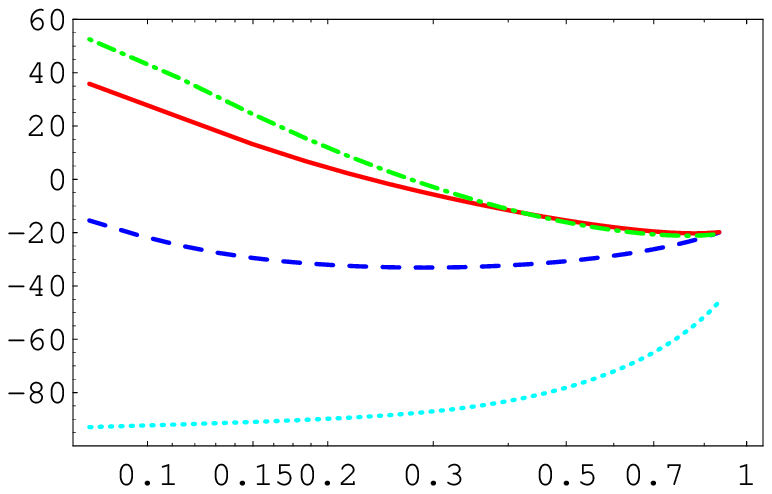}}
\put(305,0){\insertfig{7.8}{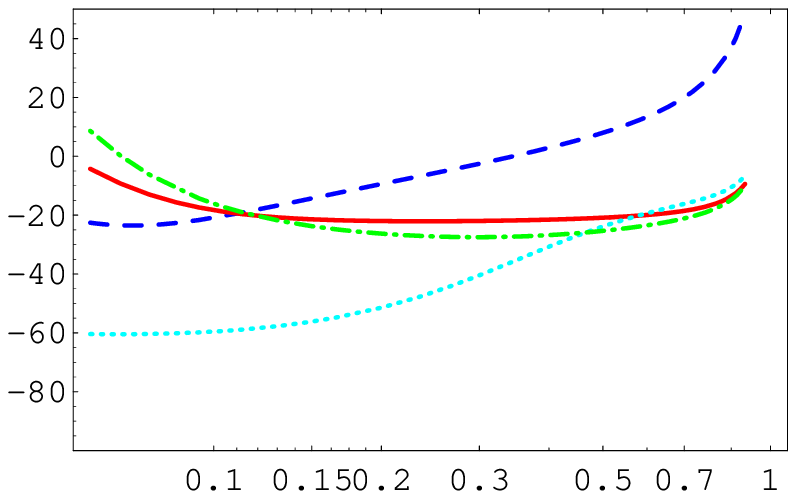}}
\put(260,120){(c)}
\put(340,120){(d)}
\put(280,-10){$\Bx$}
\put(520,-10){$\Bx$}
\end{picture}
}
\end{center}
\caption{\label{Fig-NLO-Cor} The relative NLO corrections ${\cal R}$ (see
Eq.\ \ref{RelRadCorr}) to the absolute value (a,c) and the phase (b,d) of
the CFFs ${\cal H}$ (a,b) and $\widetilde {\cal H}$ (c,d). The magnitude of
radiative corrections with neglected gluon GPD contributions are displayed
as dashed and solid curves for the factorization scale setting $\mu_F = \cQ$
and $\mu_F \approx \mu_F^{\rm DVCS}$, respectively. The complete (quark +
gluon) NLO result is given for these both scale settings by dotted and
dash-dotted lines, respectively. The coupling is set to $\alpha_s/\pi =
0.1$, the quark GPD-parameters are the same as in Fig.\ \ref{Fig-CFF-Vec},
and $b_G=2$.}
\end{figure}
%%%%%%%%%%%%%%%%%%%%%%%%%%%%%%%%%%%%%%%%%%%%%%%%%%%%%%%%%%%%%%%%%%%%%

Let us study at first the naive scale setting condition $\mu_F = \cQ$. In this
case the radiative corrections to the absolute value of the CFFs, i.e.,
\begin{equation}
\label{RelRadCorr}
{\cal R}^{\rm abs}_b
= |{\cal H}^{\rm NLO}| / |{\cal H}^{\rm LO}| - 1
\, ,
\end{equation}
are rather moderate provided gluon contributions are totally neglected. Then
for the choice of the DD-model $b$-parameter $b_{\rm sea} = 1$, one observes
that ${\cal R}^{\rm abs}_1$ is of order $- 10\%$ in both valence and
small-$\Bx$ region, as shown in Fig.\ \ref{Fig-NLO-Cor} (a) by the dashed
line. For the FPD case when $b_{\rm sea} \to \infty$, the ${\cal R}^{\rm
abs}_\infty$ is a factor of 2 larger as compared to ${\cal R}^{\rm abs}_1$,
i.e., ${\cal R}^{\rm abs}_\infty \sim -20\%$. The radiative corrections
stemming from the quark sector alone to the phase of CFFs are defined as in
Eq.\ (\ref{RelRadCorr}) but with the obvious replacements of ${\cal R}^{\rm
abs}$ by ${\cal R}^{\rm ph}$, and ${\cal H}$ by ${\rm Arg} \, {\cal H}$.
They are smaller than $20\%$ in the valence quark region, and are on a
percent level at small $\Bx$, see the dashed line in Fig.\ \ref{Fig-NLO-Cor}
(b). Qualitatively, the same features hold for $\widetilde {\cal H}$.
However, quantitatively we find here a slightly larger effect in the
valence-quark region where relative corrections are of order of $-30\%$ and
$\pm 20\%$ for the absolute value and phase, respectively, see the dashed
curves in Fig.\ \ref{Fig-NLO-Cor} (c) and (d).

To have a feeling on the magnitude of CFFs we give a few numbers. For ${\cal H}$
we have
\begin{eqnarray*}
{\cal H}^{\rm LO}
(\xi = 5 \cdot 10^{-5}, \Delta^2 = 0, \cQ^2 = 4\ \GeV^2)
\approx
(0.39 + 1.37 i)\cdot 10^5 \, ,
\end{eqnarray*}
which is considerably larger than its polarized counterpart,
\begin{eqnarray*}
\widetilde {\cal H}^{\rm LO}
(\xi = 5 \cdot 10^{-5}, \Delta^2 = 0, \cQ^2 = 4\ \GeV^2)
\approx
200 + 163 i \, ,
\end{eqnarray*}
mainly reflecting the fact that $\widetilde\alpha_i < \alpha_{\rm sea}$.

So far, at the level of amplitudes the NLO radiative corrections in the
quark sector were moderate within the model with $b_{\rm sea} = 1$. This
gives us some confidence in the applicability of the perturbative approach to
the DVCS process. Now we come to the gluonic contributions, which enter for
the first time at order ${\cal O} (\alpha_s)$. For $\mu_F = \cQ$ they are
displayed as dotted curves in Fig.\ \ref{Fig-NLO-Cor}. As we see, with our
model for gluonic GPDs their effects in the amplitudes are large. However,
the size of the product $\alpha_s/2\pi\times$ gluonic CFF, compared to the
LO result, is not a measure for the quality of perturbation theory, it
rather gives the ratio of `leading' gluonic to quark CFFs. Of course, one
cannot exclude that these large $\alpha_s$-contributions are merely generated
by an unrealistically oversized ansatz for the gluonic GPDs.

To understand the physical meaning of these large contributions, it is
instructive to study the evolution of the gluonic GPD. We use throughout
$b_G = 2$ in the profile function (\ref{DD-Ansatz}). Taking the GRV
parametrization at its intrinsic low scale $\cQ_0^2 = 0.49\ \GeV^2$,
it has been observed that the gluon GPD changes under evolution much
stronger than their non-singlet quark counterparts \cite{BelMulSchNie98a}.
This is of course also true in the forward case. However, if we take
the MRS A' parametrization at the input scale $\cQ_0^2 = 4\ \GeV^2$,
we observe a strong scale dependence of the NLO result, which is in line
of observations made in Refs.\ \cite{FreMcd01a,FreMcd01b}. Since the NLO
coefficient functions can be expressed in terms of evolution kernels
\cite{BelMul97a}, thus, a strong change under evolution implies a large
gluonic contribution in the one-loop approximation. Analogous to deeply
inelastic scattering one can study the scale dependence by taking the
logarithmic derivative of the CFFs, i.e.,
\begin{eqnarray}
\cQ^2\frac{d}{d\cQ^2} \ln {\cal H}(\xi)
=
\frac{\alpha_s(\cQ^2)}{2\pi}
\frac{
\sum_{i = u, d, s} C_i^{(-)}
\otimes
\left[
V^{qq} \otimes H_i + \frac{1}{\xi} V^{qG} \otimes H_G
\right] (\xi)
}{
{\cal H} (\xi)
}
\, ,
\end{eqnarray}
where $V^{qq}$ and $V^{qG}$ are the LO evolution kernels. This quantity is
demonstrated in Fig.\ \ref{Fig-Der-LogH} for two quite different factorization
scale settings at $\Bx \le 0.1$, where the gluonic term produces, indeed, a
dominant contribution.

%%%%%%%%%%%%%%%%%%%%%%%%%%%%%%%%%%%%%%%%%%%%%%%%%%%%%%%%%%%%%%%%%%%%%
%                          Figure
%%%%%%%%%%%%%%%%%%%%%%%%%%%%%%%%%%%%%%%%%%%%%%%%%%%%%%%%%%%%%%%%%%%%%
\begin{figure}[t]
\vspace{-0.2cm}
\begin{center}
\mbox{
\begin{picture}(-100,150)(300,0)
\put(0,0){\insertfig{8.3}{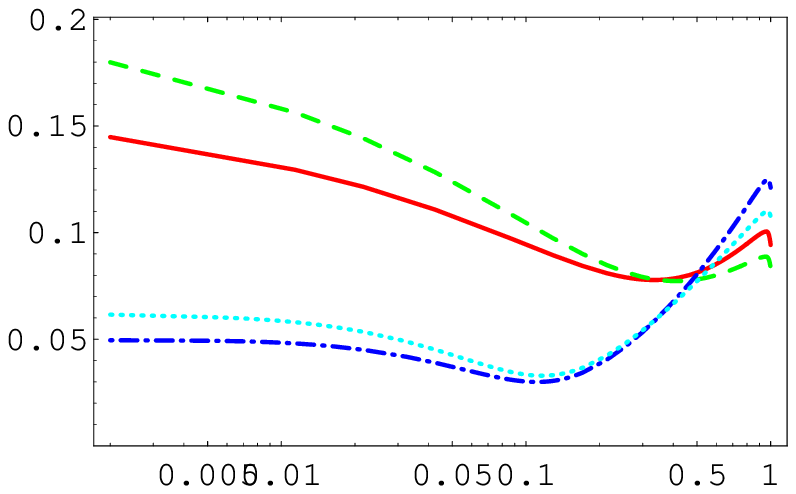}}
\put(250,0){\insertfig{8}{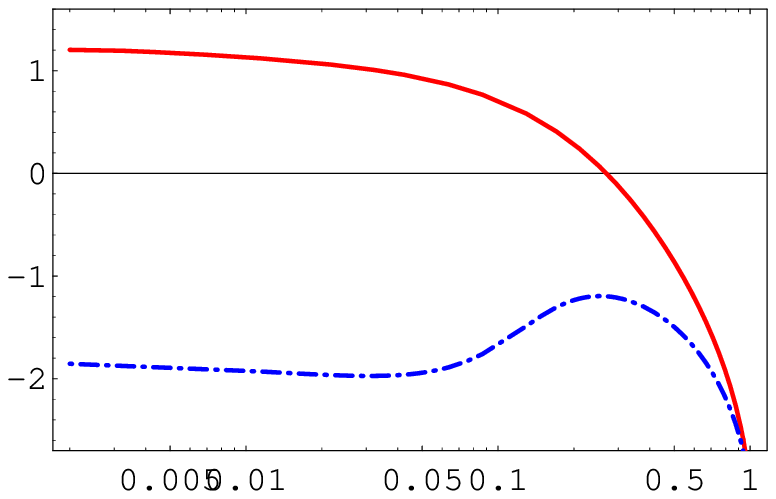}}
\put(200,120){(a)}
\put(440,120){(b)}
\put(220,-6){$\Bx$}
\put(460,-6){$\Bx$}
\end{picture}
}
\end{center}
\caption{\label{Fig-Der-LogH}
The real (a) and imaginary (b) part of $\cQ^2\frac{d}{d\cQ^2} \ln
{\cal H} $ for MRS A' parametrization with $b_{\rm val} =
b_{\rm sea} = 1$ at $\Delta^2 = 0$. The dash-dotted (dotted) line shows
the scale violation due to the evolution of quarks, while the solid
(dashed) one includes also the evolution of gluons for the scale setting
$\mu_F = 3.8 \cQ$ ($\mu_F = \cQ$). The parameters are set as in Fig.\
\ref{Fig-NLO-Cor}.
}
\end{figure}
%%%%%%%%%%%%%%%%%%%%%%%%%%%%%%%%%%%%%%%%%%%%%%%%%%%%%%%%%%%%%%%%%%%%%

Of course, it would be delightful if gluon contributions
would be also moderate. Then DVCS observables can be estimated relying
solely on LO approximation, and the phenomenology of DVCS can in the first
stage be restricted to the quark sector alone. However, even in the opposite
situation  we can minimize the size of large radiative corrections,
originated by strong evolution effects. The restriction to the
quark sector can be achieved by an appropriate factorization
scale setting, namely, in such a way that the gluonic contribution does not
enter the amplitude, e.g., at NLO
\begin{eqnarray}
{\cal H}_G (\xi, \cQ^2)
= \frac{\alpha_s}{2\pi} \sum_{i = u, d, s} Q_i^2
\int_{-1}^1 d x \; C_G( x/\xi, \cQ^2/\mu^2_F)
H_G (x, \xi, \mu^2_F)_{\mu_F = \mu_F^{\rm DVCS}} = 0 \, .
\end{eqnarray}
This scale setting procedure will go under the name of the DVCS scheme. It
requires two $(x,\xi)$-dependent factorization scales $\mu_F$ inside the
convolution, i.e., one for parity even and odd CFFs, respectively. Consequently,
this procedure provides two (one for real and one for
imaginary part) real-valued effective scales $\mu_F^{\rm eff}(\xi)$, that
depend on $\xi$, for each CFF.

An inspection of the imaginary part of the NLO coefficient in the vector
case tells us that the desired factorization scale is almost
$\xi$-independent.
For our model it is given by $\mu^{\rm DVCS}_F \approx \sqrt{2} {\rm e}
{\cal Q} \approx 3.8 {\cal Q}$. Since the real part is expressed in term of the
imaginary part by means of a dispersion relation, we expect that also the
absolute value of ${\cal H}_G$ is minimized in this way. This is demonstrated
in Fig.\ \ref{Fig-NLO-Cor}, where we display the relative NLO corrections, see
Eq.\ (\ref{RelRadCorr}) for the definition, coming from quarks (dashed, solid)
and from all partons, including gluons, (dotted, dash-dotted). We also
employ the scale found for $\cal H$ in the evaluation of $\widetilde{\cal
H}$ in the valence quark region. Here we should mention that with our
models, the CFF $\widetilde {\cal H}$ can be safely neglected in ${\cal
C}^{\rm DVCS}_{\rm unp}$ and ${\cal C}^{\cal I}_{\rm unp}$ for an
unpolarized target. In the vector (a,b) and axial-vector (c,d) cases the
gluonic contributions drastically reduce the LO quark contributions for the
naive $\mu = {\cal Q}$ scale setting (dotted line). For $\mu_F = 3.8 \cQ $,
the radiative gluonic NLO corrections are small (cf.\ dash-dotted and solid
curves). The displayed change of the quark radiative corrections (for the
absolute value from $-10\%$ to $-20\%$) should be considered as a bound,
since we did not include evolution effects. Doing so, they will compensate
the change we have displayed. Let us emphasize again, that our
quantitative considerations
are model-dependent. Taking, for instance, the FPD model, at $\mu_F = \cQ$ we
observe the value of ${\cal R}^{\rm abs}$ is twice larger in both sectors. For
the choice $\mu_F = 3.8 \cQ$, the gluonic NLO contribution is now about
$10\%$, while the quark and the whole NLO corrections are even smaller as
demonstrated in Fig.\ \ref{Fig-NLO-Cor} (a).

Let us now come to contributions that show up only in the exclusive
region. For the  D-term we find to NLO accuracy
\begin{eqnarray}
{\cal H}_{\rm D - term}
=
- 4.98
\left\{
1 + \frac{\alpha_s}{2\pi}
\left(
1.83 - 2.34 \ln\frac{\cQ^2}{\mu_F^2}
\right)
+ \cdots
\right\} \, .
\end{eqnarray}
If we set the factorization scale $\mu_F = \cQ$, the relative correction is
positive and of order of $10\%$. If the gluon model is realistic, we
should better employ the DVCS scheme, in which the D-term is reduced by
about $20\%$.

For the pion-pole contribution the scale dependence drops out in the hard
scattering part since we have chosen the asymptotic form of the pion
distribution amplitude. We get at NLO
\begin{eqnarray}
\widetilde {\cal E} (\xi,\Delta^2)
=
\frac{1}{2\xi} F_{\pi} (\Delta^2)
\left\{
1 - \frac{10}{3} \frac{\alpha_s}{2\pi}
\right\} \, ,
\end{eqnarray}
which acquires the correction of order $15\%$.

%%%%%%%%%%%%%%%%%%%%%%%%%%%%%%%%%%%%%%%%%%%%%%%%%%%%%%%%%%%%%%%%%%%%%
\subsection{Summary}
%%%%%%%%%%%%%%%%%%%%%%%%%%%%%%%%%%%%%%%%%%%%%%%%%%%%%%%%%%%%%%%%%%%%%

Let us summarize the results of this section. We introduced a simple model
with several free parameters based on the factorization of $\Delta^2$ and
$(x, \xi, \cQ^2)$ dependence. The latter is modeled according to the DD ansatz
with MRS A' and GS A parametrization for the parton densities at the input
scale $\cQ_0^2 = 4\ \GeV^2$. This oversimplification is only justified by
the fact that there is no model-independent knowledge available about the
functional dependence of GPDs on their arguments.

In this model the NLO corrections to the partonic quark CFFs are moderate.
The theoretical uncertainties entering at NLO mainly arise  from the lack of
information on the form of gluonic GPDs. Also in the case when a realistic
gluon GPD provides NLO corrections of the same order as the quark
contributions at LO (in ${\cal H}^G$ we observed a different overall sign with
respect to the unpolarized sea-quark contribution within our model), we can
exclude them from the NLO analysis by choosing the DVCS scheme. Although
this favors a rather large factorization scale, the NLO corrections in the
quark sector remain moderate. However, note that the results for the FPD
model of $\cal H$ is a factor of 2 larger than the corresponding estimate
with $b_{\rm sea} = 1$. Therefore, NLO effects can be quite sizable,
especially, in the squared DVCS term. In this product of DVCS amplitudes
they are twice larger than in the interference term, which is linear in the
them. For a rough estimate of observables, which are affected in an
uncontrollable manner by the models accepted for quantitative estimates, it
is enough to work in the quark sector at LO. However, for extraction of
model parameters from experimental data with high accuracy, a NLO analysis
will be inevitable.

Moreover, we found from general assumptions, that the parametrization of the
CFFs $ {\cal F} = \sum_{i = u, d, s} {\cal F}_i$ in the small-$\Bx$ region
is rather unique, and is in fact governed by the small $y$-behavior of the
DDs. The skewedness effect is included in the normalization of the CFFs. Our
analysis suggests that we can take the following general parametrization,
derived in the DVCS-scheme at a given input scale $\cQ^2_0 > 1 \GeV^2$, at
small $\Bx$:
\begin{eqnarray}
\label{Res-ParSmaBx}
\left\{
{
\Re{\rm e}
\atop
\Im{\rm m}
}
\right\}
{\cal H} (\xi, \Delta^2)
\!\!\!&=&\!\!\!
\left\{
{
- \cot\left( \alpha \, \pi/2 \right)
\atop
1
}
\right\}
N_{\cal H} (\Delta^2) \, \xi^{-\alpha(\Delta^2)}
\, , \\
\left\{
{
\Re{\rm e}
\atop
\Im{\rm m}
}
\right\}
{\cal E} (\xi, \Delta^2)
\!\!\!&=&\!\!\!
\left\{
{
- \cot\left( \beta \, \pi/2 \right)
\atop
1
}
\right\}
N_{\cal E} (\Delta^2) \, \xi^{- \beta (\Delta^2)}
+
\left\{ {1 \atop 0} \right\} M_{\cal E} (\Delta^2) \, \xi^{-2 p}
\, , \nonumber\\
\left\{
{
\Re{\rm e}
\atop
\Im{\rm m}
}
\right\}
\widetilde{\cal H} (\xi, \Delta^2)
\!\!\!&=&\!\!\!
\left\{
{
\tan\left( \widetilde\alpha \, \pi/2 \right)
\atop
1
}
\right\}
N_{\widetilde {\cal H}} (\Delta^2)
\xi^{-\widetilde\alpha (\Delta^2)}
\, , \nonumber\\
\left\{
{
\Re{\rm e}
\atop
\Im{\rm m}
}
\right\} \widetilde{\cal E} (\xi, \Delta^2)
\!\!\!&=&\!\!\!
\left\{
{
\tan\left( \widetilde\beta \, \pi/2 \right)
\atop
1
}
\right\}
N_{\widetilde {\cal E}} (\Delta^2)
\xi^{-\widetilde\beta (\Delta^2)}
+\left\{ {1 \atop 0} \right\} \frac{F_\pi (\Delta^2)}{2\xi}
+
\left\{ {1 \atop 0} \right\} M_{\widetilde{\cal E}} (\Delta^2)
\, \xi^{- 2 \widetilde p - 1}
\, , \nonumber
\end{eqnarray}
where we have $\alpha(\Delta^2 = 0) \sim 1.2$ and
$\widetilde\alpha(\Delta^2 = 0) \sim 0.5$ and $p, \widetilde p \ge 1$
are positive integers that reflect the appearance of isolated mesonic-like
states. So they provide the dominant contributions in the spin-flip CFFs and
could overwhelm the small-$\Bx$ behavior of the spin non-flip CFFs. Note
also that because of the small- and large-$\xi$ behavior, and the simple shape
of the (partonic) CFFs, it is obvious that one can extend the parametrization
in the whole kinematical region by allowing a weak $\xi$-dependence in the
$N$-factors and additionally in the phase, while the $M$-factors remain
$\xi$-independent.

%%%%%%%%%%%%%%%%%%%%%%%%%%%%%%%%%%%%%%%%%%%%%%%%%%%%%%%%%%%%%%%%%%%%%
\section{Quantitative estimates for DVCS observables}
\label{Sec-NumEst}
%%%%%%%%%%%%%%%%%%%%%%%%%%%%%%%%%%%%%%%%%%%%%%%%%%%%%%%%%%%%%%%%%%%%%

In this section we give numerical estimates for DVCS observables in the
twist-two and -three approximation at LO of perturbation theory for several
choices of kinematical invariants and compare them, whenever possible, with
available experimental data. To this end, we use exact expressions for the
BH propagators and $K$-factor, when the latter appears as a prefactor in
the Fourier coefficients.  In section
\ref{SubSec-NumEst-H1} we use the DVCS data by H1 collaboration to adjust
free parameters in the sea-quark sector of our model. Then we consider
asymmetries for unpolarized fixed target experiments in section
\ref{SubSec-NumEst-FixTar}. We compare our model expectations with the
beam-spin asymmetry measurements of HERMES and CLAS collaborations and give
then predictions for the E-91-023 experiment at Jefferson Lab
\cite{SteBurEloGar01}. We deliver a general discussion about the magnitude
of twist-three versus twist-two effects in these kinematical situations.
Furthermore, we present a detailed numerical analysis of the size of
twist-three effects for an unpolarized target and derive constraints for
certain charge azimuthal asymmetries. Finally, we give in section
\ref{SubSec-NumEst-LP-Tar} a few estimates for longitudinally polarized
fixed target experiments\footnote{We refer to the longitudinal
polarization in our frame. In the laboratory frame with the $z$-direction
defined with respect to the lepton beam it implies that the target has a
transverse polarization either. The reverse is also true, if in the
lab-frame the target is longitudinally polarized, in the frame Fig.\
\ref{Fig-Kin} we will have both longitudinal as well as transverse
components of the nucleon polarization vector. The projection on the first
one is easily done by integrating over the whole range of the angle
$\varphi$.}.

The reference frame used in the analyses by these experimental
collaborations differs from the one we employ in our considerations, see
Fig.\ \ref{Fig-Kin}, by the direction of the $z$-axis. Both, HERMES and CLAS
use the third direction pointing along the virtual photon's three-momentum.
Thus, their frame (distinguished below by the prime on symbols) is deduced
from ours by the rotation around the $x$-axis on the angle $\pi$. Therefore,
the proton azimuthal angles in our frame $\phi$ and their frame $\phi'_N$
are related by $\phi + \phi'_N = 2 \pi$. The same relation holds for the
real photon angles $\phi_\gamma + \phi'_\gamma = 2 \pi$. The proton and
photon angles are related to each other by the equations $\phi_\gamma = \phi
+ \pi$ and $\phi'_N = \phi'_\gamma + \pi$ in our and their frames,
respectively. Therefore, the proton and real photon azimuth is given in
terms of the angle $\phi$ by
\begin{eqnarray*}
\phi'_N = 2 \pi - \phi
\, ,\qquad
\phi'_\gamma = \pi - \phi \, .
\end{eqnarray*}

%%%%%%%%%%%%%%%%%%%%%%%%%%%%%%%%%%%%%%%%%%%%%%%%%%%%%%%%%%%%%%%%%%%%%
\subsection{Small-$\Bx$ estimates}
\label{SubSec-NumEst-H1}
%%%%%%%%%%%%%%%%%%%%%%%%%%%%%%%%%%%%%%%%%%%%%%%%%%%%%%%%%%%%%%%%%%%%%

In this paragraph we address two questions: (i) What can we learn from
DVCS in the small-$\Bx$ region? (ii) Is there a kinematical window left
to measure twist-three effects? Thanks to the small-$\Bx$ parametrization
devised in Eq.\ (\ref{Res-ParSmaBx}) and the analytical expressions for the
Fourier coefficients, computed in section \ref{Sec-AziAngDep}, both of these
questions can be answered in a straightforward manner. Although we only
consider an unpolarized proton, the analysis is rather complex and we cannot
cover all possible scenarios, which are allowed by the residual degrees
of freedom left in the parametrizations (\ref{Res-ParSmaBx}).

Let us address the first question. We will now see how one can test the
structure of CFFs and measure its small-$\Bx$ behavior in a model-independent
manner. This includes a test of the absence of isolated mesonic-like states,
the measurement of $\cal H$, and, perhaps, access to the spin-flip CFF $\cal E$.
The strategy is to rely first on the hypothesis of the sea-quark dominance in
$\cal H$ and then to derive relations between different observables. A violation
observed in such constraints will then be regarded as a signature of other
contributions involved and not accounted for in the approximation adopted here.

Employing Eqs.\ (\ref{Res-ParSmaBx}), we can now drastically simplify the
twist-two Fourier coefficients. For an unpolarized target the squared DVCS
term [see Eqs.\ (\ref{Res-Mom-DVCS-UP}) and (\ref{Def-C-Int-unp})] reads
\begin{eqnarray}
\label{Res-Mom-DVCS-unp-smaBx}
c^{\rm DVCS}_{0,{\rm unp}}
\!\!\!&=&\!\!\!
2 ( 2 - 2 y + y^2 )
\Bigg\{
\frac{1}{\sin^2\left( \alpha \, \pi/2 \right)}
\frac{
N_{\cal H}^2
}{
\xi^{2 \alpha}
}
- \frac{\Delta^2}{4M^2}
\left(
\frac{
1
}{
\sin^2\left( \beta \, \pi/2 \right)
}
\frac{N_{\cal E}^2}{\xi^{2 \beta}}
+
\frac{F_\pi^2}{4}
\right)
\Bigg\}_{|\xi = \Bx/2}
\, ,
\end{eqnarray}
where the term $F_\pi^2/4$ arises from the pion-pole
contribution\footnote{Since we assume $\widetilde\beta > -1$, we
can safely set $N_{\widetilde{\cal E}}=0$.
Note that the pion-pole contribution gives only a subleading term in
$1/\xi$, however, its numerical value can be rather large for small
$-\Delta^2$. }   in
$\widetilde {\cal E}$. The latter is absent in the interference term,
which is given by [see Eqs.\ (\ref{Res-IntTer-unp}) and (\ref{Def-C-Int-unp})]
\begin{eqnarray}
\label{Res-Mom-Int-unp-smaBx}
\left\{
{
c^{\cal I}_{1, \rm unp}
\atop
s^{\cal I}_{1, \rm unp}
}
\right\}
\!\!\!&=&\!\!\!
8 K
\left\{
{
-(2 - 2y + y^2)
\atop
\lambda y (2 - y)
}
\right\}
\\
&\times&\!\!\!
\left[
F_1
\left\{
{
-\cot\left( \alpha \, \pi/2 \right)
\atop
1
}
\right\}
\frac{N_{\cal H}}{\xi^{\alpha}}
-
\frac{\Delta^2}{4M^2} F_2
\left\{
{
- \cot\left( \beta \, \pi/2 \right)
\atop
1
}
\right\}
\frac{N_{\cal E}}{\xi^{\beta}}
\right]_{|\xi = \Bx/2}
\, . \nonumber
\end{eqnarray}
Note that the D-term can be safely neglected as emphasized above and
$\widetilde {\cal H}$ does not contribute if we assume that $
\widetilde\alpha \ll \alpha$. The last assumption is expected to hold
for parton densities in view of current experimental data. The normalization
factors can be expressed in terms of model-dependent parameters. In our case
they are determined by the partonic form factors, skewedness effects, and
normalization of parton densities. Relying on the sea-quark dominance we find
\begin{eqnarray}
\label{Res-Nor-H-smaBx}
N_{\cal H}(\Delta^2)
=
 \sum_{i = u, d, s} Q_i^2 \frac{n_{\rm sea} (1 + \gamma_{\rm sea})}{3} \,
F_1^{\rm sea} (\Delta^2)
\frac{
\pi
\Gamma(1 + b_{\rm sea} - \alpha_{\rm sea}) \Gamma(2 + 2 b_{\rm sea})
}{
2^{\alpha_{\rm sea}}
\Gamma(1 + b_{\rm sea}) \Gamma(2 + 2 b_{\rm sea} - \alpha_{\rm sea})
} \, ,
\end{eqnarray}
and analogous equation for $N_{\cal E} (\Delta^2)$ with $F_1^{\rm sea}$
being replaced by $F_2^{\rm sea}$. Here $\gamma_{\rm sea}$ contains the
SU(3) flavor-breaking effect and for the MRS parametrization it is
$\gamma_{\rm sea} = 1/10$. Note that the SU(2) flavor breaking, i.e.,
the non-zero difference $\bar d - \bar u$ behaves as a valence-quark
density and, thus, can be neglected  in the small-$\Bx$ region. In the
MRS A' parametrization the normalization factor for the sea-quark
contribution is $n_{\rm sea} = 0.956$ and $\alpha \equiv
\alpha_{\rm sea}= 1.17$. We should emphasize that at LO of perturbation
theory,
\begin{eqnarray*}
N_{F_1}
\equiv
\lim_{b_{\rm sea} \to \infty }
\frac{1}{2\pi}
N_{\cal H}(\Delta^2 = 0)
=
\sum_{i = u, d, s} Q_i^2 \,
\frac{n_{\rm sea} (1 + \gamma_{\rm sea})}{6}
\end{eqnarray*}
is given in terms of the normalization of the structure function $F_1(\Bx, \cQ^2)$
of deeply inelastic scattering, see Eq.\ (\ref{Red-for-F1}),
\begin{eqnarray*}
F_1(\Bx, \cQ^2) \approx N_{F_1}(\cQ^2) \,
\Bx^{- \alpha_{\rm sea} (\Delta^2 = 0,\cQ^2)} \, .
\end{eqnarray*}
Of course, this relation is spoiled by both radiative corrections to the
Wilson--coefficients and evolution effects.

In the coefficient $c^{\rm DVCS}_{0,{\rm unp}}$ of the squared DVCS
amplitude the pion-pole contribution induces an $\Bx$-independent term. As
we see from Eq.\ (\ref{Res-Mom-DVCS-unp-smaBx}), it can be neglected with
respect to the $\cal H$ contribution, if the condition $\xi^{\alpha}
\sqrt{-\Delta^2/4 M^2} \ll 2 N_{\cal H}/\left( F_\pi \sin\left( \frac{\alpha
\pi}{2} \right) \right)$ is fulfilled. Indeed, it is obeyed in the allowed
region of the parameter space for $B_{\rm sea}$ and $\kappa_{\rm sea}$,
provided $\Bx \le 0.01$ and $- \Delta^2 \le 0.5\ \GeV^2$. It is interesting
that the spin-flip contributions $\cal E$ could already be significant at
low momentum transfer, if $\beta > \alpha$. Note that further isolated
mesonic-like states contribute to the small-$\Bx$ behavior of $c^{\rm
DVCS}_{0,{\rm unp}}$. For instance, if there would be such a term in $\cal
E$ with $p = 1$, we will have an additional contribution proportional to
$\left(\xi^2 + \frac{\Delta^2}{4 M^2}\right) M_{\cal E}^2 \xi^{-4}$. The
effect is even more significant in the case of $\widetilde {\cal E}$, where
it would overwhelm the pion-pole contribution and induce an interference
term with $\widetilde{\cal H}$ that gives a dominant effect.

Now let us discuss the extraction of free parameters of our model making use
of experimental results from HERA. To do this, we narrow down the number of
unknowns by assuming the absence of isolated mesonic states and consider the
small momentum transfer kinematics. Thus, $\cal E$ can be neglected in the
Fourier coefficients (\ref{Res-Mom-DVCS-unp-smaBx}) and
(\ref{Res-Mom-Int-unp-smaBx}). Then the absolute value of $\cal H$ can be
extracted from the unpolarized cross section as it has been measured by H1
collaboration \cite{Adl01}. The parameter $\alpha$ can be obtained from the
$\Bx$-dependence of the measured DVCS cross section. More directly it can be
gotten from a measurement of the unpolarized azimuthal asymmetries, which
are proportional to the real part of $\cal H$:
\begin{eqnarray}
\cos^2\left(
\alpha \frac{\pi}{2}
\right)
=
\frac{
\left( c^{\cal I}_{1, {\rm unp}} \right)^2
}{
32 K^2 (2 - 2 y + y^2) c^{\rm DVCS}_{0, {\rm unp}}
}
\propto \frac{\left({\CoA}^{\rm unp}_{c(1)}\right)^2}{{\CeA}^{\rm unp}_{c(0)}}
\, .
\end{eqnarray}
Note that for small values of the lepton energy loss $y$ the squared DVCS
term is enhanced by $1/y^2$ compared to the squared BH term. Thus, a
cleaner separation of the interference and squared DVCS term can be done by
means of the charge asymmetry. Obviously, the consistency of such
measurements tests the dominance of both $\cal H$ and its sea-quark component,
since the measured value $\alpha$ should be consistent
with deeply inelastic scattering data. However, we cannot exclude that a
possible $\Delta^2$-dependence of $\alpha$ alters this relation.

If the polarized lepton beam will be available, one would probe the
imaginary part of $\cal H$ and have so the possibility to measure
$\alpha$ directly as a ratio of the real to imaginary part
of the amplitudes,
\begin{eqnarray}
\cot\left( \alpha \frac{\pi}{2} \right)
=
\frac{
y (2 - y) c^{\cal I}_{1, {\rm unp}}
}{
(2 - 2 y + y^2) s^{\cal I}_{1, {\rm unp}}
}
=
\frac{
y (2 - y) {\CoA}^{\rm unp}_{c(1)}
}{
(2 - 2 y + y^2){\CoA}^{\rm unp}_{s(1)}
}
\, .
\end{eqnarray}
Knowing the parameter $\alpha$, one has an opportunity to access the
normalization in any of these experiments and, thus, to study the effects of
skewedness, which is parametrized by $b_{\rm sea}$ in our model. Unfortunately,
also the slope $B_{\rm sea}$ of the form factor $F_1^{\rm sea}$ is unknown, so
that the skewedness effect cannot be measured in integrated asymmetries. However,
due to the extrapolation $\lim_{\Delta^2 \to 0} N_{\cal H}(\Delta^2)/N_{F_1}$, so
far it is possible.

Perhaps, the most interesting quantity is the CFF $\cal E$ since it contains
information about the orbital angular momentum fraction carried by partons.
In our simplistic model the relevant parameter is the anomalous magnetic
moment $\kappa_{\rm sea}$ of sea quarks. Unfortunately, we cannot say if
$\cal E$ is accessible from unpolarized proton beam experiments at HERA. (A
discussion about the potential of a polarized proton beam can be found in
Ref.\ \cite{BelMulNieSch00}.) From the formulae for the twist-two Fourier
coefficients (\ref{Res-Mom-DVCS-unp-smaBx}) and
(\ref{Res-Mom-Int-unp-smaBx}) one expects that $\cal E$ can be accessed at
larger momentum transfer, where, however, the statistics is low. The absence
of isolated mesonic states and negligible contribution of the spin-flip CFF
${\cal E}$ allows us to derive a relation between the Fourier coefficients,
\begin{eqnarray}
\frac{c^{\rm DVCS}_{0,{\rm unp}}}{2 - 2 y + y^2}
=
\frac{
y^2 (2 - y)^2
\left(
c^{\cal I}_{1, {\rm unp}}
\right)^2
+
(2 - 2y + y^2)^2 \left( s^{\cal I}_{1,{\rm unp}} \right)^2
}{
32 F_1^2 K^2 y^2 (2 - y)^2 (2 - 2 y + y^2)^2
}
\, ,
\end{eqnarray}
which can serve as a test of our considerations. If the spin-flip
contribution $\cal E$ is larger than the pion-pole term and compatible with
$\cal H$, the deviation from this relation is a measure of $N_{\cal E}$.
Without any further assumptions or theoretical predictions, we cannot
directly invert the three Fourier coefficients $c_1^{\cal I}$, $s_1^{\cal
I}$, and $c_0^{\rm DVCS}$ for an unpolarized target to get the four
parameters $\alpha$, $N_{\cal H}$, $\beta$, and $N_{\cal E}$. Fortunately,
if the quality of data is good enough, one can measure the $\alpha,
\beta$-parameters by fitting the slope of the $\Bx$-dependence. Then again,
one has an over-constrained set of observables, which can be used to exclude
contributions from isolated mesonic-like states.

Let us now demonstrate how the model parameters can be adjusted by using the H1
data \cite{Adl01} for unpolarized Compton scattering with the positron beam off
the proton, where the hard momentum transfer ${\cal Q}$ and the invariant energy
$W$,
\begin{equation}
W = \sqrt{ M^2 + \cQ^2 \frac{1 - \Bx}{\Bx} } \, ,
\end{equation}
varied in the range
\begin{eqnarray*}
2\ \GeV^2 < \cQ^2 <20\ \GeV^2
\, , \qquad
30\ \GeV < W < 120\ \GeV \, .
\end{eqnarray*}
In this kinematical window the lepton energy loss $y$ is not large, i.e.,
$0.01 \le y \le 0.16$. The cross section (\ref{WQ}) is integrated over the
momentum transfer squared with the restriction $|\Delta^2| < 1\ \GeV^2 $,
$\cQ^2$ or $W$, as specified, and
the azimuthal angle is integrated out entirely, $0 \leq \phi \leq 2\pi$. For
simplicity we choose $\alpha$ to be independent on $\Delta^2$, and
neglect the evolution and radiative corrections as well as we set $\kappa_{\rm
sea} = 0$. The integration over the azimuthal angle projects on the
constant terms in the squared DVCS and interference term. In the first case
this procedure is exact, while for the latter, due to the lepton propagators
involved, we only approximately pick out the power-suppressed Fourier
coefficient $c_0^{\cal I}$. Fortunately, for the H1 kinematics and within
the WW-approximation adopted in the estimate, a cancellation occurs between
contributions in the interference term, and the latter gives a
negligible effect (on a percent level) in the total differential cross
section. The ratio of BH and DVCS cross sections is proportional to
$(1-y)|\Delta^2|/y^2 \cQ^2$ and, therefore, the BH contribution is much
smaller than the DVCS one. In Fig.\ \ref{Fig-compare-H1} (a) and (b) we show
the measured cross sections $\frac{d\sigma}{d\cQ^2}$ and $\frac{d\sigma}{dW}$,
respectively, versus model predictions for different values of adjustable
parameters in our ansatz. Here we take all CFFs into account, however, as was
explained above, only the sea-quark component of $\cal H$ is dominant.

%%%%%%%%%%%%%%%%%%%%%%%%%%%%%%%%%%%%%%%%%%%%%%%%%%%%%%%%%%%%%%%%%%%%%
%                          Figure
%%%%%%%%%%%%%%%%%%%%%%%%%%%%%%%%%%%%%%%%%%%%%%%%%%%%%%%%%%%%%%%%%%%%%
\begin{figure}[t]
\vspace{-0.2cm}
\begin{center}
\mbox{
\begin{picture}(-100,150)(300,0)
\put(0,25){\rotate{$d \sigma \! / d\cQ^2\, [{\rm pb}/{\rm GeV}^2]$}}
\put(10,0){\insertfig{8}{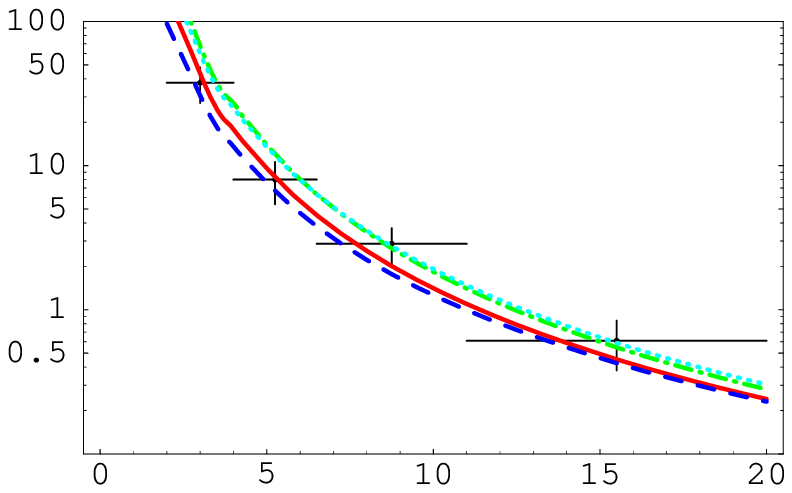}}
\put(260,29){\rotate{$d \sigma \! / dW\, [{\rm pb}/{\rm GeV}]$}}
\put(270,0){\insertfig{7.8}{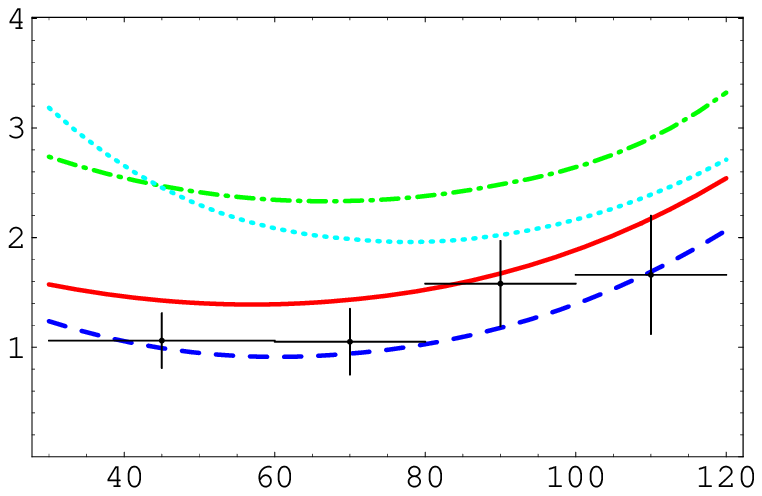}}
\put(210,115){(a)}
\put(455,115){(b)}
\put(200,-10){$\cQ^2\ [\GeV^2]$}
\put(450,-10){$W\ [\GeV]$}
\end{picture}
}
\end{center}
\caption{\label{Fig-compare-H1}
Differential cross sections  $\frac{d\sigma}{d\cQ^2}$ (a) and
$\frac{d\sigma}{d W}$ (b) for $e^+ p \to e^+ p \gamma$
measured by H1 collaboration \cite{Adl01}. The data are compared
with the LO predictions for different parametrizations: MRS A' with
$b_{\rm sea}\to \infty$ and $B_{\rm sea} = 9\ \GeV^{-2}$ (solid line)
as well as $B_{\rm sea} = 5\ \GeV^{-2}$ (dash-dotted); MRS G with
$B_{\rm sea} =9\ \GeV^{-2}$ and $b_{\rm sea}\to \infty$ (dashed) as
well as $b_{\rm sea} = 1$ (dotted).
}
\end{figure}
%%%%%%%%%%%%%%%%%%%%%%%%%%%%%%%%%%%%%%%%%%%%%%%%%%%%%%%%%%%%%%%%%%%%%

The measured normalization of the cross section $\frac{d\sigma}{dW}$, see
Fig.\ \ref{Fig-compare-H1} (b), can be achieved by tuning to a larger value
of the slope $B_{\rm sea}$, a smaller skewedness effect, or a smaller value
of $\alpha_{\rm sea}$. The best agreement we obtain with the MRS G set of
forward parton distributions with $\alpha_{\rm sea} = 1.067$ \cite{MRST98},
$B_{\rm sea} = 9\ \GeV^{-2}$, and $b_{\rm sea} \to \infty$ (dashed line in
Fig.\ (\ref{Fig-compare-H1})). Taking the MRS A' value $\alpha_{\rm sea} =
1.17$ we slightly overshoot the data (solid line). The (integrated) DVCS
cross section is roughly proportional to $1/B_{\rm sea}$, thus, for a lower
value $B_{\rm sea} = 5\ \GeV^{-1}$ we increase the theoretical estimate by
almost the factor of 2 for small $W$ (or small $y$) (dash-dotted). The
skewedness effect for the MRS G parametrization can be read off from the
normalization (\ref{Res-Nor-H-smaBx}) and provides an enhancement of the
DVCS cross section by the factor of $(1.588)^2 \approx 2.5$ (dotted line).
All of these parametrizations are consistent with the measured differential
cross section $\frac{d\sigma}{d\cQ^2}$, Fig.\ (\ref{Fig-compare-H1}) (a).
The evolution flow with ${\cal Q}^2$, which we presently neglect, will give
a moderate logarithmic change of $\cal H$ with increasing $\cQ^2$. Note that
according to Ref.\ \cite{Adl01} the data can also be fit in the aligned jet
model \cite{FraFreStr98} and by adopting the reggeon and soft pomeron
exchange mechanism \cite{DonDos00}.

Since the DVCS amplitudes enter the DVCS cross section in square, the size
of NLO corrections there is twice of what we have found for the CFF $\cal H$
in section \ref{SubSec-CFF-NLO}. They are of order $-40\%$ even in the DVCS
scheme. This requires a NLO analysis in order to extract the NLO parameters
of the GPD models. However, since our ultimate goal is to compare twist-two
and -three approximation, with the latter available only in LO of
perturbation theory, a consistent approach is to use the LO parameters we
have extracted. We will stick in the following to the MRS A' set with
$b_{\rm sea} \to \infty$ and $B_{\rm sea} = 9\ \GeV^{-2}$. Note that the
dependence of predictions on the $\kappa_{\rm sea}$ parameter is weak and
was previously neglected. However, in the following we can use it also as a
free fitting parameter.

%%%%%%%%%%%%%%%%%%%%%%%%%%%%%%%%%%%%%%%%%%%%%%%%%%%%%%%%%%%%%%%%%%%%%
%                          Figure
%%%%%%%%%%%%%%%%%%%%%%%%%%%%%%%%%%%%%%%%%%%%%%%%%%%%%%%%%%%%%%%%%%%%%
\begin{figure}[t]
\vspace{-0.2cm}
\begin{center}
\mbox{
\begin{picture}(-100,150)(300,0)
\put(0,0){\insertfig{8}{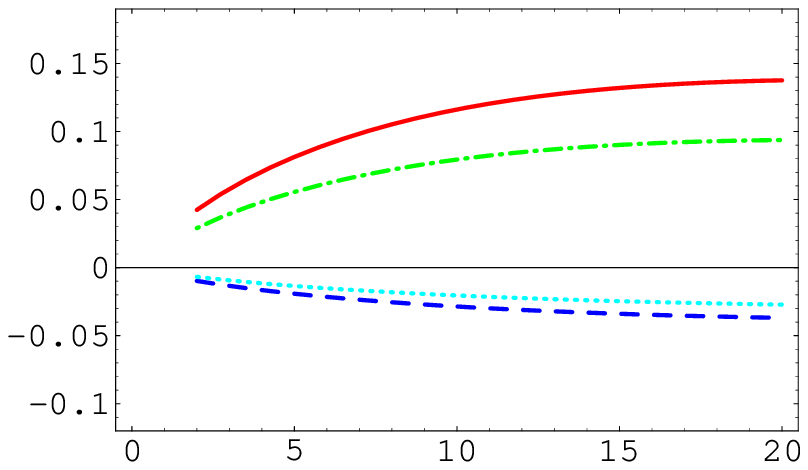}}
\put(260,0){\insertfig{8}{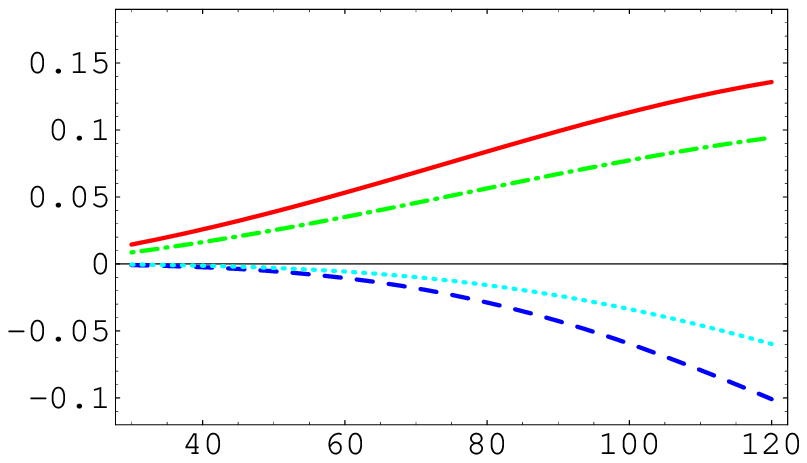}}
\put(40,120){(a)}
\put(300,120){(b)}
\put(190,-10){$\cQ^2\ [\GeV^2]$}
\put(450,-10){$W\ [\GeV]$}
\end{picture}
}
\end{center}
\caption{\label{Fig-CA-tw2-H1}
Twist-two azimuthal charge asymmetries for $\vec{e}^{\,\pm} p \to e^\pm p \gamma$
as functions of $\cQ^2$ for $W =
75\ \GeV$ and $\Delta^2 = - 0.1\ \GeV^2$ (a) as well as functions of $W$
for $\cQ^2 = 4.5\ \GeV^2$ and $\Delta^2 = - 0.1\ \GeV^2$  (b) for the
model of GPDs based on the MRS A' parametrization with $b_{\rm sea} \to
\infty$ and $B_{\rm sea} = 9\ \GeV^{-2}$. For $\kappa_{\rm sea} = 2$
($\kappa_{\rm sea} = - 3$) the asymmetries $\CoA^{\rm unp}_{c(1)}$ and
$\CoA^{\rm unp}_{s(1)}$ are plotted as solid (dash-dotted) and dashed
(dotted) lines, respectively.
}
\end{figure}
%%%%%%%%%%%%%%%%%%%%%%%%%%%%%%%%%%%%%%%%%%%%%%%%%%%%%%%%%%%%%%%%%%%%%

As in the case of cross sections, the size of asymmetries can easily be
estimated from the analytical formulae we have presented. The advantage
is obvious: (i) one gets an understanding of the size of asymmetries
depending on kinematical changes and modification of model parameters,
(ii) one can easily scan the parameter space quantitatively.

The magnitudes of the twist-two azimuthal charge asymmetries $\CoA^{\rm
unp}_{c(1)}$ and $\CoA^{\rm unp}_{s(1)}$ are displayed in Fig.\
\ref{Fig-CA-tw2-H1} for $\Delta^2 = - 0.1\ \GeV^2$. The absolute values of
the asymmetries arising from the $\cos(\phi)$- and $\sin(\phi)$-dependence
in the interference term can reach almost $15\%$ and $10\%$, respectively.
It is interesting that the asymmetries are reduced by about $30\%$ if we
vary the parameter $\kappa_{\rm sea}$ from $2$ to $-3$. This dependence
arises mainly from the interference term, where the $\cal E$-contribution is
enhanced by the normalization of $F_2$, i.e., by the anomalous magnetic
moment of the proton [cf.\ Eqs.\ (\ref{Res-Mom-Int-unp-smaBx})].
Consequently, even with a low mean value of the momentum transfer, e.g.,
$\Delta^2\approx - 0.1\ \GeV^2$, the ${\cal E}$-contribution can be
significant for charge and lepton single-spin asymmetries. Note that for
$\kappa_{\rm sea} = - 3$ in our ansatz, $\cal E$ contributes already for
$\Delta^2\approx -0.6\ \GeV^2$ with the same magnitude as $-\cal H$.
Asymmetries, averaged over the momentum transfer with the restriction
$|\Delta^2| < 1\ \GeV^2$, have almost the same size as the ones displayed in
Fig.\ \ref{Fig-CA-tw2-H1}, e.g.,
\begin{eqnarray*}
&&\CoA^{\rm unp}_{c(1)}
(W = 75\ \GeV, \cQ^2 = 4.5\ \GeV^2)
\approx
8\% \, ,
\\
&&\CoA^{\rm unp}_{s(1)}
(W = 75\ \GeV, \cQ^2 = 4.5\ \GeV^2)
\approx - 2\%
\, ,
\end{eqnarray*}
however, with a weak $\kappa_{\rm sea}$-dependence.

%%%%%%%%%%%%%%%%%%%%%%%%%%%%%%%%%%%%%%%%%%%%%%%%%%%%%%%%%%%%%%%%%%%%%
%                          Figure
%%%%%%%%%%%%%%%%%%%%%%%%%%%%%%%%%%%%%%%%%%%%%%%%%%%%%%%%%%%%%%%%%%%%%
\begin{figure}[t]
\vspace{-0.2cm}
\begin{center}
\mbox{
\begin{picture}(-100,150)(300,0)
\put(0,0){\insertfig{8}{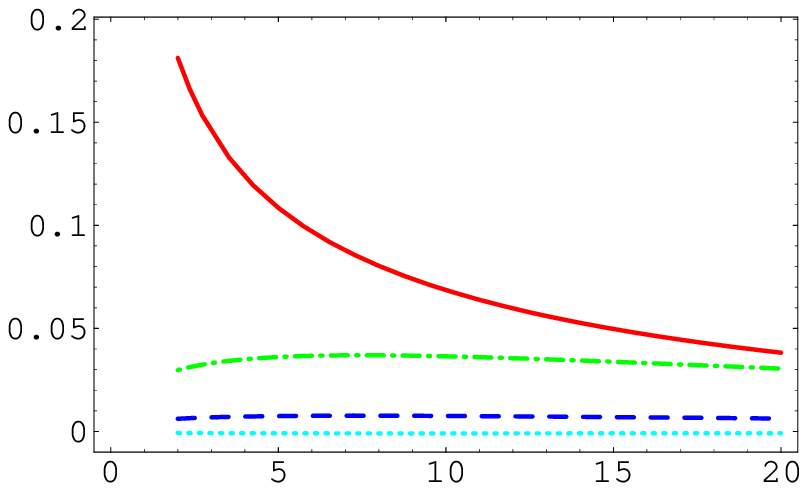}}
\put(260,0){\insertfig{8}{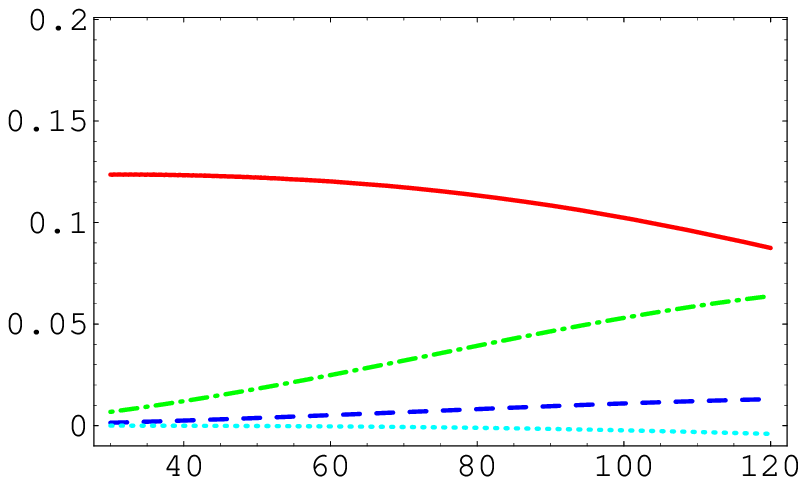}}
\put(203,120){(a)}
\put(460,120){(b)}
\put(190,-10){$\cQ^2\ [\GeV^2]$}
\put(450,-10){$W\ [\GeV]$}
\end{picture}
}
\end{center}
\caption{\label{Fig-CA-tw3-H1}
Twist-three azimuthal charge asymmetries for the same settings
as in Fig.\ \ref{Fig-CA-tw2-H1} (a) and (b), respectively, are
plotted for $\CeA^{\rm unp}_{c(1)}$ (solid), $\CoA^{\rm unp}_{c(0)}$
(dash-dotted), $\CoA^{\rm unp}_{c(2)}$ (dashed), and $\CoA^{\rm unp}_{s(2)}$
(dotted).
}
\end{figure}
%%%%%%%%%%%%%%%%%%%%%%%%%%%%%%%%%%%%%%%%%%%%%%%%%%%%%%%%%%%%%%%%%%%%%

Now we want to answer the second question we raised at the beginning of this
section about the size of twist-three effects. In the WW-approximation for our
model only the sea-quark part of $\cal H$ is dominant. Moreover, from Eqs.\
(\ref{Res-tw3eff}) and (\ref{Def-Fperb}) we see that ${\cal H}^\perp(\xi,\Delta^2)$
drops out for small $\Bx$. Thus, in the limit $b_{\rm sea} \to \infty$ we
discover from the WW-approximation of Eq.\ (\ref{Res-tw3eff})  with the help
of Eqs.\ (\ref{Res-ReImVec}) and (\ref{Res-ImGenTw3}),
\begin{eqnarray}
\label{Res-ParSmaBx-WW}
{\cal H}^{\rm eff-WW} (\xi, \Delta^2)
&=&
\left\{
2 - \alpha_{\rm sea}
\left[
\psi \left( \frac{1 + \alpha_{\rm sea}}{2} \right)
-
\psi \left( \frac{\alpha_{\rm sea}}{2} \right)
\right]
\right\}
{\cal H} (\xi, \Delta^2)
\, ,
\end{eqnarray}
a known relation, which was derived in the archive version of Ref.\
\cite{BelKirMulSch00c}. For $\alpha_{\rm sea} = 1.17$ the ratio ${\cal
H}^{\rm eff-WW} / {\cal H}$ is $\approx 0.65$, while for $b_{\rm sea} = 1$
we find $\approx 0.48$. Apparently, the size of the twist-three effects is
entirely given by the kinematical factors in front of the `universal' $\cal
C$ functions in the definition of Fourier coefficients
(\ref{Res-Mom-DVCS-UP-tw3}), (\ref{Res-IntTer-unp-c0}), and
(\ref{Res-IntTer-unp-tw3}). Thus, we realize that the twist-three harmonics
are relatively suppressed by the factor $\sqrt{- (1 - y)
\Delta_\perp^2/\cQ^2}$. For instance, the suppression is already about the
factor of 4 for $\Delta^2 = - 0.1\ \GeV^2$ and $\cQ^2 = 2 \ \GeV^2$. In
Fig.\ \ref{Fig-CA-tw3-H1} we demonstrate that this counting is not justified
however for the asymmetry $\CoA^{\rm unp}_{c(0)}$ as one can easily realize
from Eq.\ (\ref{Res-IntTer-unp-c0}). For $\Delta^2 = - 0.1\ \GeV^2$ we find
a contribution (dash-dotted line) that is of order of a few percent.
Fortunately, it is expressed in terms of twist-two CFFs and so this
twist-three azimuthal asymmetry can be employed to pin down twist-two CFFs.
Within our model we expect that the addendum (\ref{Def-C-IntAdd-unp}) does
not contribute in Eq.\ (\ref{Res-IntTer-unp-c0}) and, thus, we have the
ratio
\begin{eqnarray*}
\frac{\CoA_{c(0)}^{\rm unp}}{\CoA_{c(1)}^{\rm unp}}=
\frac{\pi}{2}
\frac{(2 - y)^3}{(2 - 2 y + y^2)}
\frac{-\Delta^2}{\cQ^2 K}
\Bigg\{
\left(1 - \frac{\Delta^2_{\rm min}}{\Delta^2} \right)
-  \frac{2(1 - y)}{(2 - y)^2}
\Bigg\}\, .
\end{eqnarray*}

The largest size among the twist-three azimuthal asymmetries has
$\CeA^{\rm unp}_{c(1)}$ (solid line), which arises from the $\cos(\phi)$ term
of the squared DVCS term. Its twist-two counterpart $\CeA^{\rm unp}_{c(0)}$
approaches the value 1 for small $W$ (small $y$), which reflects the fact
that the DVCS process dominates over BH one. If we average over the momentum
transfer squared, we still find a few percent effect for the twist-three
azimuthal charge asymmetries that is comparable with the twist-two ones,
\begin{eqnarray*}
&&\CeA^{\rm unp}_{c(1)} (W = 75\ \GeV, \cQ^2 = 4.5\ \GeV^2)
\approx
7\%
\, , \\
&&\CoA^{\rm unp}_{c(0)} (W = 75\ \GeV, \cQ^2 = 4.5\ \GeV^2)
\approx
2\% \, ,
\end{eqnarray*}
for $- \Delta^2_{\rm min } \le |\Delta^2| < 1 \ \GeV^2$.

%%%%%%%%%%%%%%%%%%%%%%%%%%%%%%%%%%%%%%%%%%%%%%%%%%%%%%%%%%%%%%%%%%%%%
\subsection{Asymmetries in unpolarized fixed target experiments}
\label{SubSec-NumEst-FixTar}
%%%%%%%%%%%%%%%%%%%%%%%%%%%%%%%%%%%%%%%%%%%%%%%%%%%%%%%%%%%%%%%%%%%%%

\begin{table}[t]
\vspace{-0.5cm}
\begin{center}
\begin{tabular}{|c|c|c|c|c|}
\hline
model
&
$b_{\rm val}$
&
$b_{\rm sea}$
&
$B_{\rm sea}$ $[{\rm GeV}^{-2}]$
&
$\kappa_{\rm sea}$
\\
\hline\hline
A
&
1
&
$\infty$
&
9
&
0
\\
\hline
B
&
$\infty$
&
$\infty$
&
$9$
&
$- 3$
\\
\hline
C
&
1
&
1
&
5
&
0
\\
\hline
\end{tabular}
\end{center}
\caption{Parameter sets for models of $H$- and $E$-type GPDs.}
\label{Tab-Models}
\end{table}

After fixing the parameters in the sea-quark sector from the H1 data, we
could hope to give better numerical estimates also for unpolarized fixed
target experiments. To have a comprehensive understanding of this issue, we
define three parameter sets for ${\cal H}$ and ${\cal E}$ based on the MRS A'
parametrization. They are summarized in Table \ref{Tab-Models}. The slope
parameter $B_{\rm D} = 3/m_{\rm D}^2$ of the D-term will be fixed later.
In $\widetilde {\cal H}$ we use the parameters
\begin{eqnarray*}
b_{\rm val} = b_{\rm sea} = 1 \, ,
\qquad
B_A = 3/m_A^2 \approx 3.7\ \GeV^{-2} \, .
\end{eqnarray*}
Obviously, in the models A and B, which are compatible with the H1 measurements,
the sea-quark contributions are relatively small. For the model C this is not the
case and our predictions overshoot the H1 data. If our oversimplified model has
some realistic features, the model C will hardly describe the beam-spin
asymmetries measured at HERMES and Jefferson Lab. The philosophy would be then
again to employ experiments to fix the model parameters left in the valence-quark
sector or to discriminate between the model A and B, or any other.

In the following section \ref{SubSubSec-NumEst-FixTar-BeaSpi} we compare
these models, used for estimates of the DVCS amplitudes calculated in the
twist-three approximation, with the experimental measurements and discuss
theoretical uncertainties in the extraction of model-dependent parameters.
Then we have in section \ref{SubSubSec-NumEst-FixTar-twist-three} a closer
look to the twist-three effects. Especially, we show that asymmetries,
usually discussed in the literature, are not quite suitable to access GPDs.
In the third section \ref{SubSec-NumEst-LP-Tar} we estimate the size and
parameter dependence of nucleon-spin asymmetries for the CLAS kinematics.

%%%%%%%%%%%%%%%%%%%%%%%%%%%%%%%%%%%%%%%%%%%%%%%%%%%%%%%%%%%%%%%%%%%%%
\subsubsection{Beam-spin asymmetry measured with HERMES and CLAS}
\label{SubSubSec-NumEst-FixTar-BeaSpi}
%%%%%%%%%%%%%%%%%%%%%%%%%%%%%%%%%%%%%%%%%%%%%%%%%%%%%%%%%%%%%%%%%%%%%

The azimuthal angular dependence of the beam-spin asymmetry, i.e.,
\begin{eqnarray}
\label{Def-ALU}
A_{\rm LU}(\phi)
=
\frac{
d\sigma^\uparrow(\phi) - d\sigma^\downarrow(\phi)
}{
d\sigma^\uparrow (\phi)+ d\sigma^\downarrow(\phi)
}\, ,
\end{eqnarray}
has been recently measured in two different fixed target experiments,
namely, by HERMES \cite{Aip01} and CLAS \cite{Ste01} collaborations with the
positron and electron beam of $27.6$ GeV and $4.25$ GeV, respectively,
scattered on hydrogen targets. At HERMES the average values of kinematical
variables are $\langle \cQ^2 \rangle = 2.6\ \GeV^2$, $\langle \Bx \rangle =
0.11$ and $\langle - \Delta^2 \rangle = 0.27\ \GeV^2$ \cite{Aip01}. The
results at CLAS are integrated over the regions $1\ \GeV^2 < \cQ^2 < 1.75\
\GeV^2 $, $0.13 < \Bx < 0.35$, and $0.1\ \GeV^2 < - \Delta^2 < 0.3\ \GeV^2$
with the condition that $W > 2\ \GeV$ \cite{Ste01}.

In both experiments we have a large average value of $\langle y \rangle$,
i.e., $\sim 0.5$ and $\sim 0.85$ for the HERMES and CLAS measurements,
respectively. Thus, we expect that the BH part of the total cross section
dominates over the DVCS one [like $(1-y)\Delta^2/y^2 \cQ^2$ as indicated by
Eqs.\ (\ref{Par-BH}-\ref{InterferenceTerm})], which is indeed the case. So
the denominator, $d\sigma^\uparrow + d\sigma^\downarrow$, of the beam-spin
asymmetry (\ref{Def-ALU}) is prevailed by the BH contribution with $c^{\rm
BH}_{0,{\rm unp}}$-coefficient being the dominant one and higher harmonics
suppressed by one or two powers of $K$. The BH part provides a much stronger
contribution for the CLAS than the for HERMES experiment. Thus, we realize
that the $\phi$-dependence of the BH propagators in $A_{\rm LU}(\phi)$
almost cancels. From this qualitative discussion we expect that in the
kinematics, we are considering, the azimuthal dependence is approximately
given by the $\sin(\phi)$ function:
\begin{eqnarray}
A_{\rm LU}(\phi)
\sim
\pm \frac{\Bx}{y}
\frac{s_{1,{\rm unp}}^{\cal I}}{c_{0,{\rm unp}}^{\rm BH}} \sin(\phi)
\propto
\Im{\rm m}
\left\{
F_1 {\cal H}
+
\frac{\Bx}{2 - \Bx} (F_1 + F_2) \widetilde {\cal H}
-
\frac{\Delta^2}{4M^2} F_2 {\cal E}
\right\} \sin(\phi) \, ,
\end{eqnarray}
where higher harmonics are suppressed by $\Delta/\cQ$. Moreover, $A_{\rm
LU}(\phi)$ in this approximation is linear in CFFs, and in the kinematics,
we are discussing, the dominant contribution arises from $\Im{\rm m}{\cal
H}$ within our model. In LO the imaginary part is directly given by the GPDs
on the diagonal $x = \pm \xi$. Thus, uncertainties due to our poor knowledge
of GPDs are minimized for the beam-spin asymmetry. Moreover, after fixing
the parameters for ${\cal H}_{\rm sea}$ from the H1 data, we can confront
our predictions with the data. Obviously, the D-term can only slightly
affect the beam-spin asymmetry. Nevertheless, we include it and set the sea-
and D-slope parameters equal $B_{\rm D} = B_{\rm sea}$.

Now let us compare our expectations at the twist-three level for LO CFFs
with the results from the two experiments shown in Fig.\ \ref{Fig-SSA}.

%%%%%%%%%%%%%%%%%%%%%%%%%%%%%%%%%%%%%%%%%%%%%%%%%%%%%%%%%%%%%%%%%%%%%
%                          Figure
%%%%%%%%%%%%%%%%%%%%%%%%%%%%%%%%%%%%%%%%%%%%%%%%%%%%%%%%%%%%%%%%%%%%%
\begin{figure}[t]
\vspace{-0.2cm}
\begin{center}
\mbox{
\begin{picture}(-100,150)(300,0)
\put(0,0){\insertfig{8}{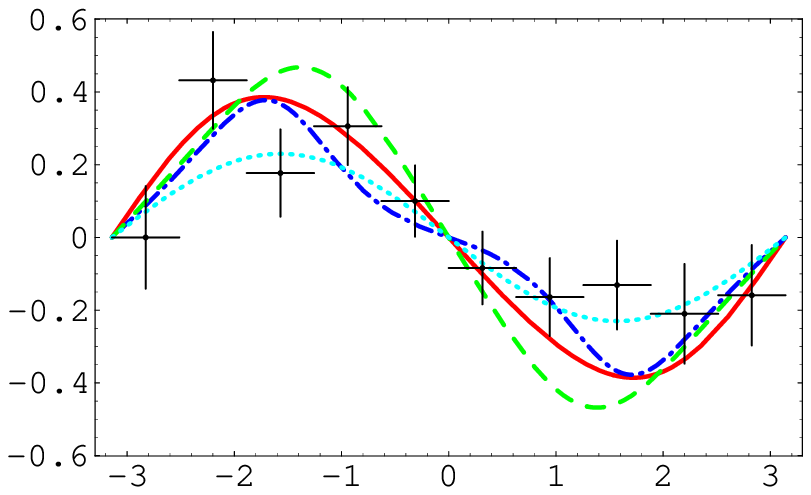}}
\put(260,0){\insertfig{8}{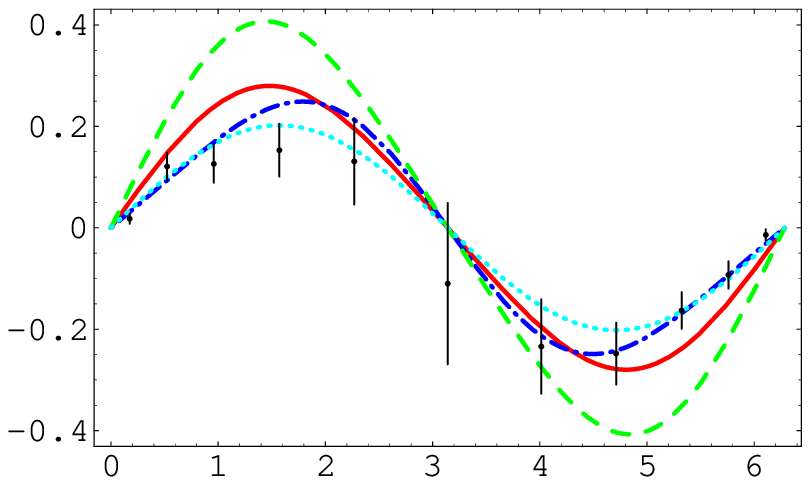}}
\put(200,110){(a)}
\put(460,110){(b)}
\put(200,-10){$\phi_\gamma^\prime$ [rad]}
\put(460,-10){$\phi_\gamma^\prime$ [rad]}
\end{picture}
}
\end{center}
\caption{\label{Fig-SSA}
The beam-spin asymmetry as function of the azimuthal angle measured in $ e^+
p\to e^+ p \gamma$ with $E = 27.6\ \GeV$ from HERMES (a) and $ e^- p\to
e^- p \gamma$ with $E = 4.25\ \GeV$ from CLAS (b) collaborations. The
dotted lines show the $\sin(\phi)$ function with the amplitude $0.23$ and
$0.202$ for HERMES and CLAS, respectively. The other curves are predicted by
the model A (solid) and C (dashed) in the WW approximation as well as the model
B beyond this approximation (dash-dotted), as specified in the text, for
$\cQ^2 = 2.6\ \GeV^2$, $\Bx = 0.11$, and $- \Delta^2 = 0.27\ \GeV^2$ in (a)
as well as for $\cQ^2 = 1.31\ \GeV^2$, $\Bx = 0.19$, and $- \Delta^2 = 0.15\
\GeV^2$ in (b).}
\end{figure}
%%%%%%%%%%%%%%%%%%%%%%%%%%%%%%%%%%%%%%%%%%%%%%%%%%%%%%%%%%%%%%%%%%%%%

The model predictions for the CLAS experiment, plotted in Fig.\ \ref{Fig-SSA} (b),
possess a much smaller deviation from the $\sin(\phi)$-like shape than the
HERMES one shown on the panel (a). For this kinematics a stronger contamination
by higher harmonics, due to the higher average $\langle y \rangle$, can be
induced by the squared DVCS term in $d \sigma^\uparrow + d \sigma^\downarrow$ as
well as twist-three effects. The model A is compatible with the HERMES data (solid
curve). For the integrated asymmetry
\begin{eqnarray}
\label{Def-intALU}
A_{\rm LU} =
\left.
2 \int_0^{2\pi}
d \phi_\gamma^\prime \sin(\phi_\gamma^\prime)
\frac{
d \sigma^\uparrow - d \sigma^\downarrow
}{
d \phi_\gamma^\prime
}
\right/
\int_0^{2\pi} d \phi_\gamma^\prime
\frac{
d \sigma^\uparrow + d \sigma^\downarrow
}{
d \phi_\gamma^\prime
}
\,
\end{eqnarray}
the experimental result from HERMES
\begin{eqnarray*}
A_{\rm LU} = - 0.23 \pm 0.04 ({\rm stat}) \pm 0.03 ({\rm syst})
\, ,
\end{eqnarray*}
is in agreement with the model-A prediction
\begin{eqnarray*}
A_{\rm LU} = - 0.27 \, .
\end{eqnarray*}
If we take a smaller slope parameter $B_{\rm sea}$ or a smaller $b_{\rm sea}$
parameter, the absolute value of the asymmetry will increase. For instance, the
model C leads to
\begin{eqnarray*}
A_{\rm LU} = - 0.37
\end{eqnarray*}
with an obvious deviation from a simple $\sin(\phi)$-shape (dashed), which
is mainly caused by purely kinematical effects. This parameter set yields a
prediction that is close to the result of Ref.\ \cite{KivPolVan00}. On the
other hand we can also decrease the absolute value of $A_{\rm LU}$ if we
take a larger value of $b_{\rm val}$, a smaller $\kappa_{\rm sea}$, or if we
go beyond the WW-approximation. To demonstrate possible effects due to the
antiquark-gluon-quark correlations, we assume a rather large contribution
proportional to the twist-two one, however, with a unique phase
difference\footnote{We have to emphasize that this is only a toy model in
which possible constraints due to sum rules and reduction to the forward
kinematics for antiquark-gluon-quark GPDs were not implemented. }
$\phi^{qGq}$,
\begin{eqnarray}
{\cal F}^{qGq}
=
\frac{1}{\xi} |{\cal F}|
\exp
\left\{
i\ {\rm arg}({\cal F}) + i \phi^{qGq}
\right\}
\, , \mbox{\ \ \ for\ \ \ }
{\cal F}
=
\left\{
{\cal H},
{\cal E},
\widetilde {\cal H}
\right\} \, .
\end{eqnarray}
Here we also included an extra factor $1/\xi$ since the convolution given in
Eq.\ (\ref{Res-tw3eff}) looks rather singular. Now the model B with
$\phi^{qGq} = - \pi/3$ results in a smaller asymmetry,
\begin{eqnarray*}
A_{\rm LU} = - 0.16 \, ,
\end{eqnarray*}
which is also compatible with the data (dash-dotted curve). In this case, we
observe the appearance of $\sin(2\phi)$ and $\sin(3\phi)$ harmonics, which arise
from the interplay of twist-three effects in the interference term and purely
kinematical effects. Note that the amplitude of the $\sin(\phi)$-harmonic in
$A_{\rm LU}(\phi)$ is insensitive to the twist-three sector\footnote{Such
dependence arises mainly from twist-three effects in the denominator on the
right-hand side of Eq.\ (\ref{Def-ALU}), which is, however, dominated by the
BH cross section.}. However, for the azimuthal beam-spin
asymmetry (\ref{Def-intALU}) this does not necessarily hold true.

Now we come to the measurements of CLAS collaboration, which can be fit by
\begin{eqnarray*}
A_{\rm LU}(\phi_\gamma^\prime)
=
\alpha \sin(\phi_\gamma^\prime) + \beta \sin(2\phi_\gamma^\prime)
\, ,
\end{eqnarray*}
with\footnote{Do not mix $\alpha$ and $\beta$ in this parametrization with
$y$ and $\xi$ exponents of parton densities and GPDs, respectively, discussed
in the preceding sections.} \cite{Ste01}
\begin{eqnarray*}
\alpha \!\!\!&=&\!\!\! 0.202 \pm 0.028 ({\rm stat}) \pm 0.013 ({\rm syst})
\, , \\
\beta \!\!\!&=&\!\!\! -0.024 \pm 0.021 ({\rm stat}) \pm 0.009 ({\rm syst})
\, .
\end{eqnarray*}
The $\beta$-parameter is obviously compatible with zero. In the case of the
model C (dashed curve) and now also for the model A (solid curve) one fails
to describe the data since one gets
\begin{eqnarray*}
\alpha = 0.4 \, , \qquad \beta = 0.028 \, ,
\end{eqnarray*}
and
\begin{eqnarray*}
\alpha = 0.28 \, , \qquad \beta = 0.014\, ,
\end{eqnarray*}
respectively. Taking now the model B and going beyond the WW-approximation by
inclusion of antiquark-gluon-quark effects with the phase difference $\phi^{qGq}
= \pi$, we find a result compatible with the data (dash-dotted curve), namely,
\begin{eqnarray*}
\alpha = 0.24 \, , \qquad \beta = - 0.03 \, .
\end{eqnarray*}
As we mentioned before, the value of $\alpha$ is almost insensitive to the
antiquark-gluon-quark contribution. Indeed, in the WW-approximation we have
for the model B the same value of $\alpha$ but quite different $\beta$, i.e.,
\begin{eqnarray*}
\alpha = 0.24 \, , \qquad \beta = 0.014 \, ,
\end{eqnarray*}
which is slightly out of the experimental range. Our model-dependent
analysis shows that the CLAS measurement of the beam-spin asymmetry is
sensitive to antiquark-gluon-quark contributions. However, due to the large
experimental uncertainties, this statement might be premature. On the
theoretical side, the legitimacy of this assertion depends on the answers to
the following two crucial questions: (i) Can we be sure that our equations
are valid at low $\cQ^2$ as of order of 1 $\GeV^2$? (ii) If yes, is the sign
of $\beta$ model-independent in the WW-approximation?

To address the first question, we give now a more qualitative discussion
about our present understanding of corrections, which are expected to modify
our predictions. We provide here some hints rather than answers. In both
experiments the perturbative NLO effects in the DVCS scheme may be of order
of $-20\%$ or so (cf.\ Fig.\ \ref{Fig-NLO-Cor}). Furthermore, we expect that
the evolution effects can alter predictions on a few percent level at most.
The power corrections involved in the analysis are of two kinds $- \Delta^2
/\cQ^2$ and $M^2/\cQ^2$. While the first correction in our model is naively
expected to be of $\pm 10\%$ or so in both experiments, the latter is
certainly larger for the CLAS settings and, therefore, their quantitative
estimate is desirable. To this end, one can adopt the partial result of
Ref.\ \cite{BelMul01a}. In contrast to the forward case, where each power of
the target mass $M^2$ is accompanied by a suppression factor $\Bx^2$, the
situation in the DVCS kinematics is not so obvious and requires detailed
numerical studies. So far it is likely that besides radiative corrections
also power-suppressed contributions are important for the interpretation of
the CLAS experiment. This is especially crucial for the interpretation of
the $\sin(2\phi)$ harmonic.

%%%%%%%%%%%%%%%%%%%%%%%%%%%%%%%%%%%%%%%%%%%%%%%%%%%%%%%%%%%%%%%%%%%%%
%                          Figure
%%%%%%%%%%%%%%%%%%%%%%%%%%%%%%%%%%%%%%%%%%%%%%%%%%%%%%%%%%%%%%%%%%%%%
\begin{figure}[t]
\vspace{-0.2cm}
\begin{center}
\mbox{
\begin{picture}(-100,150)(300,0)
\put(0,0){\insertfig{8}{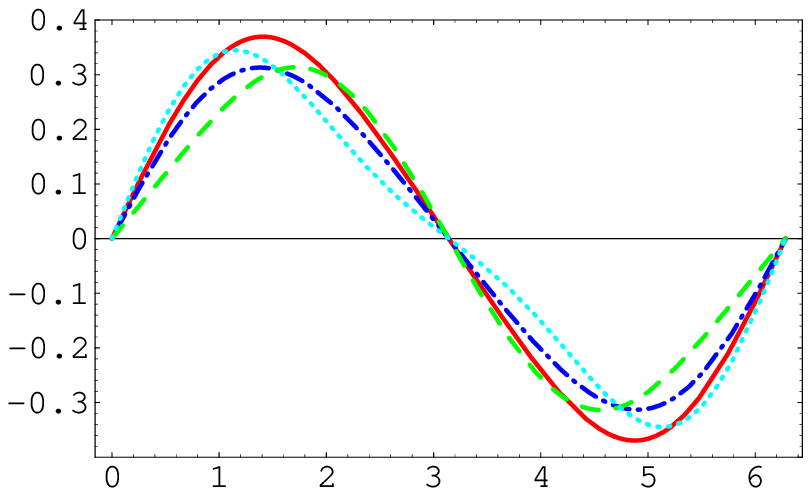}}
\put(260,0){\insertfig{8}{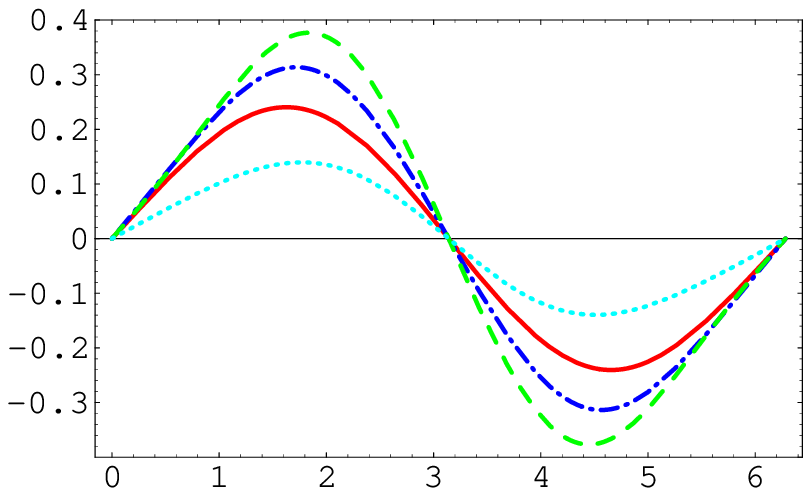}}
\put(200,110){(a)}
\put(460,110){(b)}
\put(200,-10){$\phi_\gamma^\prime$ [rad]}
\put(460,-10){$\phi_\gamma^\prime$ [rad]}
\end{picture}
}
\end{center}
\caption{
\label{Fig-JLAB-BeaSpi} Estimates of the beam spin asymmetry $A_{\rm LU}(\phi)$
for the E-91-023 experiment at Jefferson Lab \protect\cite{SteBurEloGar01} with
the $E = 6\ \GeV$ electron beam scattered on a proton target.
Predictions for the model A (solid) in the
WW-approximation and B (dash-dotted) in the same approximation as well as with
the antiquark-gluon-quark contribution with $\phi^{qGq} = \pi$ (dashed) and
$\phi^{qGq}= 0$ (dotted) are plotted in (a) for $\Bx = 0.3$, $\Delta^2 = - 0.25$,
and $\cQ = 2.5\ \GeV^2$. Here the D-term has been taken into account with
$B_{\rm D}=B_{\rm sea}$.  In (b) we show the model B estimates with $\phi^{qGq} =
\pi $ for $\Bx = 0.35$, $\cQ^2 = 3\ \GeV^2$ (solid), $\Bx = 0.3$, $\cQ^2 = 2.5\
\GeV^2$ (dash-dotted), $\Bx = 0.25$, $\cQ^2 = 2\ \GeV^2$ (dashed), and $\Bx
= 0.19$, $\cQ^2 = 2\ \GeV^2$ (dotted), where in all cases $\Delta^2 = - 0.25 \
{\rm GeV}^2$.
}
\end{figure}
%%%%%%%%%%%%%%%%%%%%%%%%%%%%%%%%%%%%%%%%%%%%%%%%%%%%%%%%%%%%%%%%%%%%%

Keeping these reservations in mind, we finally estimate the beam-spin
asymmetry $A_{\rm LU}(\phi)$ that we expect from future measurements at
Jefferson Lab with the electron beam energy $E = 6\ {\rm GeV}$ in E-91-023
experiment with CLAS. Our expectations are shown in Fig.\ \ref{Fig-JLAB-BeaSpi}
(a) and (b) for the models A and B in different approximations as well as for
different average values of the kinematical variables. They are chosen so, that
$ 0.7 < y < 0.76$ and, thus, the DVCS amplitude is kinematically suppressed.
We have a similar situation as discussed above, however, for larger $\cQ^2$.
This may offer a possibility to investigate a deviation from the predicted
asymmetries, expected in the WW-approximation, in more detail. To clarify the
issue whether this deviation is caused by the antiquark-gluon-quark or the mass
corrections, it is necessary to fix $\Bx$ and vary $\cQ^2$. For instance, our
expectation for the average value of $\Bx = 0.19$ (as in  Fig.\ \ref{Fig-SSA} (b)),
are shown as the dotted line in Fig.\ \ref{Fig-JLAB-BeaSpi} (b) for $\cQ^2 = 2\
{\rm GeV}^2$. For this kinematics, we have only a tiny deviation from the
$\sin(\phi)$-shape in the WW-approximation, i.e.,
\begin{eqnarray*}
A_{\rm LU}(\phi^\prime_\gamma)
=
0.136 \sin(\phi^\prime_\gamma) + 0.005 \sin(2\phi^\prime_\gamma) \, .
\end{eqnarray*}
A shape closer to a $\sin(\phi)$-like form in the E-91-023 measurement would
indicate that the source of higher harmonics in the CLAS experiment at lower
beam energy is caused by mass corrections, since they are now relatively
suppressed by the factor of $\approx 0.65$. Obviously, a definite statement
requires higher statistics in order to decrease the error on $\beta$.

%%%%%%%%%%%%%%%%%%%%%%%%%%%%%%%%%%%%%%%%%%%%%%%%%%%%%%%%%%%%%%%%%%%%%
\subsubsection{A closer inspection of twist-three effects}
\label{SubSubSec-NumEst-FixTar-twist-three}
%%%%%%%%%%%%%%%%%%%%%%%%%%%%%%%%%%%%%%%%%%%%%%%%%%%%%%%%%%%%%%%%%%%%%

The essential feature of the twist-three contributions is the appearance of
new coefficients including a $\phi$-independent harmonic in the interference
term. It can be seen from the equations given in sections
\ref{SubSec-AziAngDep-DVCS} and \ref{SubSec-AziAngDep-INT} that
for longitudinally polarized target higher
Fourier coefficients in the squared DVCS as well as interference term are
relatively suppressed by the factor $K \propto \sqrt{(1 - y)} \Delta_\perp
/{\cal Q} $ with an additional multiplicative $y$-dependence. Thus, higher
harmonics vanish faster at the kinematical boundaries $\Delta_\perp \to 0$
and $y \to 1$ than the twist-two ones. The nonperturbative dependence of the
twist-three CFFs is given by the universal $\cal C$ coefficients. In the
WW-approximation these CFFs have a magnitude similar to the twist-two ones
(see Fig.\ \ref{Fig-CFF-Vec}). Although for the longitudinally polarized
target the constant interference terms $c_0^{\cal I}$, defined in Eqs.\
(\ref{Res-IntTer-unp-c0}) and (\ref{Res-IntTer-LP-c0}), are suppressed by
$1/\cQ$, they do not vanish at the kinematical boundaries. Therefore, we
expect the existence of phase space regions where the interference term is
strongly contaminated by this purely kinematical twist-three effect. Note
that this is not the case for $c_{0,\rm TP}^{\cal I}$, see Eq.\
(\ref{Res-IntTer-TP-c0}).

Let us study the twist-three effects for the longitudinally polarized fixed
target experiment in more detail. For $\Delta^2 = - 0.25 \ \mbox{GeV}^2$ and
${\cal Q}^2 = 4 \ \mbox{GeV}^2$, we expect a suppression of the higher
harmonics, arising beyond the twist-two approximation, roughly by the factor
of 4. Note that $\Delta^2/{\cal Q}^2 \sim 0.06$ and the target mass effects,
i.e., $M^2/{\cal Q}^2$ could be suppressed by a similar factor since they
are combined with an extra multiplicative $\xi$-dependence. So in the future
precise DVCS measurements there will hopefully be a kinematical window left,
where twist-three effects are accessible and $1/\cQ^2$-power corrections are
quantitatively not important. We set for the following numerical estimates
$E_{\rm beam} = 27.6\ \GeV$, and take the model A with the pion pole and
D-term, having equal slopes $B_{\rm D} = B_{\rm sea}$. For the squared DVCS
term we find, indeed, the anticipated suppression in the WW-approximation.
For the longitudinally polarized target and polarized lepton beam, the
Fourier series for $\Bx = 0.1$ ($y \approx 0.77$) reads
\begin{eqnarray}
\label{Res-CS-DVCS-num}
\frac{d \sigma^{\rm DVCS}}{d \Bx d |\Delta^2| d {\cal Q}^2 d \phi}
\!\!\!&\approx&\!\!\!
\bigg\{
0.0023 \left[1 + 0.19 \cos(\phi) - 0.09 \lambda \sin(\phi) \right]
\\
&&\!\!\!+
0.0004 \Lambda
\left[
\lambda +0.16 \lambda \cos(\phi) - 0.04 \sin(\phi)
\right]
\bigg\}
\frac{{\rm nb}}{\mbox{GeV}^4}.
\nonumber
\end{eqnarray}
Let us comment on these numbers. The normalization of the cross section
strongly depends  on the parameter set we have chosen. For instance, for the
model C we find an enhancement in the normalization almost by the factor of
3. However, it depends only weakly on the pion-pole and D-term
contributions. The suppression of terms involving target polarization arises
rather from the destructive interference of $\cal H$ and $\widetilde {\cal
H}$ and should be considered as a reflection of the model taken for the
estimate. For the $\cos(\phi)$ harmonics we have the expected damping, while
the strong suppression of the $\sin(\phi)$ terms arises from the adopted
WW-approximation. The main ingredient of the latter leading to this feature
is the fact that the twist-three CFFs have a phase similar to the twist-two
ones. Thus, the imaginary part of their interference is roughly zero.
A strong deviation from such a small value would indicate an
antiquark-gluon-quark contribution having quite a different phase
structure as compared to the twist-two CFFs.

The interference term for the positron scattering reads
\begin{eqnarray}
\label{Res-CS-INT-num}
&&\frac{d\sigma^{\cal I}}{d\Bx d|\Delta^2| d{\cal Q}^2 d\phi }
\approx
\frac{1}{1 + 0.93 \cos(\phi) + 0.08 \cos(2\phi)}
\\
&&\qquad\qquad
\times
\bigg\{
- 0.007 \left[ 0.56 + \cos(\phi) + 0.17 \cos(2\phi) \right]
- 0.02 \lambda \left[ \sin(\phi) + 0.06 \sin(2\phi) \right]
\nonumber\\
&&\qquad\qquad
+ 0.002 \lambda \Lambda \left[ 0.52 + \cos(\phi) + 0.03 \cos(2 \phi) \right]
- 0.006 \Lambda \left[ \sin(\phi) + 0.09 \sin(2 \phi) \right]
\bigg\}
\frac{{\rm nb}}{\mbox{GeV}^4} \, ,
\nonumber
\end{eqnarray}
with the same kinematical settings as above. Again, the overall
normalization strongly depends on the GPD models. The suppression of higher
harmonics is of the magnitude we have expected or even larger, by the same
token as before. The constant term is rather large. As already mentioned,
this is a purely kinematical twist-three term, i.e., it is completely
determined by the twist-two CFFs. So we see that such a term can alter the
cross section in a much stronger manner than the others. However, it will
not spoil the separation of the dynamical twist-two and -three sectors.
Thus, it is helpful for both experimental consistency checks and access to
the real part of CFFs. Let us also note that the $\phi$-dependence arising
from the BH-propagators is rather strong for the kinematics, we have chosen.

%%%%%%%%%%%%%%%%%%%%%%%%%%%%%%%%%%%%%%%%%%%%%%%%%%%%%%%%%%%%%%%%%%%%%
%                        Figure
%%%%%%%%%%%%%%%%%%%%%%%%%%%%%%%%%%%%%%%%%%%%%%%%%%%%%%%%%%%%%%%%%%%%%
\begin{figure}[t]
\vspace{-0.5cm}
\begin{center}
\mbox{
\begin{picture}(-10,300)(300,0)
\put(265,270){(a)}
\put(40,160){\rotate{\small $d \sigma \! /
(d\cQ^2 d\Delta^2 d\Bx d\phi)\, [{\rm nb}/{\rm GeV}^4]$}}
\put(52,150){\insertfig{8.3}{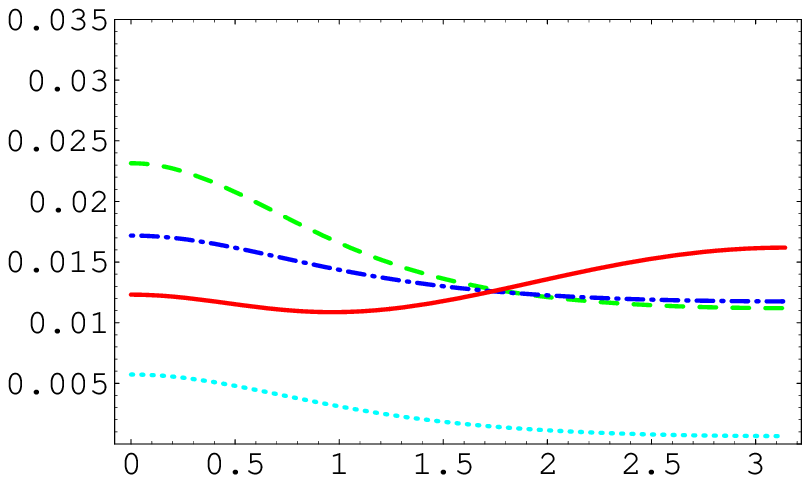}}
\put(340,270){(b)}
\put(295,165){\rotate{\small $(d\sigma^+\! - d\sigma^-)/
(d\sigma^+\! + d\sigma^-) $}}
\put(305,150){\insertfig{8}{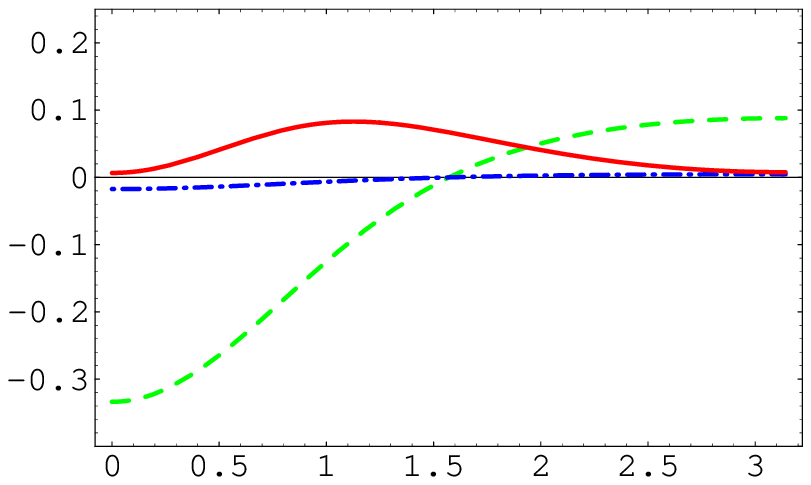}}
\put(40,-5){\rotate{\small $d \sigma \! /
(d\cQ^2 d\Delta^2 d\Bx d\phi)\, [{\rm nb}/{\rm GeV}^4]$}}
\put(295,70){\rotate{\small $A_{\rm LU}$}}
\put(60,-3){\insertfig{8.05}{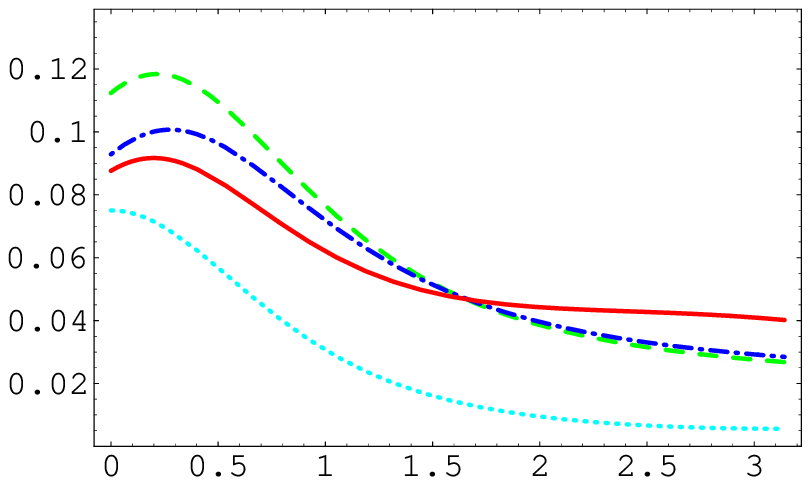}}
\put(305,0){\insertfig{8}{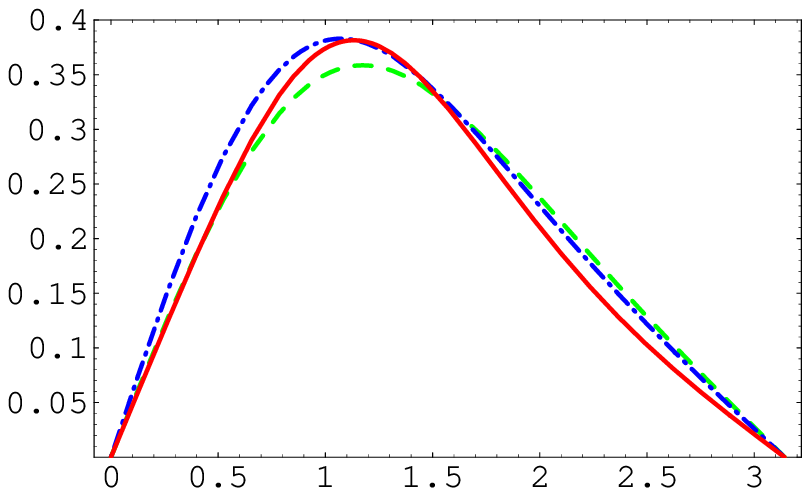}}
\put(265,118){(c)}
\put(340,118){(d)}
\put(260,-10){$\phi^\prime_\gamma$ [rad]}
\put(500,-10){$\phi^\prime_\gamma$ [rad]}
\end{picture}
}
\end{center}
\caption {\label{Fig-CroSec} The unpolarized $e^- p \to e^- p \gamma$
cross section (a), the charge
asymmetry (b) for ${\cal Q}^2 = 2.5\ {\rm GeV}^2$, $\Delta^2 = - 0.25\
{\rm GeV}^2$ and $\Bx = 0.3$ as well as the polarized electron-beam  cross
section (c), and the single-spin asymmetry (d) for ${\cal Q}^2 = 2.5\
{\rm GeV}^2$, $\Delta^2 = - 0.25\ {\rm GeV}^2$ and $\Bx = 0.15$ are shown
for the BH term (dotted), the twist-two without the D-term (dashed), twist-two
 (dash-dotted) and -three (solid) with the D-term for electron
beam within the model C.
}
\end{figure}
%%%%%%%%%%%%%%%%%%%%%%%%%%%%%%%%%%%%%%%%%%%%%%%%%%%%%%%%%%%%%%%%%%%%%

It is instructive to compare our results in the WW-approximation with the
one given in \cite{KivPolVan00}. The sets of GPDs used there presently
differ by the choice of the forward parton densities and partonic form
factors. To be close to the settings of Ref.\ \cite{KivPolVan00}, we take
the model C with $B_{\rm D} = B_{\rm sea} = 5\ \GeV^{-2}$. Note that our $
H_{\rm sea}$ increases less rapidly at small $\xi$ than in the
parametrization from Ref.\ \cite{MRST98}, and we choose a stronger fall-off
with $\Delta^2$ of the D-term. In Fig.\ \ref{Fig-CroSec}, we show the same
predictions for the $\phi$-dependence of the cross section as given in
Figs.\ 2 and 4 of Ref.\ \cite{KivPolVan00}. Our estimates for the
unpolarized cross section (a) and the polarized beam $(\lambda=1)$ cross
section (b) for electron scattering on the unpolarized proton target with
${\cal Q}^2 = 2.5\ {\rm GeV}^2$ and $\Delta^2 = - 0.25\ {\rm GeV}^2$ look
quite similar to \cite{KivPolVan00}. We can clearly observe large
differences of the cross sections and charge asymmetry, i.e., $\{d\sigma^+
(\phi) - d\sigma^- (\phi)\} / \{d\sigma^+ (\phi) + d\sigma^- (\phi)\}$,
between twist-two (dash-dotted curve) and -three (solid curve)
contributions. As we have shown previously in Eq.\ (\ref{Res-CS-INT-num}),
the main effect comes here from the interplay of the constant interference
term with the $\phi$-dependence of the BH propagators.

The most significant differences of both results are prominent in the charge
azimuthal asymmetry, displayed in Fig.\ \ref{Fig-CroSec} (b). Since in our
parametrization the positive sea-quark contribution is larger, we find
without the D-term a large negative value of this asymmetry at
$\phi^\prime_\gamma = 0$ in the twist-two approximation (dashed curve).
Adding the D-term contribution we get the asymmetry, which is compatible
with zero in the same approximation (dash-dotted curve), while the
parametrization of \cite{KivPolVan00} yields a large positive value at
$\phi^\prime_\gamma = 0$. So we conclude that due to many parameters
involved in the modeling of GPDs there is no unequivocal experimental
signature of the D-term. Moreover, at the twist-three level without the
D-term we have a similar magnitude of the asymmetry as displayed by the
solid curve in Fig.\ \ref{Fig-CroSec} (b), which includes the D-term. This
tells us that there is a large cancellation between the twist-two and -three
effects induced by the D-term. Consequently, to access the twist-two GPDs
one has to separate the twist-two and -three sectors. Note that the
cancellation between sea- and valence-quarks as well as the D-term
contributions induces a rather strong dependence on the parametrization for
the unpolarized beam-charge asymmetry, and, thus, may also induce a stronger
contamination by radiative corrections as was observed for separate
contributions.

The beam-spin asymmetry in Fig.\ \ref{Fig-CroSec} (d) is only slightly
modified by the twist-three contributions in the WW-approximation. These
corrections enter mainly in the denominator of this asymmetry, while the
contamination of the numerator by new harmonics remains small as we have
discussed above in Eqs.\ (\ref{Res-CS-DVCS-num}) and (\ref{Res-CS-INT-num}).
Again, for this asymmetry we expect only moderate changes due to radiative
corrections within our ansatz of order of 20\% or so.

%%%%%%%%%%%%%%%%%%%%%%%%%%%%%%%%%%%%%%%%%%%%%%%%%%%%%%%%%%%%%%%%%%%%%
\subsubsection{Estimates and parameter dependence of charge asymmetries}
\label{SubSubSec-NumEst-FixTar-ChaAsy}
%%%%%%%%%%%%%%%%%%%%%%%%%%%%%%%%%%%%%%%%%%%%%%%%%%%%%%%%%%%%%%%%%%%%%

As we have demonstrated in the previous two paragraphs for the beam-spin and
charge asymmetries, an important phenomenological issue is the separation of
the twist-two and -three sectors. This problem is completely resolved by the
introduction of charge asymmetries via Eqs.\
(\ref{Def-CA-Mom-c-0dd}-\ref{CoAsn}) and partially for the single-spin
asymmetry defined by Eq.\ (\ref{Def-SSA-Mom}). We emphasize again that the
charge asymmetries allow also to access the imaginary part of the
interference term in a clean manner, which is hardly possible for the
beam-spin asymmetries. Differences encountered due to various definitions of
the charge and beam-spin asymmetries in the twist-two and -three
approximations are shown in Fig.\ \ref{Fig-CA-SSA}. The asymmetries $A_{\rm
C}$ (a), $A_{\rm SL}$ and $\SSA_{1}$ (b) are normalized to the unpolarized
cross section that does not or only weakly does depend on the dynamical
twist-three contributions. However, as was pointed out above, the constant
term in the interference may generate an important effect. From the
differences of the twist-three (dotted) and twist-two (dash-dotted)
approximation one observes in the panel (b) that the twist-three
contamination is quite essential for smaller values of $\Bx$, i.e., larger
value of $y$, where the BH propagator ${\cal P}_1(\phi)$ induces a rather
strong $\phi$-dependence. In the case of the beam-spin asymmetry an
improvement is necessary if considerable antiquark-gluon-quark correlations
are present. As it can be seen by comparison of the twist-two (short-dashed)
and -three (dashed) approximation for $\SSA_{1}$, a clean separation is only
achieved when the DVCS amplitude is suppressed with respect to the BH one,
i.e., $\sqrt{-(1-y)\Delta^2}/y\cQ \ll 1$. Note also that the twist-two
approximation of $\SSA_{1}$ is close to the charge asymmetry $\CoA^{\rm
unp}_{s(1)}$. The deviation of $\CoA^{\rm unp}_{s(1)}$ from $\SSA_{1}$
arises only from the interference term in the denominator.

%%%%%%%%%%%%%%%%%%%%%%%%%%%%%%%%%%%%%%%%%%%%%%%%%%%%%%%%%%%%%%%%%%%%%
%                          Figure
%%%%%%%%%%%%%%%%%%%%%%%%%%%%%%%%%%%%%%%%%%%%%%%%%%%%%%%%%%%%%%%%%%%%%
\begin{figure}[t]
\vspace{-0.5cm}
\begin{center}
\mbox{
\begin{picture}(-100,150)(300,0)
\put(0,0){\insertfig{8}{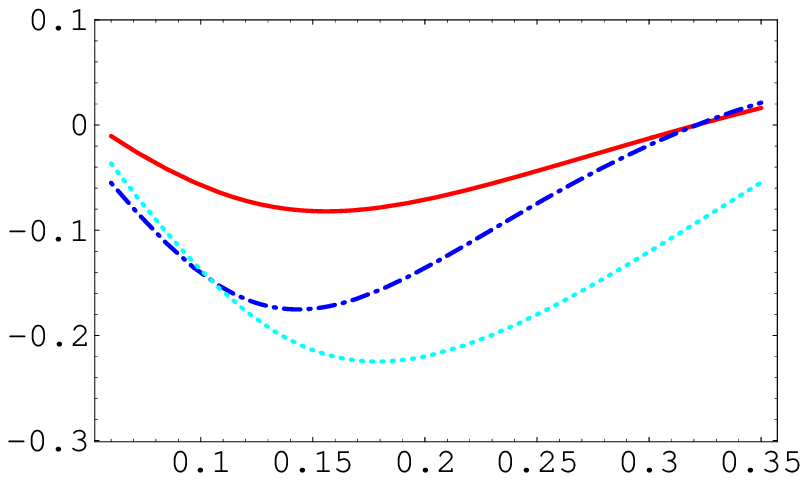}}
\put(260,0){\insertfig{8}{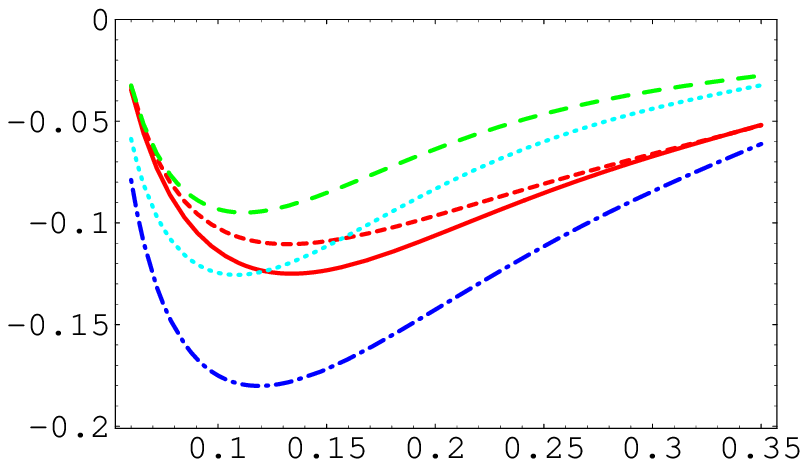}}
\put(195,115){(a)}
\put(302,115){(b)}
\put(210,-5){$\Bx$}
\put(470,-5){$\Bx$}
\end{picture}
}
\end{center}
\caption{\label{Fig-CA-SSA}
The azimuthal charge asymmetry $A_{\rm C}$ (a) and the beam-spin asymmetry
$A_{\rm SL}$ (b) are plotted in twist-two (dash-dotted) and twist-three
(dotted) for the HERMES kinematics with ${\cal Q}^2 = 2.5\ \GeV^2$ and
$\Delta^2 = - 0.25$.  The solid curves show the predictions of the
improved charge asymmetries $\CoA^{\rm unp}_{c(1)}$  and $\CoA^{\rm
unp}_{s(1)}$. In (b) the short-dashed and dashed curves display $\SSA_1$ in
twist-two and -three approximation, respectively. All curves are given for
model A with $B_{\rm D}=9\ \GeV^2$ and includes an antiquark-gluon-quark
correlation with $\phi^{qGq} = - \pi/3$.}
\end{figure}
%%%%%%%%%%%%%%%%%%%%%%%%%%%%%%%%%%%%%%%%%%%%%%%%%%%%%%%%%%%%%%%%%%%%%

Unfortunately, the improved definitions reduce the size of asymmetries. This
disadvantage is unavoidable provided one is willing to extract in a clean way
information on the twist-two and -three GPDs.

Now let us discuss in more detail the magnitude of the improved charge
azimuthal asymmetries, defined in Eqs.\
(\ref{Def-CA-Mom-c-0dd}-\ref{Def-CA-Mom-s-Even}) for the charge-odd and
-even parts of the cross section, in dependence on the twist-two
parametrizations and antiquark-gluon-quark correlations. For the twist-two
sector, we present our estimates in Fig.\ \ref{Fig-CA-Mom-Tw2} with the
models A and C. As was mentioned above in section
\ref{SubSubSec-NumEst-FixTar-twist-three}, in contrast to the $\sin (\phi)$
harmonic the size and the sign of the $\cos (\phi)$ term crucially depends
on the parametrization of the valence- and sea-quark densities as well as
the D-term contribution. According to our model, the sea quarks give a
positive real part of ${\cal H}$, while the valence quarks and D-term
provide a negative effect. The $\CA_{c(1),{\rm unp}}^{\rm odd}$, plotted in
panel (a), is now positive if it is dominated by the sea quarks, like in the
model C without the D-term (dashed), contrary to all other cases.
There is also a possibility that those contributions may compensate each
other and this will lead to the almost vanishing asymmetry. Since the model
C (dashed) is ruled out by the beam-spin measurement, we rather expect a
sign alternating asymmetry [the model A without the D-term (solid)] or a
negative one [the model A with $B_{\rm D} = 5\ \GeV^{-2}$ (dash-dotted)].
Its magnitude can be of order of $\pm 10\%$, however, it can also be
compatible with zero. In panel (b) we plot the leading
twist-two asymmetry $\CeA_{c(0)}^{\rm unp}$, which shows that the BH
amplitude dominates for small $\Bx$. Also for HERMES there is a kinematical
region in which the DVCS cross section overwhelms the BH one. This happens
for the model A (B) at $\Bx\approx 0.2$ ($\Bx\approx 0.17$). The DVCS cross
section only weakly depends on the D- and pion-pole term.

%%%%%%%%%%%%%%%%%%%%%%%%%%%%%%%%%%%%%%%%%%%%%%%%%%%%%%%%%%%%%%%%%%%%%
%                          Figure
%%%%%%%%%%%%%%%%%%%%%%%%%%%%%%%%%%%%%%%%%%%%%%%%%%%%%%%%%%%%%%%%%%%%%
\begin{figure}[t]
\vspace{0cm}
\begin{center}
\mbox{
\begin{picture}(0,270)(300,0)
\put(47,150){\insertfig{8}{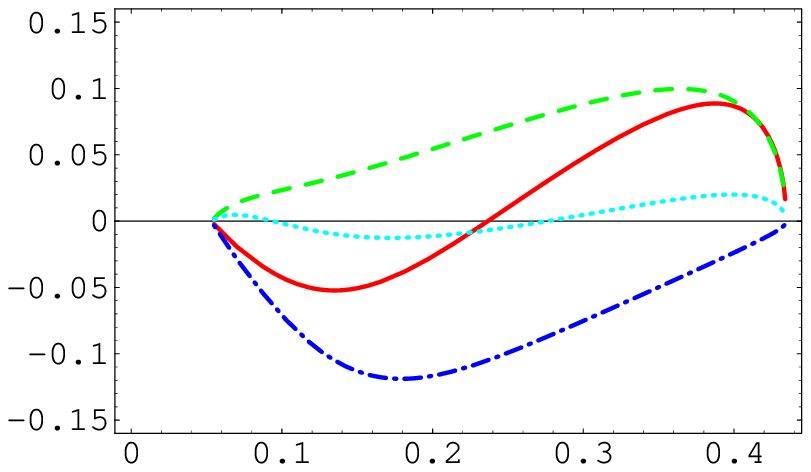}}
\put(252,175){(a)}
\put(330,175){(b)}
\put(300,153){\insertfig{7.8}{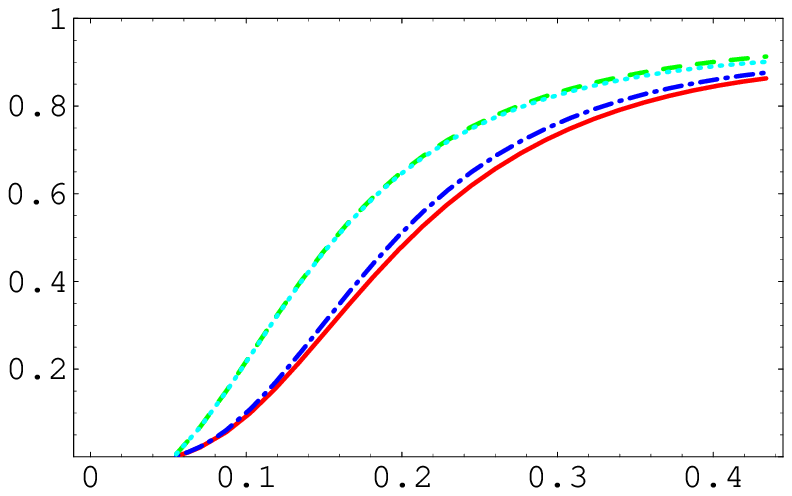}}
\put(47,0){\insertfig{8}{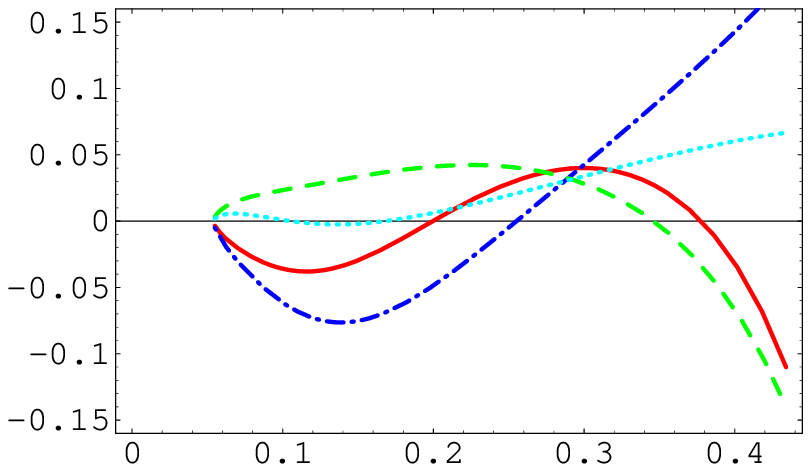}}
\put(297,3){\insertfig{7.8}{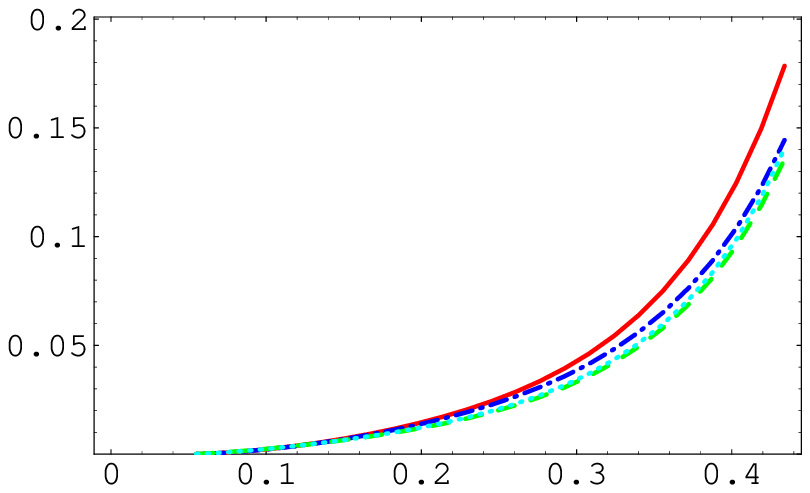}}
\put(252,120){(c)}
\put(330,120){(d)}
\put(265,-10){$\Bx$}
\put(505,-10){$\Bx$}
\end{picture}
}
\end{center}
\caption {\label{Fig-CA-Mom-Tw2} Azimuthal charge asymmetries of
$\cos$-harmonics $\CoA_{c(1)}^{\rm unp}$ (a), $\CeA_{c(0)}^{\rm unp}$ (b),
$\CoA_{c(0)}^{\rm unp}$ (c), and $\CoA_{\Delta}^{\rm unp}$ (d) are plotted
for $E_{\rm beam} = 27.6\ \GeV$, $\Delta^2=-0.25,$ and $\cQ^2 =2.5\ \GeV^2$.
The model A (C) with the D-term, $B_{\rm D} = 5\ \GeV^{-2}$, and neglected
one are displayed as dash-dotted (dotted) and solid (dashed) curves,
respectively.}
\end{figure}
%%%%%%%%%%%%%%%%%%%%%%%%%%%%%%%%%%%%%%%%%%%%%%%%%%%%%%%%%%%%%%%%%%%%%

In Fig.\ \ref{Fig-CA-Mom-Tw2} (c) we show the twist-three asymmetry
$\CoA_{c(0),{\rm unp}}$, which can be rather sizeable, especially, at the
kinematical boundary $\Delta_\perp \to 0$. The complexity of its shape
partly emerges from the kinematical factors in Eq.\
(\ref{Res-IntTer-unp-c0}). According to this equation and from the
definition of the charge asymmetries in section \ref{SubSec-PhyObs-ChaAsy},
we have the constraint
\begin{eqnarray}
\CoA_{\Delta}^{\rm unp}
&\!\!\!\equiv\!\!\!&
\CoA_{c(0)}^{\rm unp}
-
\frac{\pi}{2}
\frac{(1 - \Bx)(2 - y)^3}{(2 - 2 y + y^2)}
\frac{-\Delta^2}{\cQ^2 K}
\Bigg\{
\left(1 - \frac{\Delta^2_{\rm min}}{\Delta^2} \right)
-  \frac{(2 - \Bx)(1 - y)}{(1 - \Bx)(2 - y)^2}
\Bigg\}
\CoA_{c(1)}^{\rm unp}
\nonumber\\
&\!\!\!=\!\!\!&
\frac{\pi}{2}
\frac{(2 - y)}{(2 - 2 y + y^2)} \frac{\sqrt{-\Delta^2}}{\sqrt{\cQ^2} K}
(1 - y)(2 - \Bx) \times \dots \, ,
\end{eqnarray}
where the ellipsis stand for terms generated by the addendum $\Delta {\cal
C}_{\rm unp}^{\cal I}$, defined in Eq.\ (\ref{Def-C-IntAdd-unp}). For large
$y$ this difference should vanish. Although, $\Delta {\cal C}_{\rm
unp}^{\cal I}$ is suppressed by at least one power of $\Bx$, we see in Fig.\
\ref{Fig-CA-Mom-Tw2} (d) that it starts to be sizeable for $\Bx > 0.3 $.
This effect arises from the fact that for this kinematics in our model
$|\widetilde{\cal H}| > |{\cal H}|$, and $\widetilde{\cal H}$ can contribute
considerably both to $\CA_{c(1),{\rm unp}}^{\rm odd}$ and $\CoA_{c(0)}^{\rm
unp}$. If one would measure a large deviation of $\CoA_{\Delta}^{\rm unp}$
from zero, it will indicate that $\Re{\rm e} {\cal H}$ is small and does not
give the dominant contributions in the even harmonics.

Now we come to the dynamical twist-three effects, which show up in the
charge asymmetries $\CoA_{c(2)}^{\rm unp}$ and $\CeA_{c(1)}^{\rm unp}$
displayed in Fig.\ \ref{Fig-CA-Mom-Tw3} (a) and (b), respectively. In the
WW-approximation with the model A (dash-dotted and solid curves) we find
again the expected magnitude of asymmetries: the size of the charge-even
asymmetry $\CeA_{c(1)}^{\rm unp}$ is larger than the twist-two charge-odd
one $\CoA_{c(1)}^{\rm unp}$. Furthermore, we see that the shape of
$\CoA_{c(2)}^{\rm unp}$ is correlated with the one of $\CoA_{c(1)}^{\rm
unp}$. A large $\CoA_{c(2)}^{\rm unp}$ asymmetry can be generated by
multi-parton correlation effects shown as dashed and dotted lines,
respectively. Such correlation can even enhance the asymmetry
$\CeA_{c(1)}^{\rm unp}$.

%%%%%%%%%%%%%%%%%%%%%%%%%%%%%%%%%%%%%%%%%%%%%%%%%%%%%%%%%%%%%%%%%%%%%
%                          Figure
%%%%%%%%%%%%%%%%%%%%%%%%%%%%%%%%%%%%%%%%%%%%%%%%%%%%%%%%%%%%%%%%%%%%%
\begin{figure}[t]
\vspace{-0.5cm}
\begin{center}
\mbox{
\begin{picture}(-100,150)(300,0)
\put(0,0){\insertfig{8}{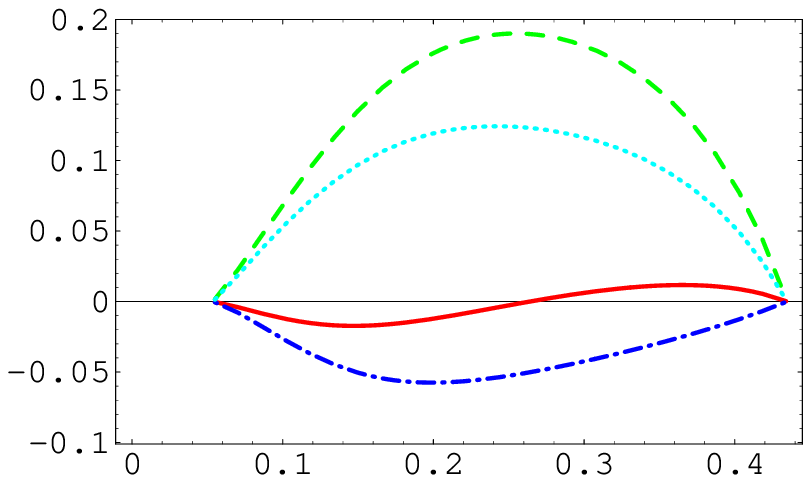}}
\put(260,0){\insertfig{8}{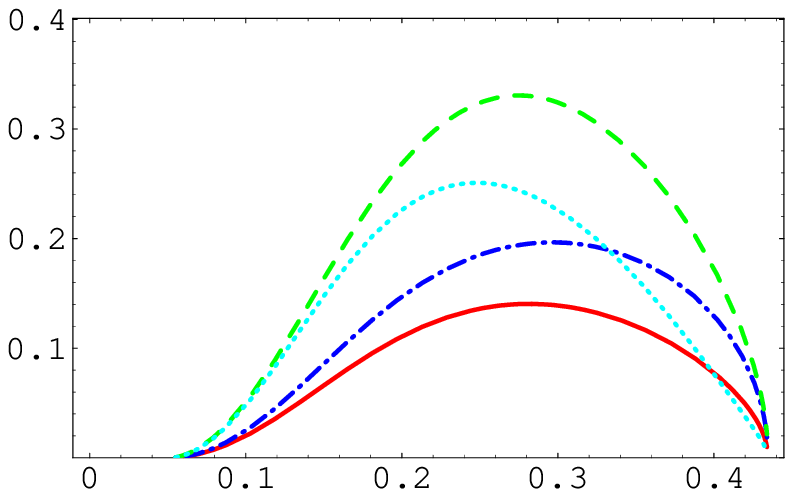}}
\put(40,120){(a)}
\put(295,120){(b)}
\put(220,0){$\Bx$}
\put(480,0){$\Bx$}
\end{picture}
}
\end{center}
\caption{\label{Fig-CA-Mom-Tw3}
Twist-three azimuthal charge asymmetries of $\cos$-harmonics
$\CoA_{c(2)}^{\rm unp}$ (a) and $\CeA_{c(1)}^{\rm unp}$ (b) are plotted for
the same kinematics as in Fig.\ \ref{Fig-CA-Mom-Tw2}. The model A with
(without) the D-term in the WW-approximation and with the
antiquark-gluon-quark correlation ($\phi^{qGq} = - \pi/3$) are displayed as
dash-dotted (dotted) and solid (dashed) curves, respectively. }
\end{figure}
%%%%%%%%%%%%%%%%%%%%%%%%%%%%%%%%%%%%%%%%%%%%%%%%%%%%%%%%%%%%%%%%%%%%%

Let us add that the parameter dependence in the twist-two sector and
twist-three effects for the polarized beam can also be discussed in a clear
manner in terms of the charge asymmetries $\CoA_{s(1)}^{\rm unp}$,
$\CoA_{s(2)}^{\rm unp}$, and $\CeA_{c(1)}^{\rm unp}$. However, since
qualitatively they display the same effects as it has already been discussed
for the beam-spin asymmetry in section \ref{SubSubSec-NumEst-FixTar-BeaSpi},
we will not repeat these considerations here.

%%%%%%%%%%%%%%%%%%%%%%%%%%%%%%%%%%%%%%%%%%%%%%%%%%%%%%%%%%%%%%%%%%%%%
\subsection{Numerical estimates for longitudinally polarized target}
\label{SubSec-NumEst-LP-Tar}
%%%%%%%%%%%%%%%%%%%%%%%%%%%%%%%%%%%%%%%%%%%%%%%%%%%%%%%%%%%%%%%%%%%%%

%%%%%%%%%%%%%%%%%%%%%%%%%%%%%%%%%%%%%%%%%%%%%%%%%%%%%%%%%%%%%%%%%%%%%
%                          Figure
%%%%%%%%%%%%%%%%%%%%%%%%%%%%%%%%%%%%%%%%%%%%%%%%%%%%%%%%%%%%%%%%%%%%%
\begin{figure}[t]
\vspace{-0.5cm}
\begin{center}
\mbox{
\begin{picture}(-100,150)(300,0)
\put(0,0){\insertfig{8}{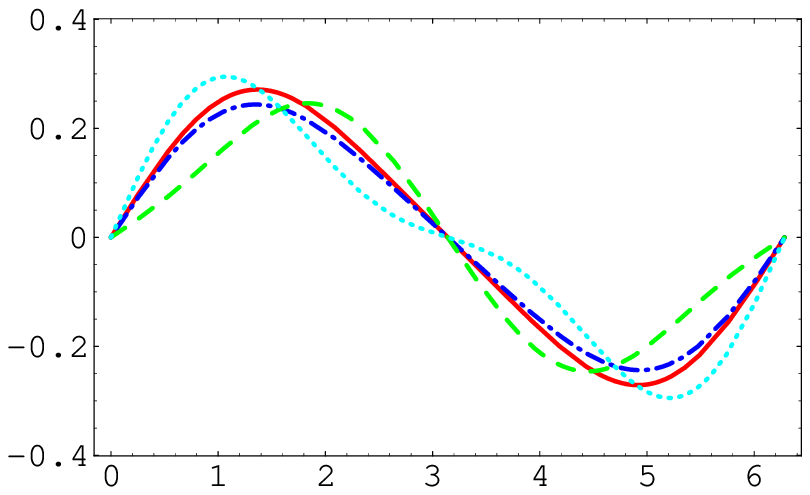}}
\put(260,0){\insertfig{8}{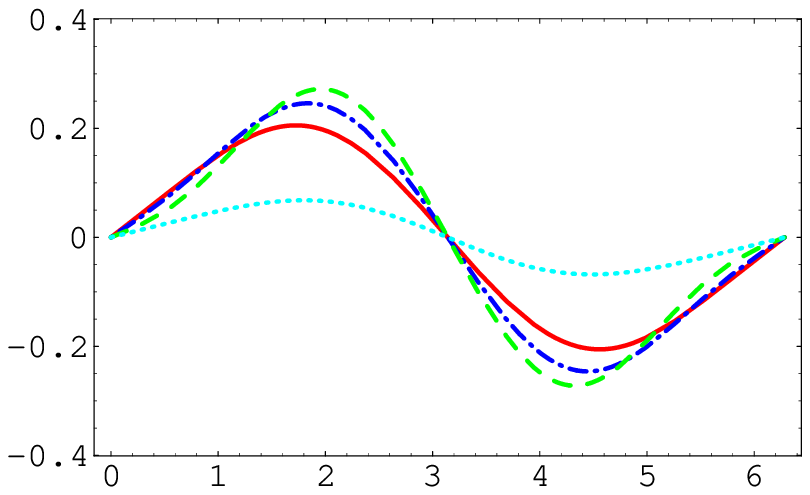}}
\put(190,120){(a)}
\put(445,120){(b)}
\put(200,-5){$\phi_\gamma^\prime$ [rad]}
\put(460,-5){$\phi_\gamma^\prime$ [rad]}
\end{picture}
}
\end{center}
\caption{\label{Fig-LongNuc-E113}
Estimates of the target spin asymmetry $A_{\rm UL}(\phi)$
for the E-91-023 experiment at Jefferson Lab \protect\cite{SteBurEloGar01}
with the $E = 6\ \GeV$ electron beam. Same models and kinematics as in
Fig.\ \ref{Fig-JLAB-BeaSpi} (a) and (b).
}
\end{figure}
%%%%%%%%%%%%%%%%%%%%%%%%%%%%%%%%%%%%%%%%%%%%%%%%%%%%%%%%%%%%%%%%%%%%%

Finally, we provide quantitative estimates for the cross sections of
polarized electron scattering off the longitudinally polarized proton
target. In these settings there are two further observables, which can be
used to unravel GPDs from  data.

The first one is the target-spin asymmetry with unpolarized lepton beam
\begin{equation}
A_{\rm UL} (\phi)
=
\frac{
d\sigma^\Uparrow (\phi) - d\sigma^\Downarrow (\phi)
}{
d\sigma^\Uparrow (\phi) + d\sigma^\Downarrow (\phi)
} \, ,
\end{equation}
where the up (down) arrow stands for $\Lambda = + (-) 1$ longitudinal
polarization. Completely analogous to the beam-spin asymmetry, the BH cross
section cancels and one is left with the leading- and higher-twist
contributions from the
interference term and power-suppressed effects from the squared DVCS
amplitude. From the explicit expressions for the Fourier coefficients one
can qualitatively understand the magnitude of the asymmetry. For instance,
comparing the main term in the asymmetry ${\cal C}^{\cal I}_{\rm LP}$, see
Eq.\ (\ref{Def-C-Int-LP}), with ${\cal C}^{\cal I}_{\rm unp}$, previously
analyzed in section \ref{SubSubSec-NumEst-FixTar-BeaSpi}, one observes that
the CFF ${\cal H}$, dominating the latter at
moderate and small $\Bx$, now enters the amplitude with an additional power
of $\Bx$. It becomes parametrically of the same order as the parity-odd CFF
$\widetilde {\cal H}$: $|\widetilde {\cal H}| \sim \Bx |{\cal H}|$. Thus
both of them play a distinctive role in building up the nucleon-spin
asymmetry which is displayed in Fig.\ \ref{Fig-LongNuc-E113} for the
kinematics of the E-91-023 experiment at Jefferson Lab with $E = 6 \ \GeV$
electron beam \cite{SteBurEloGar01}. On the left panel (a) we give the
estimates with the models A and B for the kinematical variables chosen as
$\Bx = 0.3$, $\Delta^2 = - 0.25 \ \GeV^2$, and ${\cal Q}^2 = 2.5 \ \GeV^2$.
While on the panel (b) we display the asymmetry for several settings of
the kinematical variables with the model B, which gave the best agreement
with the unpolarized proton data.

%%%%%%%%%%%%%%%%%%%%%%%%%%%%%%%%%%%%%%%%%%%%%%%%%%%%%%%%%%%%%%%%%%%%%
%                          Figure
%%%%%%%%%%%%%%%%%%%%%%%%%%%%%%%%%%%%%%%%%%%%%%%%%%%%%%%%%%%%%%%%%%%%%
\begin{figure}[t]
\vspace{-0.5cm}
\begin{center}
\mbox{
\begin{picture}(-100,150)(300,0)
\put(0,0){\insertfig{8}{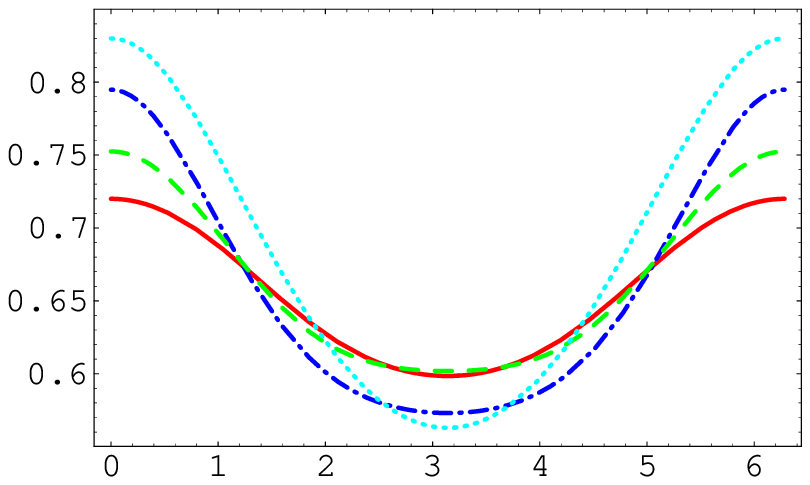}}
\put(260,3){\insertfig{8}{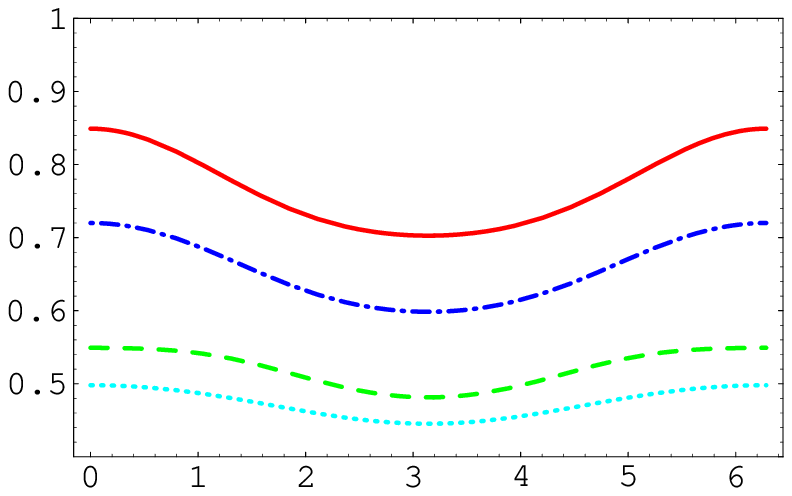}}
\put(190,120){(a)}
\put(445,120){(b)}
\put(200,-5){$\phi_\gamma^\prime$ [rad]}
\put(460,-5){$\phi_\gamma^\prime$ [rad]}
\end{picture}
}
\end{center}
\caption{\label{Fig-LongLeptNuc-E113}
Estimates of the double spin asymmetry $A_{\rm LL}(\phi)$ for the E-91-023
experiment at Jefferson Lab \protect\cite{SteBurEloGar01} with a $E = 6\
\GeV$ electron beam. Predictions for the model A (solid) with $B_{\rm D}=
B_{\rm sea}$ in the WW-approximation  and model B  in the same
approximation without D-term (dash-dotted) as well as with D-term and
antiquark-gluon-quark contributions with $\phi^{qGq} = \pi$ (dashed)
are plotted in (a) for $\Bx = 0.3$, $\Delta^2 = - 0.25$,
and $\cQ = 2.5\ \GeV^2$. The dotted  curve shows the BH contribution alone.
 In (b) we show the model A estimates for the same kinematics as in
Fig.\ \ref{Fig-JLAB-BeaSpi}(b).
}
\end{figure}
%%%%%%%%%%%%%%%%%%%%%%%%%%%%%%%%%%%%%%%%%%%%%%%%%%%%%%%%%%%%%%%%%%%%%

The second interesting observable is the double lepton-nucleon spin asymmetry.
The combination of the cross section which picks out only the $\lambda
\Lambda$-part, bilinear in the nucleon and lepton helicity, is achieved by the
following combination of the leptoproduction cross sections
\begin{equation}
A_{\rm LL} (\phi)
=
\frac{
d\sigma^{\uparrow \Uparrow} (\phi)
-
d\sigma^{\downarrow \Uparrow} (\phi)
-
d\sigma^{\uparrow \Downarrow} (\phi)
+
d\sigma^{\downarrow \Downarrow} (\phi)
}{
d\sigma^{\uparrow \Uparrow} (\phi)
+
d\sigma^{\downarrow \Uparrow} (\phi)
+
d\sigma^{\uparrow \Downarrow} (\phi)
+
d\sigma^{\downarrow \Downarrow} (\phi)
} \, .
\end{equation}
Since the double-spin asymmetry is not protected from the contribution of
the BH process, one expects that it overwhelms the DVCS signal
inasmuch as  it dominates the cross section for the CLAS kinematics. Indeed,
this is what one finds in Fig.\ \ref{Fig-LongLeptNuc-E113} (a), where the BH
cross section alone generates a large asymmetry\footnote{Note that we set in
the asymmetry $A_{\rm LL}$ the DVCS amplitude in both the numerator and
denominator to zero.}, displayed as dotted curve. Obviously, the
interference and squared DVCS term produce a significant signal on its
background. Due to rather good knowledge of the BH process its subtraction
from the data should not present an obstacle of principle. However,
what is more troublesome is the contamination of the asymmetry by the DVCS
cross section and
the fact that three CFFs, i.e., ${\cal H}$, $\widetilde{\cal H}$, and the
pion-pole term of $\widetilde{\cal E}$, can contribute on the same footing
to the interference term. This makes the disentanglement of separate
components and, therefore, GPDs unmanageable from such an experiment alone.
The expectations for the same kinematical settings as in Fig.\
\ref{Fig-LongNuc-E113} (b) are given in Fig.\ \ref{Fig-LongLeptNuc-E113} (b)
for the model A.

%%%%%%%%%%%%%%%%%%%%%%%%%%%%%%%%%%%%%%%%%%%%%%%%%%%%%%%%%%%%%%%%%%%%%
\section{Summary and conclusions}
%%%%%%%%%%%%%%%%%%%%%%%%%%%%%%%%%%%%%%%%%%%%%%%%%%%%%%%%%%%%%%%%%%%%%

Our present study has resulted into a complete analytical structure of
the real-photon leptoproduction cross section with power accuracy.
Although these results are quite complex, they allow for a qualitative
discussion of observables in dependence of the kinematics and GPD
parameters. Thus, in contrast to a purely numerical approach, our analysis
tolerates a careful examination of experimental data. The most accurate
information on GPDs can be deduced from the part of the total cross section,
which stems from the interference of the BH process with the
DVCS amplitude. The main tool of experimental exploration of the latter is
the use of diverse asymmetries involving charge and spin.

The leading contributions in the light-cone expansion of the DVCS amplitude
give access to diverse properties of the nucleon constituents due to a
non-zero orbital momentum carried by them. These include, the parton
angular momentum, the helicity-flip polarized glue uncontaminated by quarks,
the pion cloud of the proton etc., just to name a few.

Twist expansion of the DVCS amplitude finds its reflection in the dependence
of the cross section on the azimuthal angle of the outgoing real photon (or
recoiled proton). One of the main results of our analysis is the finding
that the Fourier harmonics of the separate charge-even and -odd components
of the complete cross section, expressed in terms of the lowest-twist GPDs,
stay uncontaminated even if $1/{\cal Q}$-power effects are switched on.
Thus, power corrections to the Fourier coefficients will be suppressed by
$1/\cQ^2$. This gives a hope for a clean exploration of individual
contributions, provided the same settings will become available in
experiments. This can be achieved by a combining use of charge and spin
asymmetries together with a precise extraction of different Fourier
coefficients. These procedures enable a separate measurement of all
coefficients of the interference and squared DVCS term. Extraction of the
latter requires the subtraction of the squared BH amplitude from data.

For a lepton beam with a definite charge, high precision data may also
result into a separation of the interference and squared DVCS terms made
feasible by a Fourier analysis. Albeit, in this case the leading twist $\cos
(\phi)/ \sin (\phi)$ dependence in the cross section, stemming from the
interference term, gets corrected by the twist-three contribution of the
squared DVCS amplitude.

The magnitude of higher-order corrections strongly depends on the models
assumed for GPDs. The minimization of the radiative effects is possible
within the so-called DVCS scheme. It is defined by the condition of
absorption of gluonic higher-order corrections into the factorization scale,
dividing short and long distances. This procedure results into a moderate
modification of LO predictions, advocating the neglect of NLO terms for a
crude estimate. Once a more accurate data will become available, a NLO
analysis, similar to the one that is customarily performed nowadays for
deeply inelastic scattering, will be necessary. Moreover, by measuring the
scale dependence at small $\Bx$, one can get a deeper insight into the
structure of gluonic GPDs.

There are several theoretical problems that should be answered in the future.

The most crucial one is the magnitude of higher-twist corrections,
especially, of the target mass corrections. Let us mention that the
resummation of target mass effects from the twist-two operators performed in
Ref.\ \cite{BelMul01a} takes into account only a part of the entire tower of
mass corrections, which are generated also by multi-particle operators in
the off-forward Compton amplitude. They can be transformed into operators
containing total derivatives by means of QCD equations of motion. Therefore,
at any given $n$th order in the mass expansion, operators up to
twist-$n$ will contribute to the total correction. Therefore, the
resummation cannot be handled in this situation, and it is sensible to
address the problem order by order in the hadron mass.

A deeper insight into the structure of perturbative series can be achieved
by means of conformal symmetry \cite{Mue97a,BelMul97a}. This formally
requires to set the QCD $\beta$-function to zero and to choose a special
renormalization scheme, the so-called conformal scheme. With these
simplifications the Wilson coefficients of deeply inelastic scattering,
known in next-to-next-to-leading order (NNLO), allow a numerical evaluation
of the twist-two DVCS amplitude to NNLO in the valence quark region.
Moreover, the corrections proportional to the $\beta$-function have been
calculated to the same order and can, therefore, be taken into account
\cite{BelSch98}.

If experiments can measure twist-three effects, one certainly needs their
better theoretical understanding. Let us mention that the LO evolution of
the antiquark-gluon-quark GPDs is governed by known two-particle kernels.
Since the one-loop corrections to the short distance twist-two coefficient
functions are important, one could anticipate that NLO effects can be as
sizable in the twist-three amplitudes. This requires a detailed study of
radiative corrections to the latter along the line similar to the
computations done for the transversely polarized structure function $g_2$ of
deeply inelastic scattering in Refs.\
\cite{JiLuOsbSon00,BelJiLuOsb00,BraKorMan00,JiOsb01}.

After these issues are addressed, their consequences and impact on the
amplitudes have to be quantitatively analyzed in order to demonstrate the
regions of the phase space, which are less sensitive to their presence and
where an accurate determination of the parameters of the twist-two and
-three GPDs can be performed. Of course, such studies are model-dependent.
Thus, one should incorporate as much as possible information from
perturbative QCD. This requires modeling the GPDs at a low normalization
point, e.g., $\cQ_0 \sim 0.5\ \GeV$, and their evolution upwards to
experimental scales.

In the most favorable situation, which means high enough ${\cal Q}^2$
required to discard power suppressed contributions and limit the analysis to
the twist-two approximation in LO, the imaginary part of DVCS with real
photon can give access to the GPDs on the diagonal $x = \pm \xi$ only, i.e.,
it becomes a function of a single scaling variable $F (\xi, \xi, \Delta^2)$.
The only way one can measure it as a function of two variables, momentum
fraction and skewedness, is in the deeply leptoproduction (virtual
photoproduction) of a lepton pair,
\begin{eqnarray*}
N \gamma^\star \to N' \gamma^\star \to N' \ell \bar\ell \, ,
\end{eqnarray*}
since the generalized Bjorken variable and skewedness are independent in
this case, $\xi \neq \eta$.

To conclude, experimental facilities having electron and positron beams is an
ideal place to disentangle and study GPDs, and, thus,
to extract fundamental information about the spin structure of the nucleon.
However, if one of them is not available, an accurate measurement with high
momentum transfer probes will serve the purpose and will result into
systematic tests of our understanding of the quark-gluon content of the
nucleon.

%%%%%%%%%%%%%%%%%%%%%%%%%%%%%%%%%%%%%%%%%%%%%%%%%%%%%%%%%%%%%%%%%%%%%
\section*{Acknowledgements}
%%%%%%%%%%%%%%%%%%%%%%%%%%%%%%%%%%%%%%%%%%%%%%%%%%%%%%%%%%%%%%%%%%%%%

We would like to thank S. Stepanyan for providing us experimental data on
the DVCS measurement with CLAS and communications on the Jefferson Lab
experiment E-91-023, and H. Avakian for help to access the HERMES data on
the beam-spin asymmetry in DVCS. We are grateful to V. Burkert, M.
McDermott, M. Diehl, A. Freund, W.D. Nowak, A.V. Radyushkin, A. Sch\"afer,
and M. Vanderhaeghen for discussions. This work was supported by the US
Department of Energy under contract DE-FG02-93ER40762 (A.B.) and
Studienstiftung des deutschen Volkes (A.K.). A.B. would like to thank the
theory group at the University of Wuppertal for its hospitality.

%%%%%%%%%%%%%%%%%%%%%%%%%%%%%%%%%%%%%%%%%%%%%%%%%%%%%%%%%%%%%%%%%%%%%
%                           References
%%%%%%%%%%%%%%%%%%%%%%%%%%%%%%%%%%%%%%%%%%%%%%%%%%%%%%%%%%%%%%%%%%%%%

\end{document}